\documentclass[manuscript, authorversion, screen]{acmart} %

\usepackage{tabularx}
\usepackage{listings}
\usepackage[inline]{enumitem}
\usepackage[most]{tcolorbox}

\usepackage{placeins}
\usepackage{pdflscape}
\usepackage{subcaption}

\let\code\texttt

\newcommand\pipelinescale{0.9}
\newcommand\frameworkscale{0.75}
\newcommand\plotscale{1.0}
\newcommand\tablescale{1.3} %

\lstdefinelanguage{js}{
    sensitive,
    morekeywords={break,continue,delete,else,for,function,if,in,new,return,this,typeof,var,void,while,with}, %
    morekeywords={await,async,case,catch,class,const,default,do,enum,export,extends,finally,from,implements,import,instanceof,let,static,super,switch,throw,try}, %
    morekeywords={false,null,true,boolean,number,string,undefined,Array,Boolean,Date,Math,Number,String,Object}, %
    morecomment=[l]{//},
    morecomment=[s]{/*}{*/},
    morestring=[b]",
    morestring=[b]',
    morestring=[b]`,
}

\lstdefinelanguage{json}{
    sensitive,
    morekeywords={true,false},
    morestring=[b]",
}

\lstdefinelanguage{oas}{
    sensitive,
    morekeywords={summary,description,servers,parameters}, %
    morekeywords={openapi,info,paths,components,security,tags}, %
    morekeywords={title,version}, %
    morekeywords={url}, %
    morekeywords={\$ref,get,put,post,delete,patch}, %
    morekeywords={tags,operationId,requestBody,responses,deprecated,security}, %
    morekeywords={\$ref}, %
    morekeywords={name,in,required,deprecated,style,explode,schema}, %
    morekeywords={type,properties,items,enum,format,maximum,minimum}, %
    morekeywords={content,required}, %
    morekeywords={schemas,responses,requestBodies,headers,securitySchemes}, %
    morekeywords={type,name,in,scheme,bearerFormat,flows}, %
    morestring=[b]",
}

\setcopyright{acmlicensed}
\copyrightyear{2026}
\acmYear{2026}
\acmDOI{XXXXXXX.XXXXXXX}

\acmJournal{TOSEM}
\acmVolume{37}
\acmNumber{4}
\acmArticle{111}
\acmMonth{8}

\begin{document}

\title[Mitigating Errors in LLM-Generated Web API Invocations]{Mitigating Errors in LLM-Generated Web API Invocations via Retrieval-Augmented Generation and Constrained Decoding}

\author{Daniel Maninger}
\orcid{0009-0005-0649-4958}
\affiliation{
  \institution{Technische Universität Darmstadt}
  \city{Darmstadt}
  \country{Germany}
}
\affiliation{
  \institution{Hessian Center for Artificial Intelligence (hessian.AI)}
  \city{Darmstadt}
  \country{Germany}
}
\email{daniel.maninger@tu-darmstadt.de}

\author{Leon Chemnitz}
\orcid{0009-0004-9409-7794}
\affiliation{
  \institution{Pariton AI}
  \city{Berlin}
  \country{Germany}
}
\affiliation{
  \institution{Technische Universität Darmstadt}
  \city{Darmstadt}
  \country{Germany}
}
\email{leon.chemnitz@pariton.ai}

\author{Jannis Brugger}
\orcid{0000-0002-7919-4789}
\affiliation{
  \institution{Technische Universität Darmstadt}
  \city{Darmstadt}
  \country{Germany}
}
\affiliation{
  \institution{Hessian Center for Artificial Intelligence (hessian.AI)}
  \city{Darmstadt}
  \country{Germany}
}
\email{jannis.brugger@tu-darmstadt.de}

\author{Tushar Lamba}
\affiliation{
  \institution{Technische Universität Darmstadt}
  \city{Darmstadt}
  \country{Germany}
}
\email{tushar.lamba@stud.tu-darmstadt.de}

\author{Amir Molzam Sharifloo}
\affiliation{
  \institution{Technische Universität Darmstadt}
  \city{Darmstadt}
  \country{Germany}
}
\email{amir.molzam@tu-darmstadt.de}

\author{Mira Mezini}
\orcid{0000-0001-6563-7537}
\affiliation{
  \institution{Technische Universität Darmstadt}
  \city{Darmstadt}
  \country{Germany}
}
\affiliation{
  \institution{Hessian Center for Artificial Intelligence (hessian.AI)}
  \city{Darmstadt}
  \country{Germany}
}
\affiliation{
  \institution{National Research Center for Applied Cybersecurity ATHENE}
  \city{Darmstadt}
  \country{Germany}
}
\email{mezini@cs.tu-darmstadt.de}

\renewcommand{\shortauthors}{Maninger et al.}

\begin{abstract}

Integration of web APIs is a cornerstone of modern software systems, yet writing correct web API invocation code remains challenging due to complex and evolving API specifications. Although LLMs are increasingly used for code generation, previous work has empirically shown that their ability to generate correct web API integrations is limited. At the same time, mitigation techniques and their effectiveness for this setting remain insufficiently understood.

In this paper, we propose and systematically evaluate retrieval‑augmented generation~(RAG) and constrained decoding~(CD) as two complementary approaches to improving LLM‑generated web API invocation code. For RAG, we design a retriever that processes OpenAPI specifications and retrieves compact endpoint representations to inject into model prompts. For CD, we introduce an automatic translation from OpenAPI specifications to regex‑based constraints enforced during generation.

We evaluate both approaches on WAPIIBench's existing synthetic dataset and on a new real‑world dataset derived from GitHub repositories. Our results show that RAG reduces hallucinations and improves correctness when generating full API invocations but reduces it when the endpoint is already provided as it encourages the generation of unnecessary parameters. In contrast, CD reliably prevents illegal URLs, HTTP methods, and arguments and substantially improves overall correctness for both starter codes.

\end{abstract}

\begin{CCSXML}
<ccs2012>
   <concept>
       <concept_id>10011007.10011074.10011092.10011782</concept_id>
       <concept_desc>Software and its engineering~Automatic programming</concept_desc>
       <concept_significance>500</concept_significance>
       </concept>
   <concept>
       <concept_id>10010147.10010257.10010293.10010294</concept_id>
       <concept_desc>Computing methodologies~Neural networks</concept_desc>
       <concept_significance>500</concept_significance>
       </concept>
   <concept>
       <concept_id>10002951.10003260.10003304.10003306</concept_id>
       <concept_desc>Information systems~RESTful web services</concept_desc>
       <concept_significance>500</concept_significance>
       </concept>
 </ccs2012>
\end{CCSXML}

\ccsdesc[500]{Software and its engineering~Automatic programming}
\ccsdesc[500]{Computing methodologies~Neural networks}
\ccsdesc[500]{Information systems~RESTful web services}

\keywords{artificial intelligence, software engineering, large language models, code generation, web APIs, benchmarks, retrieval-augmented generation, constrained decoding}

\maketitle

\section{Introduction}
\label{sec:introduction}

Web APIs provide services and functionality through a standardized, HTTP-based interface. Developers leverage them to rapidly create applications that seamlessly integrate with a wide range of online services (powered by \emph{API integration code}). The market surrounding API integrations is experiencing explosive growth~\cite{marketsandmarkets_api_2024, postman_state_2024}. However, writing correct API integration code is a tedious and challenging task. Developers need to understand and correctly utilize a significant amount of information to be able to write an \emph{API invocation}\footnote{We differentiate between ``API \emph{integration} (code)'', which refers to the general concept of connecting an arbitrary number of different web APIs with each other through code, and ``API \emph{invocation} (code)'', which refers to the code snippet that calls a single web API. An API integration consists of one or more API invocations plus the code surrounding them.} in compliance with its specification~\cite{robillard_what_2009, rauf_systematic_2019, wickert_python_2021}. Moreover, the task of ensuring correct API usage is further complicated by the fact that APIs evolve and change over time.

Large language models~(LLMs) have great potential to boost productivity in software development, e.g., by automatically generating program code from natural language descriptions~\cite{cui_effects_2025}. However, LLMs may also hallucinate and make mistakes, raising concerns about correctness~\cite{dou_whats_2024, tambon_bugs_2025}, security~\cite{pearce_asleep_2022, perry_users_2023}, and general quality~\cite{gitclear_coding_2025} of LLM-generated code.

While the capabilities of LLMs have been investigated for many different software engineering tasks, including non-web API invocations~\cite{zhuo_pop_2023}, their aptitude for web API integrations remained unexplored, until recently, when we introduced WAPIIBench~\cite{maninger_benchmarking_2025}, a new benchmark and pipeline for evaluating LLMs on web API invocation code generation, and used it to study state-of-the-art LLMs. Our study revealed significant limitations: LLMs generate incorrect results most of the time with manyfold error types, ranging from selecting the wrong endpoint to leaving out required arguments or hallucinating entirely illegal arguments.

\paragraph{Research Gap}

Yet how the issues revealed by our previous study~\cite{maninger_benchmarking_2025} can be mitigated remains an open question. While several techniques, including retrieval-augmented generation~(RAG)~\cite{lewis_retrieval-augmented_2020}, constrained decoding~(CD)~\cite{hokamp_lexically_2017, deutsch_general-purpose_2019}, and fine-tuning, have been shown to effectively mitigate LLM generation errors in other domains, their efficacy has not been investigated for web API invocations. 
Such invocations present distinct challenges: the interface is defined externally in an OpenAPI specification rather than in the source code, the invocation must be constructed explicitly by assembling a URL, HTTP method, and parameters across multiple different locations (path, query, header, body), and the specification itself evolves independently of the client code (see also Section~\ref{subsec:openapi}). None of these challenges apply to the simpler function-call or tool-use settings for which RAG and CD have been studied.

To close this gap, in this paper, we present the results of a systematic study of the effectiveness of RAG, CD, and their combination for improving the performance of LLMs on web API invocation tasks. More specifically, we formulate and answer the following research questions:

\pagebreak

\begin{tcolorbox}[coltitle=black, colbacktitle=gray!10, colback=gray!10, boxrule=0pt, colframe=gray!10, top=2pt, bottom=2pt, left=4pt, right=4pt, breakable, pad at break=2pt]
\textbf{RQ1:} \textit{To what extent does RAG improve the correctness of LLM-generated web API invocation code?}
\end{tcolorbox}

\begin{tcolorbox}[coltitle=black, colbacktitle=gray!10, colback=gray!10, boxrule=0pt, colframe=gray!10, top=2pt, bottom=2pt, left=4pt, right=4pt, breakable, pad at break=2pt]
\textbf{RQ2:} \textit{To what extent does CD improve the correctness of LLM-generated web API invocation code?}
\end{tcolorbox}

\begin{tcolorbox}[coltitle=black, colbacktitle=gray!10, colback=gray!10, boxrule=0pt, colframe=gray!10, top=2pt, bottom=2pt, left=4pt, right=4pt, breakable, pad at break=2pt]
\textbf{RQ3:} \textit{To what extent does combining RAG and CD improve the correctness of LLM-generated web API invocation code compared to each technique in isolation?}
\end{tcolorbox}

We leave fine-tuning out of consideration for two main reasons. (a) Fine-tuning requires a corpus of training data and substantial computational resources;
it is unclear how we could obtain sufficient amounts of high-quality training data for the specific kind of web API invocation tasks we are targeting. (b) Improvements achieved by fine-tuning can quickly become obsolete when the underlying data changes; this is especially relevant in the context of our work, as API specifications are constantly evolving.

\paragraph{Methodology}

To answer Research Questions 1--3, we (a) developed RAG and CD approaches tailored to the specific requirements of web API invocation tasks, along with a combined approach designed to explore their potential synergies, and (b) evaluate their effectiveness against the respective baselines across two complementary settings.
 
In the first setting, we use the synthetic dataset from WAPIIBench~\cite{maninger_benchmarking_2025}. In the second, we evaluate on a new real-world dataset that we constructed in the context of the work presented here by mining public GitHub repositories for JavaScript files containing Axios-based web API invocations and manually converting them into evaluation tasks. To accommodate file-level code context, we extend WAPIIBench accordingly. The resulting dataset covers tasks for 11 real-world APIs, with OpenAPI specifications ranging from a few hundred to over 76,000 lines.

These two evaluation settings offer complementary strengths and limitations. The synthetic dataset provides systematic, controlled coverage of all endpoints for a curated set of APIs, but may not fully reflect the diversity of real-world integration scenarios. The real-world dataset, by contrast, captures common usage patterns of widely adopted APIs, yet is necessarily incomplete; it does not exhaustively cover all endpoints of any given API.

\paragraph{Results}

Our experiments show that RAG reduces hallucinations and can improve the correctness of generated API invocations. On the synthetic dataset, depending on the provided starter code, it increases average correctness by 113\% when generating complete API invocations but decreases it by 3\% when the endpoint is already provided. On the real-world dataset, the impact of RAG on correctness is almost negligible, but the reduction of hallucinations is consistent across both datasets. The reason for the ambivalent results on correctness is that RAG encourages models to use more parameters than required. The retriever itself is able to identify the sought-after endpoint among all available endpoints with very high precision, implying that remaining errors are largely due to the models' ability to make proper use of the retrieved information.

The study of constrained models confirms that decoding with our constraints enforces compliance with an API's specification when generating an invocation, reliably reducing the amount of illegal URLs, HTTP methods, and arguments to zero. Moreover, across models, CD significantly improves the overall correctness of the generated code, at relative gains of +209\% and +143\% on average on the synthetic dataset, depending on the provided starter code. Similar positive trends can be observed on the real-world dataset.

The combination of RAG and CD achieves high average gains in correctness---on the synthetic dataset +332\% and +111\%, depending on the starter code---and reduces hallucinations to zero, similar to CD. However, compared to CD, the gains are less consistent, as several models show a drop in correctness below the baseline.

\paragraph{Contributions}

In summary, we make the following contributions:
\begin{description}
    \item[A specialized retrieval method] for OpenAPI specifications that provides models with relevant in-context information about the used API. \textrightarrow~Section~\ref{sec:impl-rag}
    \item[An automatic constraint generator] that translates OpenAPI specifications to regex-based constraints, which ensure that generated code is specification-complaint. \textrightarrow~Section~\ref{sec:impl-cd}
    \item[A real-world dataset] to validate and complement the results on WAPIIBench's original synthetic dataset. \textrightarrow~Section~\ref{sec:dataset}
    \item[A systematic empirical evaluation] of the ability of LLMs to generate correct and hallucination-free web API invocation code with retrieval-augmented generation and constrained decoding. \textrightarrow~Section~\ref{sec:evaluation}
\end{description}

Artifacts are available on GitHub\footnote{\url{https://github.com/stg-tud/WAPIIBench}}, including the RAG and CD implementations presented in this paper as well as the new real-world dataset. In addition, we provide all model-generated codes and evaluation results on Zenodo\footnote{\url{https://doi.org/10.5281/zenodo.13758414}}. Additional details and comprehensive result tables are provided in the appendix at the end of this paper.

\section{Background}
\label{sec:background}

In the following, we will introduce some concepts fundamental to our work.

\subsection{Web APIs and OpenAPI}
\label{subsec:openapi}

\begin{lstlisting}[caption={General structure of an API invocation using Axios and JavaScript}, label=lst:invocation-example, language=js, float]
const axios = require('axios');
axios.<method>('https://server.com/path/to/endpoint/arg1', {
    arg2: 'request body parameters'
}, {
    headers: {
        arg3: 'header parameters'
    },
    params: {
        arg4: 'query parameters'
    }
});
\end{lstlisting}

Unlike local function calls, web API invocations require constructing an HTTP request rather than invoking a named function. APIs expose \emph{endpoints}, identified by the combination of an HTTP method and a URL, and requests may include parameters in multiple locations, such as the path, query string, headers, and request body. As illustrated in Listing~\ref{lst:invocation-example}, this leads to more complex invocation structures with multiple parameter groups, nested data types, and externally defined interfaces.

Most modern web APIs are described using the \emph{OpenAPI} specification\footnote{\url{https://www.openapis.org/}}, the de facto standard for machine-readable API documentation~\cite{smartbear_state_2020}. OpenAPI specifications, written in JSON or YAML, describe available endpoints, their parameters, request and response schemas, authentication requirements, and other metadata. In contrast to local function signatures, this information is maintained separately from the client code and must be interpreted to construct valid requests.

This paper focuses on JavaScript, one of the most widely used programming languages~\cite{statista_most_2024}, and the Axios library\footnote{\url{https://axios-http.com/}}, a popular HTTP client for JavaScript~\cite{devographics_state_2022}. The evaluation framework, WAPIIBench, is built around this setup and represents web API invocations as Axios requests to endpoints defined through OpenAPI specifications.

\subsection{WAPIIBench}
\label{subsec:wapiibench}

\begin{figure*}
    \centering
    \includegraphics[width=\pipelinescale\textwidth]{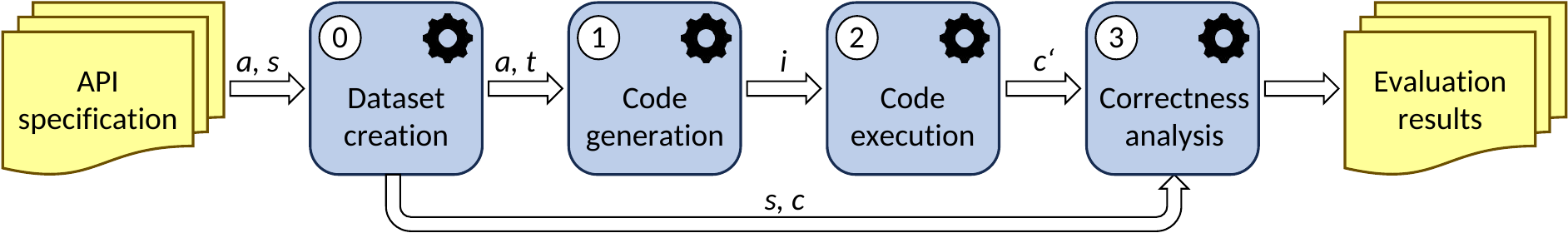}
    \Description{Flow diagram visualizing the evaluation pipeline described in the caption.}
    \caption{Benchmark design for evaluating the capabilities of LLMs in generating web API invocation code. (0)~Based on an API~$a$ and its specifications~$s$, an API invocation tasks~$t$ and corresponding correct request configurations~$c$ are created. (1)~For each $t$, the LLM under evaluation generates an API invocation~$i$. (2)~$i$ is executed in a controlled environment, yielding a request configuration~$c'$. (3)~$c'$ is compared to $c$ and validated against $s$ to obtain various metrics.}
    \label{fig:eval-pipeline}
\end{figure*}

For our studies, we rely on WAPIIBench, a benchmark for web API integration code generation that we previously introduced~\cite{maninger_benchmarking_2025}. It includes a dataset of web API invocation tasks and an execution-based evaluation pipeline. A schematic overview is provided in Figure~\ref{fig:eval-pipeline}.

WAPIIBench's dataset comprises 395 tasks across four real-world APIs: Asana, Google Calendar, Google Sheets, and Slack (one task for each endpoint). Each task includes a natural language \emph{task description} and a \emph{request configuration}, which is used as the ground-truth solution of the task and contains the expected URL, HTTP method, and arguments. An example task description can be found in Figure~\ref{lst:task-synth}. The dataset was synthetically generated based on the OpenAPI specifications of the respective APIs and then subjected to rigorous automated and manual vetting and correction process. This way, we ensured that each task is well-defined and unambiguously solvable.

The evaluation pipeline has three stages:
\begin{enumerate*}
    \item \emph{Code generation},
    \item \emph{Code execution}, and
    \item \emph{Correctness analysis}.
\end{enumerate*}

(1) In the code generation stage, we insert the API name and the task description (as a line comment) into a prompt that instructs the model under evaluation to solve the given task by completing the starter code that is part of the prompt (cf. Listing~\ref{lst:prompt-synth} in the appendix). The prompt also includes some clarifications to avoid instances of misalignment and recurring error patterns observed in preliminary experiments.
Experiments can be conducted with two different starter code variants, called \emph{setups}. The \emph{full completion} setup includes the beginning of the API invocation (\code{axios.}). It is used to assess the ability of the model to identify the correct endpoint (i.e., the combination of URL and HTTP method) and to solve the task as a whole. The \emph{argument completion} setup already includes the correct endpoint (\code{axios.<method>('<url>', }) and aims to evaluate whether the model can use this endpoint in compliance with the API's specification by selecting only permitted parameters.

(2) In the code execution stage, the model-generated API invocation is cut out of the surrounding code and inserted into a controlled environment, which enables the safe execution the generated API invocation without sending requests to external servers. Additionally, this environment serializes and saves the request's configuration, so it can be used in the following correctness analysis.

(3) In correctness analysis stage, the captured request configuration is compared to the ground-truth configuration and validated against the corresponding API specification. The analysis happens in element-wise and aggregated forms to obtain various fine-grained metrics explained in Table~\ref{tab:metrics-full} in the appendix.

\subsection{Retrieval-Augmented Generation}
\label{subsec:background-rag}

Retrieval augmented generation~(RAG)~\cite{lewis_retrieval-augmented_2020} is a commonly used technique to improve model performance on knowledge-intensive tasks. The general idea is to search a knowledge base or collection of documents for relevant pieces of information, so-called \emph{chunks}, and augment the prompt with them before generating a response. This way, the model does not need to rely solely on memorized knowledge and can base its response on the retrieved information, which enhances reliability. However, the effectiveness of RAG depends on the retriever's ability to find relevant chunks, as well as on the model's ability to leverage the information from the retrieved chunks.

RAG can be implemented in many ways. The main degrees of freedom are: how documents are split into chunks, which embedding model is used to store chunks in a vector database, which algorithm is used for retrieving relevant chunks from the vector database, and how many chunks to retrieve and include in the model's prompt. Commonly, a larger amount of chunks is retrieved and a so-called \emph{reranker} model is used to determine the top chunks to pass on to the generator model.

\subsection{Constrained Decoding}
\label{subsec:background-cd}

Constrained decoding~(CD)~\cite{hokamp_lexically_2017, deutsch_general-purpose_2019} (also known as \emph{structured} or \emph{guided} generation) is a general technique that modifies an LLM's next-token predictions based on predefined constraints to prevent the generation of tokens that would lead to undesirable output sequences. This technique operates without requiring model modifications or prompt adjustments. How constraints are obtained is highly application-specific. For representing constraints, regular expressions or context-free grammars are commonly used~\cite{willard_efficient_2023, beurer-kellner_guiding_2024, dong_xgrammar_2025}.

The general constrained decoding algorithm (cf.~\cite{poesia_synchromesh_2022, ugare_syncode_2025, mundler_type-constrained_2025} extends an LLM's regular autoregressive generation loop with a so-called \emph{completion engine}~\cite{poesia_synchromesh_2022, wei_copiloting_2023, mundler_type-constrained_2025}, which dynamically filters possible next token to enforce compliance with given constraints. More concretely, in each generation step, the current token sequence $s_1, \dots, s_i$ and the token vocabulary $t_1, \dots, t_n$ are passed to the completion engine, which checks for each token $t_j$ whether $s_1, \dots, s_i, t_j$ satisfies the constraints and would therefore be a valid continuation. The completion engine returns a token mask $m \in \{0, 1\}^n$ which is used to set the probability $p_j$ of all tokens $t_j$ to zero that failed the constraint check. Based on the masked probability distribution, the next token $s_{i+1}$ is sampled. Note that this process can be optimized in many ways~\cite{poesia_synchromesh_2022, willard_efficient_2023, ugare_syncode_2025, dong_xgrammar_2025}.

\section{Retrieval-Augmented Generation for Web API Invocations}
\label{sec:impl-rag}

\begin{figure*}
    \centering
    \includegraphics[width=\frameworkscale\textwidth]{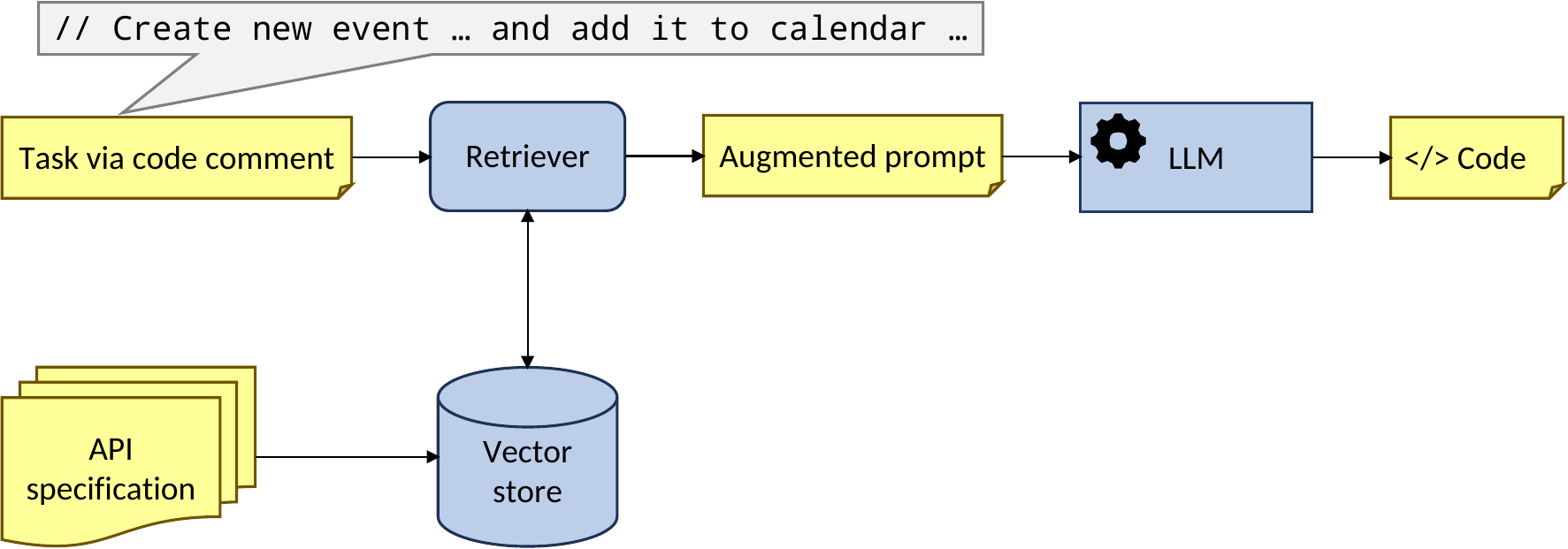}
    \Description{Flow diagram visualizing the retrieval-augmented generation framework described in the caption.}
    \caption{Retrieval-augmented generation framework for generating web API invocations. Instead of passing the prompt directly to the language model, it is first augmented with information about potentially relevant API endpoints. Chunks of information are extracted from the API specification and persisted in a vector store. The retriever performs a similarity search to find chunks that match the given task description.}
    \label{fig:rag-framework}
\end{figure*}

In order to apply RAG to web API invocations, we retrieve chunks from OpenAPI specifications using the respective task description as query, as shown in Figure~\ref{fig:rag-framework}. While RAG systems typically split documents into chunks based on a predetermined length or sentence boundaries, our chunks encompass the complete documentation of a single endpoint. We do this because API specifications are highly context-dependent---retrieving smaller chunks out of context would be detrimental. For example, the documentation of a parameter is useless if it is unclear which endpoint it belongs to. To ensure each chunk is self-contained and contains all relevant information, we preprocess OpenAPI specifications prior to chunking, inlining all references and copying API-wide security schemes and path-level parameters into each endpoint's parameter list.

To embed, store, and retrieve chunks, our retriever uses the \code{all-MiniLM-L6-v2}\footnote{\url{https://huggingface.co/sentence-transformers/all-MiniLM-L6-v2}} embedding model, a Chroma vector database, and approximate nearest neighbor search, respectively. To increase the hit rate, each chunk is stored with up to five different embeddings, created by including and excluding different parts of the endpoint documentation in the string that is fed into the embedding model. For example, one variant contains only the endpoint's path and description, while other variants additionally contain the parameter list or the response schema. Across all tasks in the dataset, our retriever achieves 75.7\% top-1 accuracy and 95.2\% top-5 accuracy (after deduplicating chunks). We also implemented reranking but did not use it, because it could not improve the retriever performance any further.

Before injecting the retrieved endpoint documentation into the prompt, it is formatted and truncated in a postprocessing step. The goal of the truncation is to avoid exceeding a model's context length. We cut the output before the line in which the character threshold is exceeded and append a \code{[TRUNCATED]} marker. To reduce the need for truncation, we format the documentation in a way that is compact but also easy to understand for LLMs. We explored different formatting options such as JSON and YAML, but they turned out to be token-inefficient due to a lot of indentation and verbose constructs such as \code{name: <name>}, where simply stating the parameter name would be sufficient. Instead, we designed a custom format inspired by TypeScript type declarations, as shown in Listing~\ref{lst:chunk-example}.

\begin{lstlisting}[caption={Example of a retrieved endpoint chunk in a TypeScript-inspired format}, label=lst:chunk-example, language=js, float]
// GET /tags/{tag-name}/media/recent
// Get a list of recently tagged media
type GetTagsMediaRecentRequest = {
  path: {
    "tag-name": string // The tag name
  }
  query: {
    count?: number // Count of tagged media to return
    min_tag_id?: string // Return media before this `min_tag_id`
    max_tag_id?: string // Return media after this `max_tag_id`
    access_token?: string // API key security scheme
  }
  header: {
    Authorization?: string // OAuth2 security scheme
  }
}
\end{lstlisting}

\section{Constrained Decoding for Web API Invocations}
\label{sec:impl-cd}

\begin{figure*}
    \centering
    \includegraphics[width=\frameworkscale\textwidth]{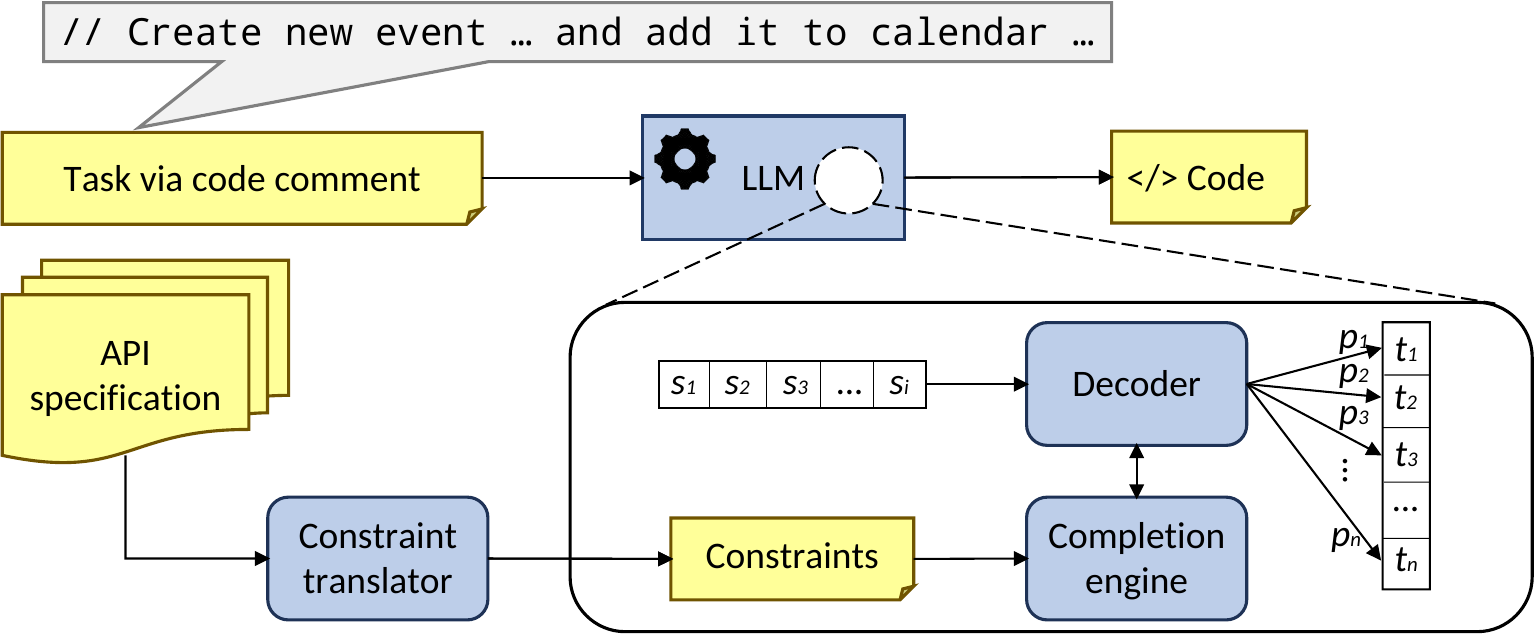}
    \Description{Flow diagram visualizing the constrained decoding framework described in the caption.}
    \caption{Constrained decoding framework for generating web API invocations. In regular autoregressive generation, a decoder takes the token sequence $s_1, \dots, s_i$ and predicts the probabilities $p_j$ for each possible next token $t_j$, based on which $s_{i+1}$ is selected. Constrained decoding augments this loop with a constraint check that sets the probability of tokens that fail the check to zero. We derive our constraints from OpenAPI specifications, and they ensure that generated API invocations comply with the specification.}
    \label{fig:cd-framework}
\end{figure*}

The main challenge we have to address when applying constrained decoding in our setting is constraint generation. Since every web API requires different constraints, we implement an automatic translation from OpenAPI specifications to constraints that are represented as regular expressions. When generating the constraints, the special characteristics and complexities of web API invocations need to be taken into account. Besides respecting the general JavaScript and the specific Axios syntax (cf. Listing~\ref{lst:invocation-example}), our constraints need to handle interdependencies between allowed URLs, HTTP methods, and different kinds of parameters. Moreover, the number, ordering, and nesting of parameters can be arbitrary, while some parameters are required and some are optional, making advanced constraint formulations necessary.

We implemented a simple, model-agnostic constrained decoding framework for our investigations, which is depicted in Figure~\ref{fig:cd-framework}. The constraints are automatically derived from OpenAPI specifications and represented as regular expressions (regex). They are designed to ensure that generated API invocations are always compliant with the API's specification. Note that our contribution here lies in this transformation of specifications into constraints, and not the completion engine, which is a straightforward implementation of the standard CD algorithm sketched in Section~\ref{subsec:background-cd}. We developed a custom completion engine to maintain full control over the system, explore features not supported by existing completion engines (e.g., advanced regex features and constraint orchestration as described below), and add some debugging functionality. For professional use cases, migrating to an off-the-shelf engine (cf. Section~\ref{subsec:rw-cd}) would be expedient to enhance performance and robustness.

A main challenge when translating OpenAPI specifications to regular expressions in a generalizable and scalable way is the sheer size and complexity of real-world APIs, which is also reflected in the respective specification files. Each API can have dozens if not hundreds of endpoints (in the specifications we worked with, up to 523 in the GitHub API), and each endpoint can have dozens of parameters, distributed on different locations (body, header, path, query). Parameters can be optional or required, and they can be passed in arbitrary order. Additionally, many parameters have complex data types (objects and arrays) with a strictly defined nesting structure. There are also interdependencies between different parts of an API invocation, e.g., only for certain HTTP methods, parameters can be placed in the request body. In summary, some elements of an API invocation must be followed strictly, while other elements are flexible.

Additional challenges arise at the syntax-level of the language the API invocation is implemented in. Our regexes should accept valid JavaScript while not limiting variations permitted by the language, such as formatting or the use of different quotation marks. This is important as different LLMs show different stylistic preferences---forcing them away from their preferred style may degrade performance. There are also a lot of edge cases that need to be handled carefully by the constraints, e.g., an escaped quotation mark must not terminate a string literal, and many other seemingly minor details---our experiments showed that models will find and exploit such loopholes.

We address the challenges by structuring constraints hierarchically from the HTTP method and the URL to the different parameters. Once the method is matched, the search space is reduced to URLs for which an endpoint with the respective method is defined. Next, once the URL is matched, the search space is reduced to the parameters of the identified endpoint. Finally, once the parameter name is matched, the search space is reduced to values with the right data type. Within the parameter lists of a given endpoint, we use capture groups\footnotemark{} to keep track of parameters that have been defined. Conditional patterns\footnotemark[\value{footnote}] are then used to check the existence of a specific group and thus guard against the double definition of a parameter. Moreover, conditionals are used in conjunction with negative lookahead assertions\footnotemark[\value{footnote}] to make it impossible to close a parameter list before all required parameters have been defined.\footnotetext{\url{https://docs.python.org/3/library/re.html}}

While we could represent all our constraints as one monolithic regex, we opted for a more modular and flexible approach\footnote{They could still be joined if needed for compatibility with other frameworks.}. We organize constraints as a set of so-called \emph{generation rules}. Each generation rule has a regex-encoded \emph{start condition}, \emph{stop condition}, and \emph{body}. Once the start condition matches the code generated so far, we constrain the generation according to the body until the stop condition matches.

There is one specialized generation rule for each HTTP method, which is triggered when the corresponding Axios invocation is generated (e.g., \code{axios.post(}, with the opening parenthesis being part of the trigger). As multiple generation rules may be active at the same time (a token is only admissible if no rule rejects it), we can leverage overlapping constraints. We do so via a special generation rule that becomes active before all other rules and guides the generation towards the start condition of the other rules, ensuring one of them is going to be activated (thus avoiding using methods not supported by the given API). Moreover, we let generation rule bodies overlap with their stop conditions to ensure they properly terminate. This approach can also be generalized to many other applications of constrained decoding to orchestrate multiple constraints in a flexible and efficient manner.

Given the breadth of API designs that can be described through OpenAPI specifications plus the complexity of the resulting constraints, we implemented a test suite to ensure that (a)~any specification documents can be processed and translated to constraints and (b)~resulting constraints allow and disallow exactly the structures they are supposed to. For (a), we successfully parse 34 real-world OpenAPI specifications, including those used in the synthetic and real-world dataset as well as additional ones. For (b), we created 125 test cases, comprising positive and negative code examples, that our regex constraints must, respectively, match or not match---all these test cases pass. Further implementation details about our constraints are discussed in Appendix~\ref{app:cd-details}.

\section{Real-World Dataset Creation}
\label{sec:dataset}

WAPIIBench, introduced in our previous work~\cite{maninger_benchmarking_2025}, is based on a synthetic dataset. This dataset enables controlled, large-scale experiments with precise ground truth. However, it may not fully capture the range of real-world API integration scenarios. In this paper, we therefore complement the original synthetic benchmark with a newly constructed dataset derived from real-world repositories. 

To construct the new dataset we addressed several challenges that were left open in our earlier work. Real-world code rarely contains sufficient information to serve directly as an unambiguous task description, which makes task reconstruction labor-intensive and limits the number of tasks that can realistically be included. Moreover, even when a plausible task can be identified, manual inspection is still necessary to determine whether the code is correct, or at least complete and plausible, which parts are relevant to the task, and which API version it assumes. 

The extension of WAPIIBench with the real-world dataset allows us to evaluate in two complementary settings. The synthetic dataset provides comprehensive and systematic coverage of all endpoints for a limited number of APIs under tightly controlled conditions. However, it may not reflect the real-world web API integration scenarios. In contrast, being derived from repository code, the tasks in the real-world dataset reflect common use cases of widely used APIs. The real-world dataset provides less comprehensive---unlike the synthetic dataset, it is necessarily incomplete and does not cover all endpoints of a given API---but more diverse and realistic coverage, including file-level code context of varying complexity.

\begin{table}
    \centering
    \caption{Number of tasks per API in the real-world dataset}
    \label{tab:api-dist-real}
    \begin{tabular}{lr}
        \toprule
        API & \# tasks \\
        \midrule
        Etherscan & 3 \\
        FrankerFaceZ & 2 \\
        GitHub & 8 \\
        Google Maps Platform & 2 \\
        Instagram & 2 \\
        JSONPlaceholder & 3 \\
        npm Registry & 2 \\
        Slack Web & 1 \\
        Telegram Bot & 1 \\
        YouTube Data & 2 \\
        Zephyr Cloud & 2 \\
        \midrule
        Total & 28 \\
        \bottomrule
    \end{tabular}
\end{table}

To derive the real-world dataset, we queried BigQuery\footnote{\url{https://cloud.google.com/bigquery/}} for GitHub JavaScript files that import the Axios library, yielding an initial sample of 277 files. We manually screened each file to identify web API invocations suitable for being converted into a task. We excluded files if they lacked actual Axios invocations, obscured the server address (making the target API unidentifiable), or interacted with APIs lacking an OpenAPI specification. Finally, to prevent skewing the dataset’s API distribution, we partially excluded two files---effectively web API wrapper libraries---that contained disproportionately high numbers of invocations to a single API.
After filtering, 18 files remained, from which we constructed a dataset consisting of 28 tasks (some files yielded more than one). The dataset spans 11 APIs, with the task distribution shown in Table~\ref{tab:api-dist-real}.
All retrieved code files plus the reason for exclusion and other notes can be found in our artifact. 

\begin{figure}
    \centering
    \begin{subfigure}{0.48\textwidth}
        \begin{lstlisting}[language=js]
// Save the project with gid '778899' as a template with the name 'New Project Template' and make it public to its team.

const axios = require('axios');

axios.
        \end{lstlisting}
        \caption{Synthetic dataset}
        \label{lst:task-synth}
    \end{subfigure}
    \hfill
    \begin{subfigure}{0.48\textwidth}
        \begin{lstlisting}[language=js]
const axios = require('axios');
const User = require('../users/userModel').User;

module.exports = {
  getUniqueTagPics(req, res) {
    const hashtag = req.body.hashtag;
    User.where({ id: req.body.userId })
    .fetch()
    .then((user) => {
      if (!user) {
        res.status(404).send('User not found');
      } else {
        const accessToken = user.get('instagram_token');
        if (accessToken) {
          // Get a list of media recently tagged with the value in `hashtag`. Authenticate with the access token in `accessToken`.
          axios.
        \end{lstlisting}
        \caption{Real-world dataset}
        \label{lst:task-real}
    \end{subfigure}
    \caption{Comparison of a task and starter code from the synthetic dataset~\cite{maninger_benchmarking_2025} to a task and starter code from the real-world dataset (Sec.~\ref{sec:dataset}). The former targets the Asana API, the latter the Instagram API. The task is included in the starter code as a line comment.}
    \Description{Two JavaScript code listings.}
    \label{fig:task-comparison}
\end{figure}

To convert a file into one or more tasks we performed the following manual steps: localizing the API invocation in the source code; identifying the corresponding endpoint in the API specification; authoring a natural-language task description; creating a ground-truth request configuration; preparing starter code grounded in the original code context; and capturing relevant definitions within that starter code. An example task is shown in Listing~\ref{lst:task-real}; 

\emph{Starter Code Preparation:} By default, the starter code is a slice of the original file from its beginning up to the API invocation site, giving the model realistic file-level context\footnote{The length of starter codes in the dataset varies between 4 and 163 lines.}. However, a few targeted adjustments are necessary. (a)~Removing giveaways: If the file defines constants such as API keys or the request URL, those definitions are stripped out; leaving them in would hand the model part of the answer. (b)~Removing similar calls: Code that closely resembles the target invocation (e.g., other calls to the same API) is also removed, since it would encourage the model to copy rather than reason.

\emph{Handling Variables in Real-World Tasks:} Unlike the synthetic dataset where all API parameters are plain literals, the real-world tasks use file-level starter code, meaning some parameters are passed as variables or constants. This introduces two challenges.

(a) Injecting definitions at execution time. Because our pipeline only executes the isolated API invocation (not the surrounding starter code), any definitions that must be in scope are captured in advance and injected into the execution environment. For variables with statically unknown values, such as function parameters, we generate random instances of the appropriate data type and schema. This prevents the model from guessing a literal value and forces it to reference the correct variable. These random values are used only during execution and correctness analysis (cf. Section~\ref{sec:background}); the model never sees them during code generation.

(b) Handling unexpected variable references. A model may attempt to use variables from the starter code that were not part of the intended solution. Without those definitions present, execution would throw a \code{ReferenceError}. Since we would rather count such cases as an incorrect argument value than a non-executable implementation, we iteratively expand the set of injected definitions until all remaining \code{ReferenceError} in the execution logs stem exclusively from genuinely hallucinated identifiers.

Beyond handling task-specific starter codes and variable definitions, we extended the evaluation pipeline as well as the retriever and constraint generator to support a larger set of OpenAPI features present in the specifications of the APIs in the new dataset.

For instance, to support \emph{server variables} (similar to path parameters but in the URL's server part), we needed to (a)~normalize model-generated URLs before matching them against the specification to find the endpoint they selected, (b)~merge the list of server variables with the list of the endpoint's path parameters, (c)~extract the values of all server variables and path parameters from the URL, and (d)~add the resulting name--value pairs to the request configuration, so they can be properly analyzed later in the evaluation pipeline.

Moreover, the extraction of Axios invocations from the generated codes in preparation of executing them was reworked to remove any assumptions about the structure of the starter code, as those assumptions would not hold for the real-world dataset.

Finally, since many tasks in the new dataset require passing variables from the starter code as path parameters, we implemented constraints that allow inserting variables into URL strings via string interpolation as well as concatenation.

The outlined extensions have made WAPIIBench much more flexible w.r.t. task structure and robust w.r.t. API-specific peculiarities, facilitating reuse and further development. Adding a new task to a dataset is as simple as adding a new entry to a JSON file---it will automatically be evaluated together with the previously existing tasks. Adding an entirely new datasets requires less than 70 lines of code (mostly boilerplate), as evidenced by our script to run the evaluation with the new real-world dataset. If a new task or dataset involves a new API, its OpenAPI specification has to be placed in a dedicated directory and a mapping from the API's file name to the API's proper name has to be added to an existing dictionary.

\section{Evaluation}
\label{sec:evaluation}

This section covers our experimental setup and results for different generation settings as well as a discussion about their efficiency and answers the research questions formulated in Section~\ref{sec:introduction}.

\subsection{Experimental Setup}
\label{sec:setup}

To assess their efficacy for web API invocation code generation and answer the research questions posed in Section~\ref{sec:introduction}, we put our retriever (Section~\ref{sec:impl-rag}) and constraints (Section~\ref{sec:impl-cd}) into action using WAPIIBench's dataset and evaluation pipeline (Section~\ref{subsec:wapiibench}) as well as the new real-world dataset (Section~\ref{sec:dataset}).

\paragraph{Experiment Variations}

We evaluate four different \emph{generation settings}:
\begin{description}[noitemsep]
    \item[Vanilla] Generation based only on the prompt and starter code
    \item[RAG] Retrieval-augmented generation
    \item[CD] Constrained decoding
    \item[RAG + CD] Retrieval-augmented generation combined with constrained decoding
\end{description}

Furthermore, conduct all experiments with two different starter code variants, a.k.a. \emph{setups} (cf. also Section~\ref{subsec:wapiibench}):
\begin{description}[noitemsep]
    \item[Full completion] The model has to select the correct endpoint (URL plus HTTP method) as well as request arguments in order to solve the given task
    \item[Argument completion] The correct endpoint is already provided, the model has to select the correct arguments in order to solve the given task
\end{description}

\begin{table}
    \centering
    \caption{Evaluated LLMs and their sizes}
    \label{tab:models}
    \begin{tabular}{ll}
        \toprule
        Model & Size \\
        \midrule
        CodeT5+~\cite{wang_codet5_2023} & 6B, 16B \\
        StarCoder~\cite{li_starcoder_2023} & 1B, 3B, 7B, 15.5B \\
        StarCoder2~\cite{lozhkov_starcoder_2024} & 3B, 7B, 15B \\
        DeepSeek-Coder~\cite{guo_deepseek-coder_2024} & 1.3B, 6.7B, 33B \\
        Qwen2.5-Coder~\cite{hui_qwen25-coder_2024} & 0.5B, 1.5B, 3B, 7B, 14B, 32B \\
        Llama~3.1~\cite{dubey_llama_2024} & 8B, 70B \\
        Code Llama~\cite{roziere_code_2023} & 7B, 13B, 34B, 70B \\
        GPT-4o mini~\cite{openai_gpt-4o-mini_2024} & N/A \\
        GPT-4o~\cite{openai_gpt-4o_2024} & N/A \\
        \bottomrule
    \end{tabular}
\end{table}

\paragraph{Models}

We study the performance of 24 state-of-the-art open-source LLMs, listed in Table~\ref{tab:models}.
These models were selected due to their frequent appearance in coding leaderboards\footnote{E.g., \url{https://evalplus.github.io/leaderboard.html}}. To obtain an approximate upper bound on performance, we also evaluated OpenAI's commercial GPT-4o and GPT-4o mini models. As they are closed-source, we cannot apply constrained decoding to the GPT models. Due to a model-internal bug, Code Llama~(34B) cannot use constrained decoding either.

For all experiments, greedy decoding is used to ensure deterministic and reproducible results. In the RAG experiments, five endpoint chunks are retrieved per task. If the total length of the retrieved text exceeds 8000 characters, it is truncated. This threshold is set to strike a balance between supporting models with shorter context lengths and truncating as few retriever outputs as possible. Only for CodeT5+, we have to reduce the threshold to 4000 to avoid inference-time crashes and skip some tasks from the real-world dataset as their starter code alone already exceeds the model's maximum context length (we count this as a generation-time error). The prompts used for the synthetic and the real-world tasks are given in Listing~\ref{lst:prompt-synth} and~\ref{lst:prompt-real} in the appendix.

\paragraph{Metrics}

We focus our discussion on the following four key metrics; a multitude of additional metrics is provided in Appendix~\ref{app:results}:
\begin{description}[noitemsep]
    \item[Correct implementations] Ratio of generated codes that are executable and correctly solve the given task, i.e., their request configuration matches the ground-truth
    \item[Hallucinated endpoints] Ratio of generated codes that use a combination of URL and HTTP method that is not defined in the API's specification\footnote{While we use the intuitive term ``hallucinated'' for this error category, it subsumes any mismatch with the API specification, including cases where the model's intention may be correct, but its spelling is incorrect.} (only applicable to full completion)
    \item[Hallucinated implementations] Ratio of generated codes that use at least one request parameter that is not defined in the API's specification (only applicable to argument completion)
    \item[Executable implementations] Ratio of generated codes that are executable, i.e., complete, syntactically correct, and free of runtime errors
\end{description}
Here, all metrics will be reported relative to the total number of tasks in the respective dataset. Metrics relative to only the tasks that resulted in executable code are provided in the appendix. For a fair comparison between generation settings compatible and incompatible with closed-source models, we always exclude the GPT-4o models when aggregating metrics. Otherwise, vanilla and RAG would get an advantage over CD and RAG~+~CD due to the contribution of the powerful GPT-4o models to the aggregate values. Additionally, we exclude models from average calculations if they did not generate any executable implementation.

\begin{figure}
    \centering
    \includegraphics[width=\plotscale\linewidth]{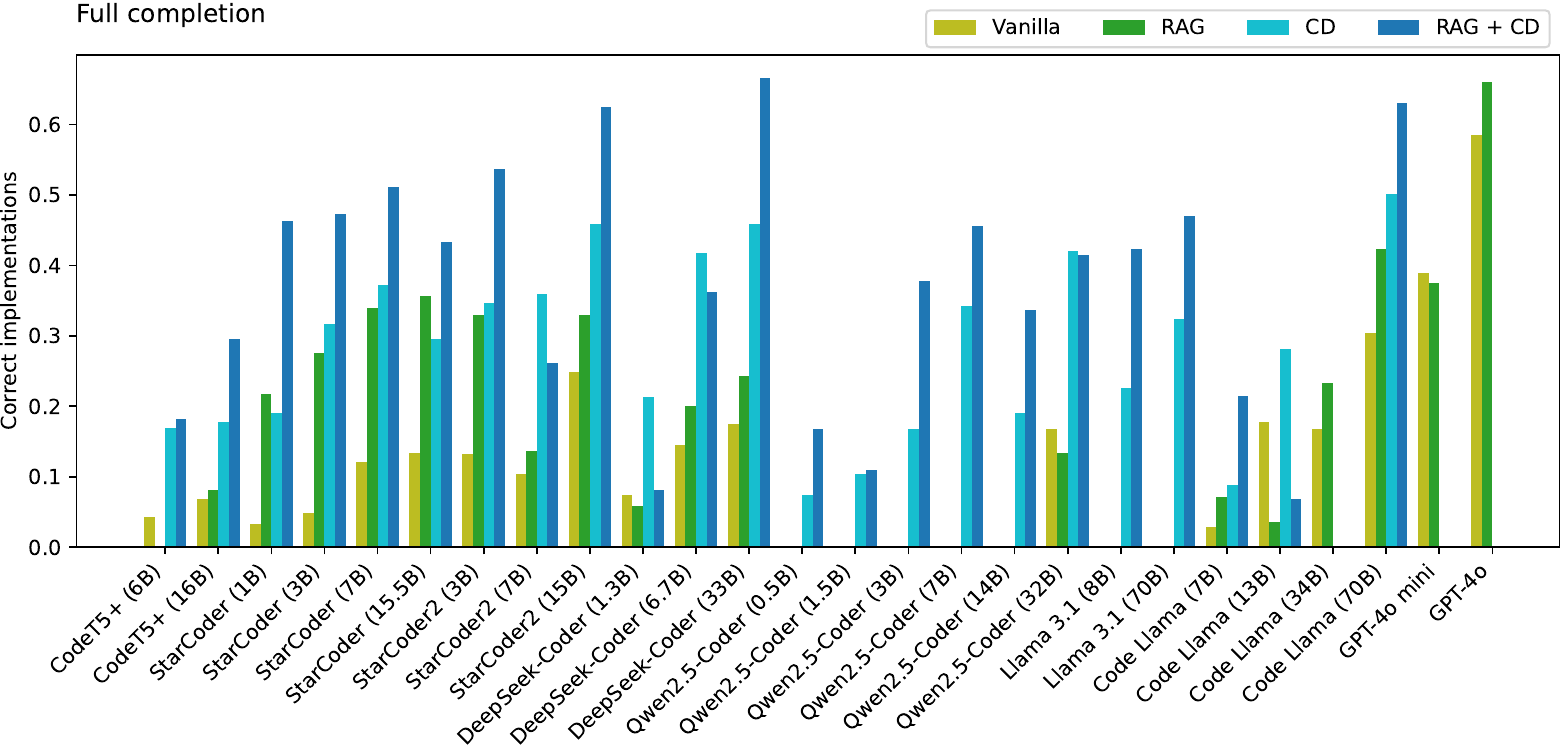}
    \includegraphics[width=\plotscale\linewidth]{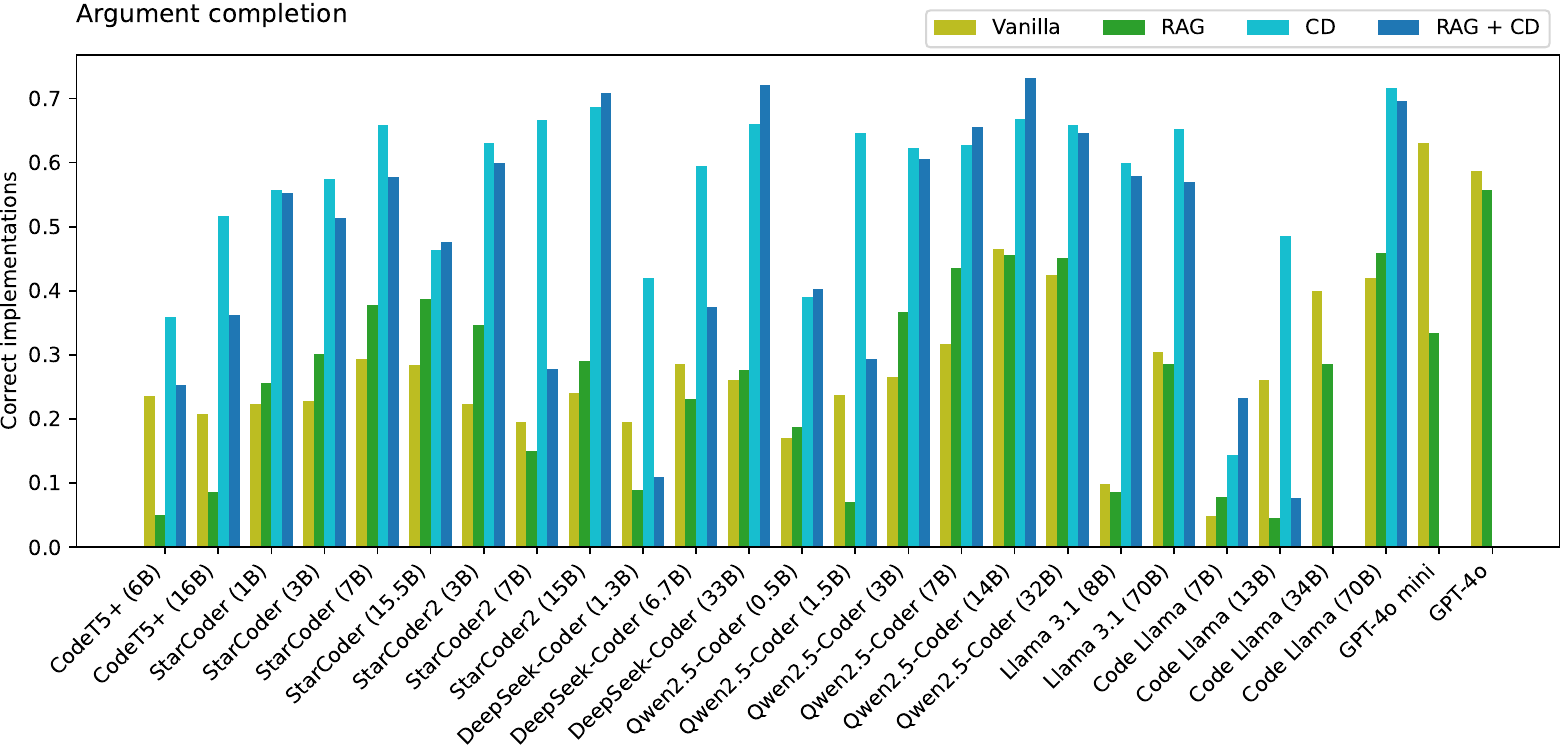}
    \Description{Grouped bar chart visualizing the performance difference between vanilla generation, RAG, constrained decoding, and RAG + constrained decoding.}
    \caption{Comparison of \emph{correct implementations} between generation settings on the \emph{synthetic dataset}}
    \label{fig:res-synth-correct}
\end{figure}

\begin{figure}
    \centering
    \includegraphics[width=\plotscale\linewidth]{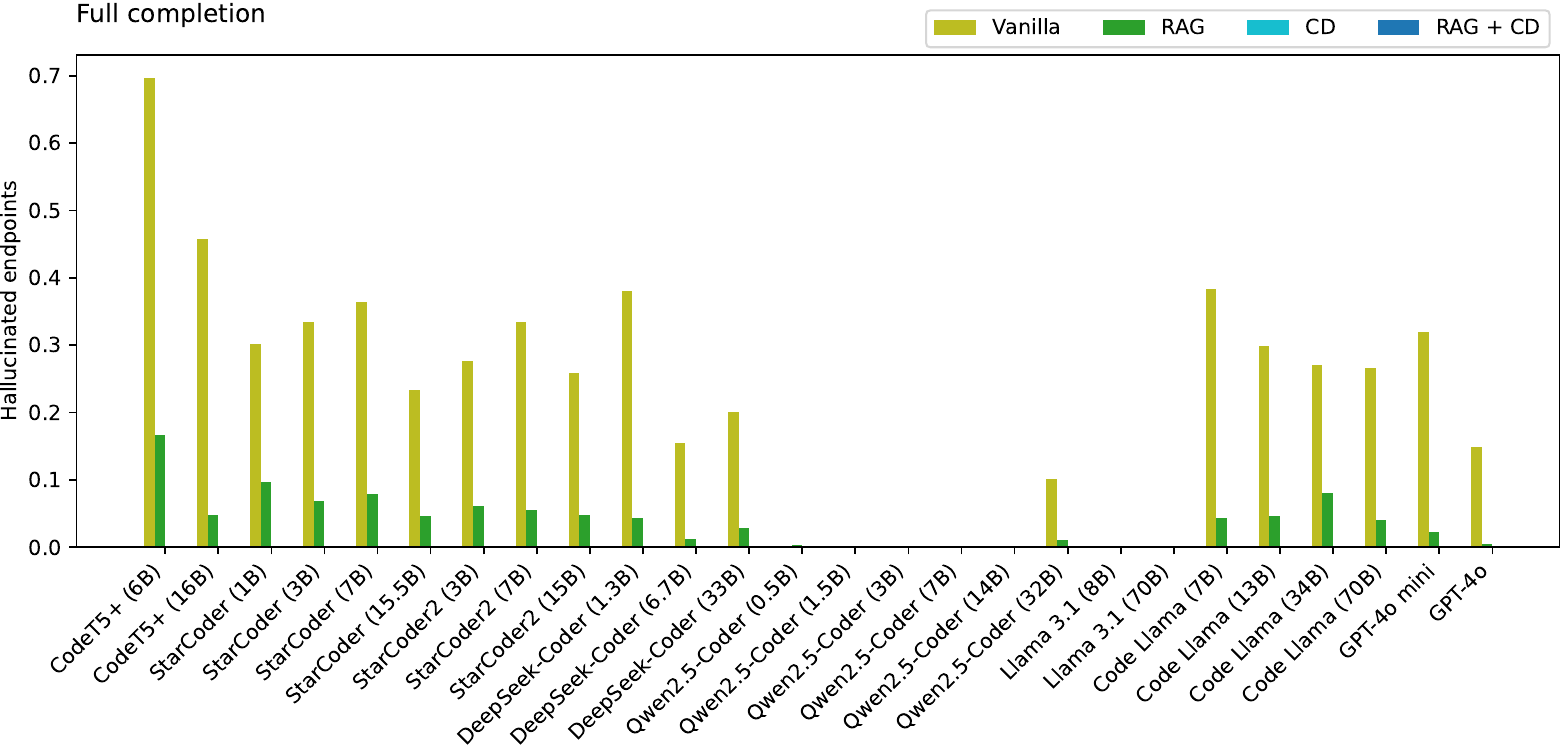}
    \includegraphics[width=\plotscale\linewidth]{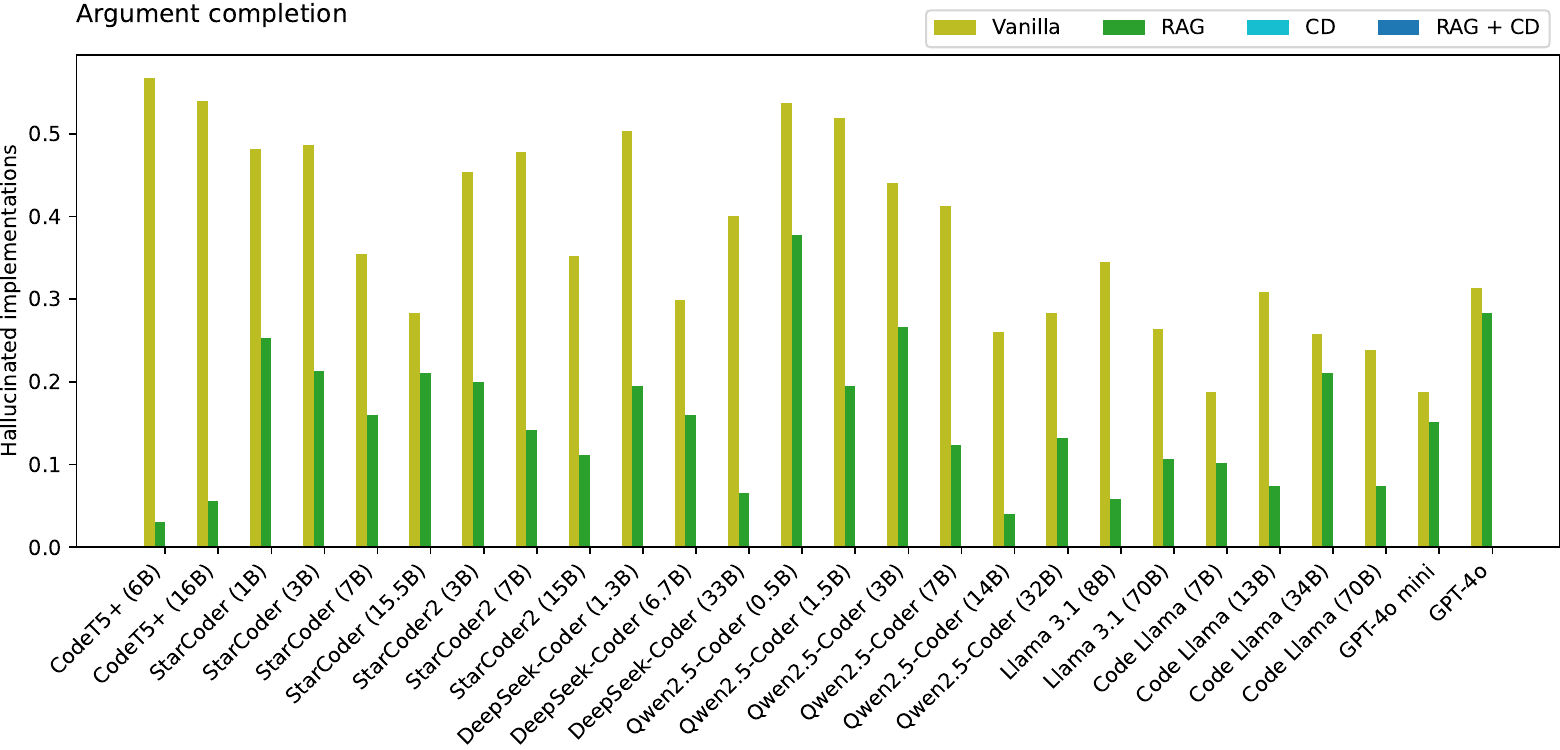}
    \Description{Grouped bar chart visualizing the performance difference between vanilla generation, RAG, constrained decoding, and RAG + constrained decoding.}
    \caption{Comparison of \emph{hallucinated endpoints} between generation settings on the \emph{synthetic dataset}}
    \label{fig:res-synth-illegal}
\end{figure}

\begin{figure}
    \centering
    \includegraphics[width=\plotscale\linewidth]{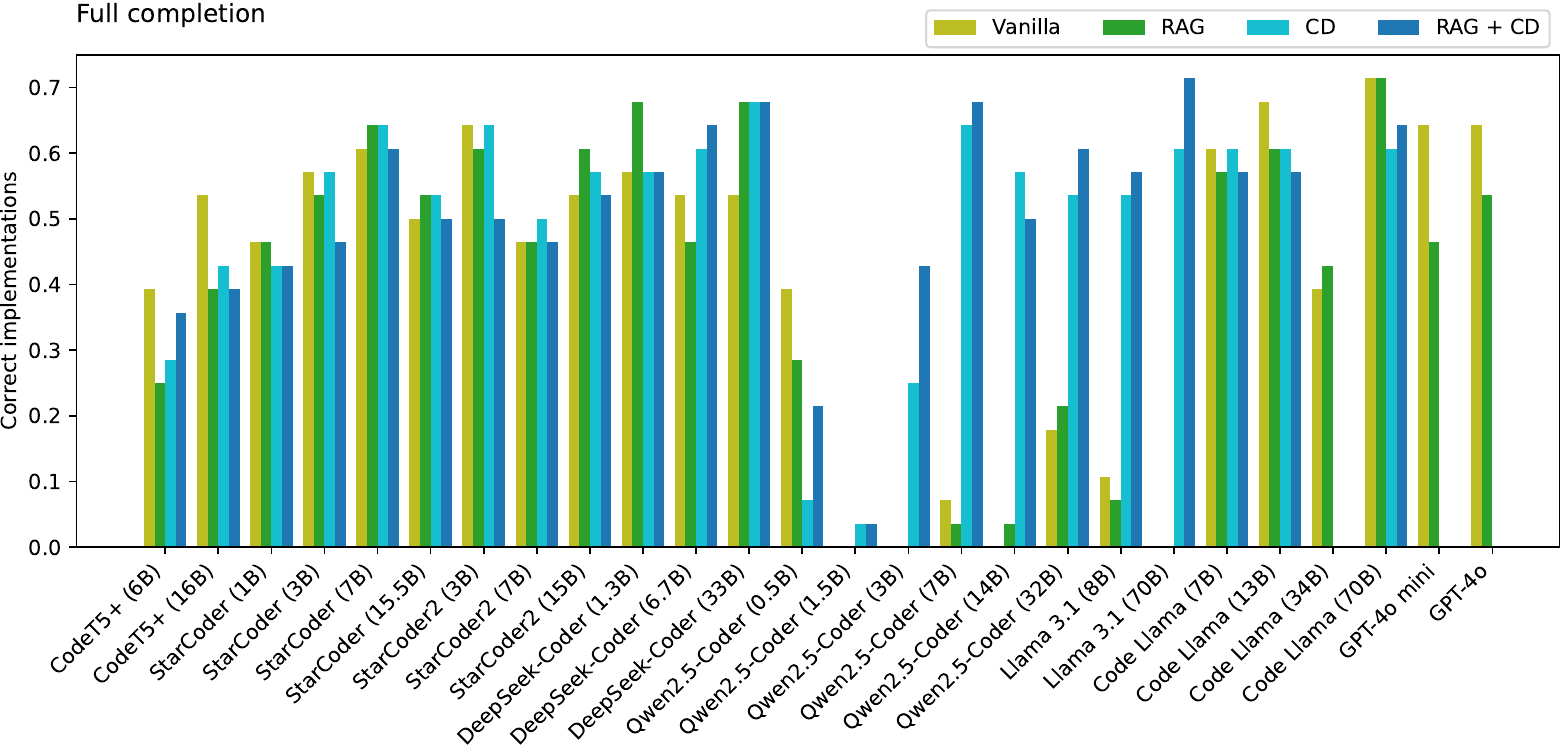}
    \includegraphics[width=\plotscale\linewidth]{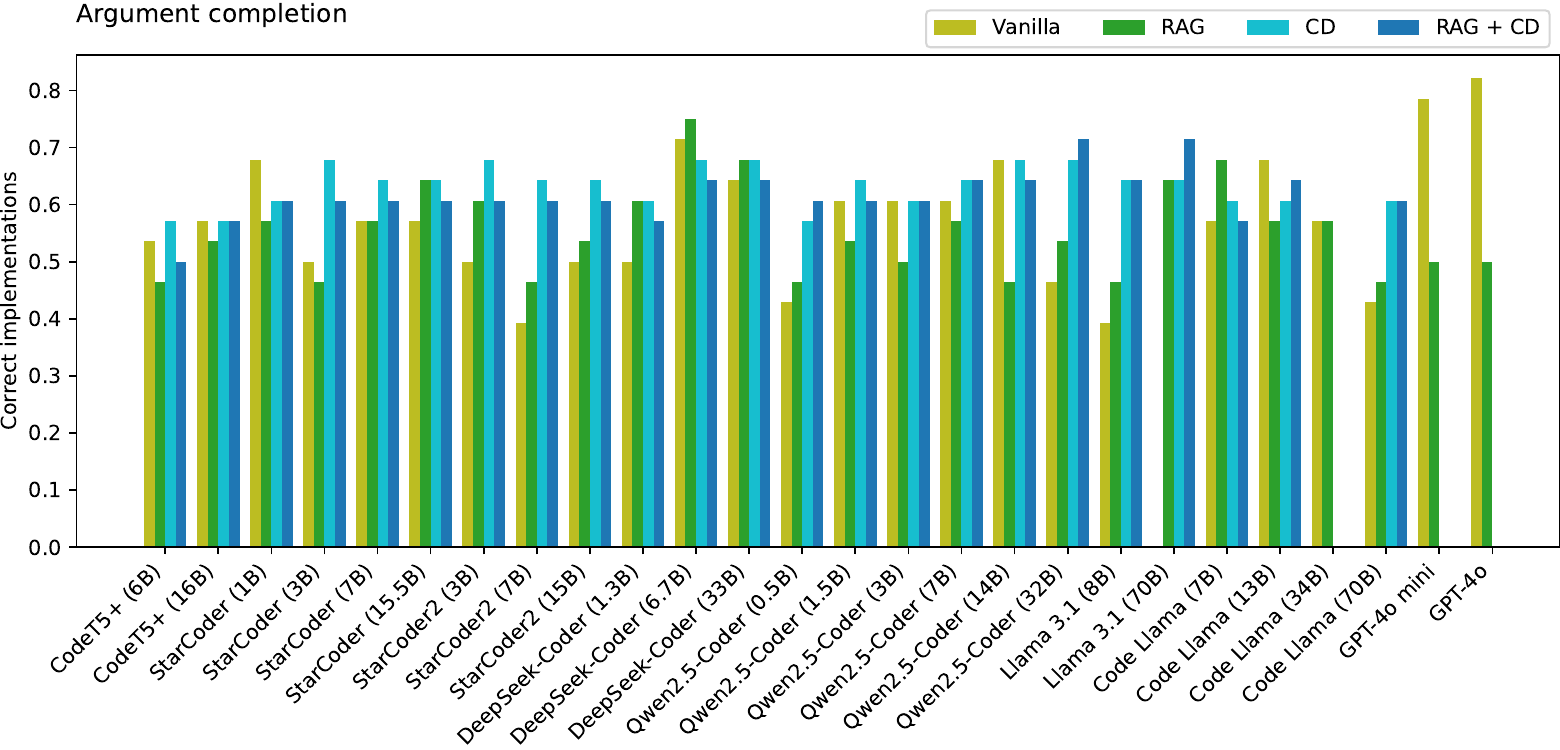}
    \Description{Grouped bar chart visualizing the performance difference between vanilla generation, RAG, constrained decoding, and RAG + constrained decoding.}
    \caption{Comparison of \emph{correct implementations} between generation settings on the \emph{real-world dataset}}
    \label{fig:res-real-correct}
\end{figure}

\begin{figure}
    \centering
    \includegraphics[width=\plotscale\linewidth]{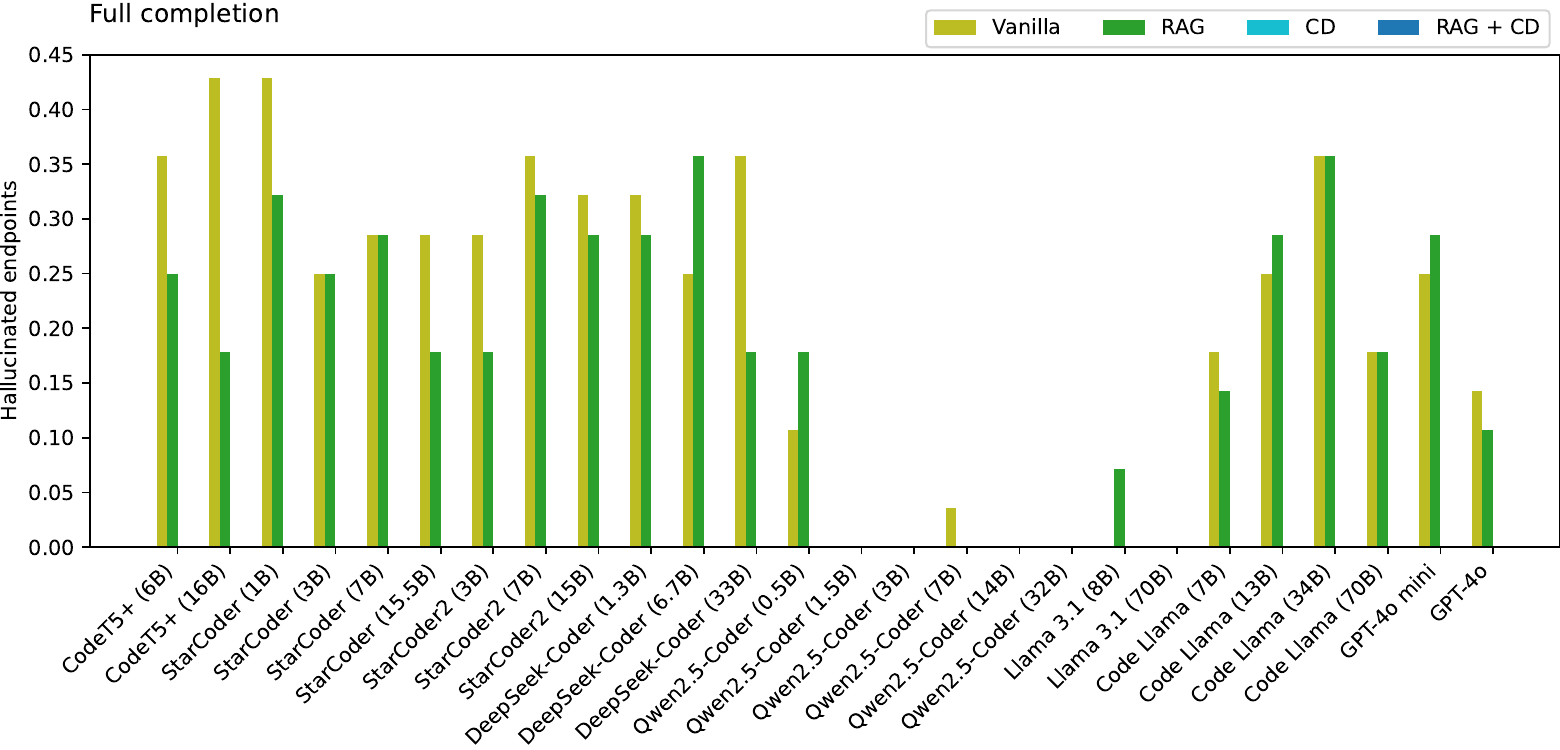}
    \includegraphics[width=\plotscale\linewidth]{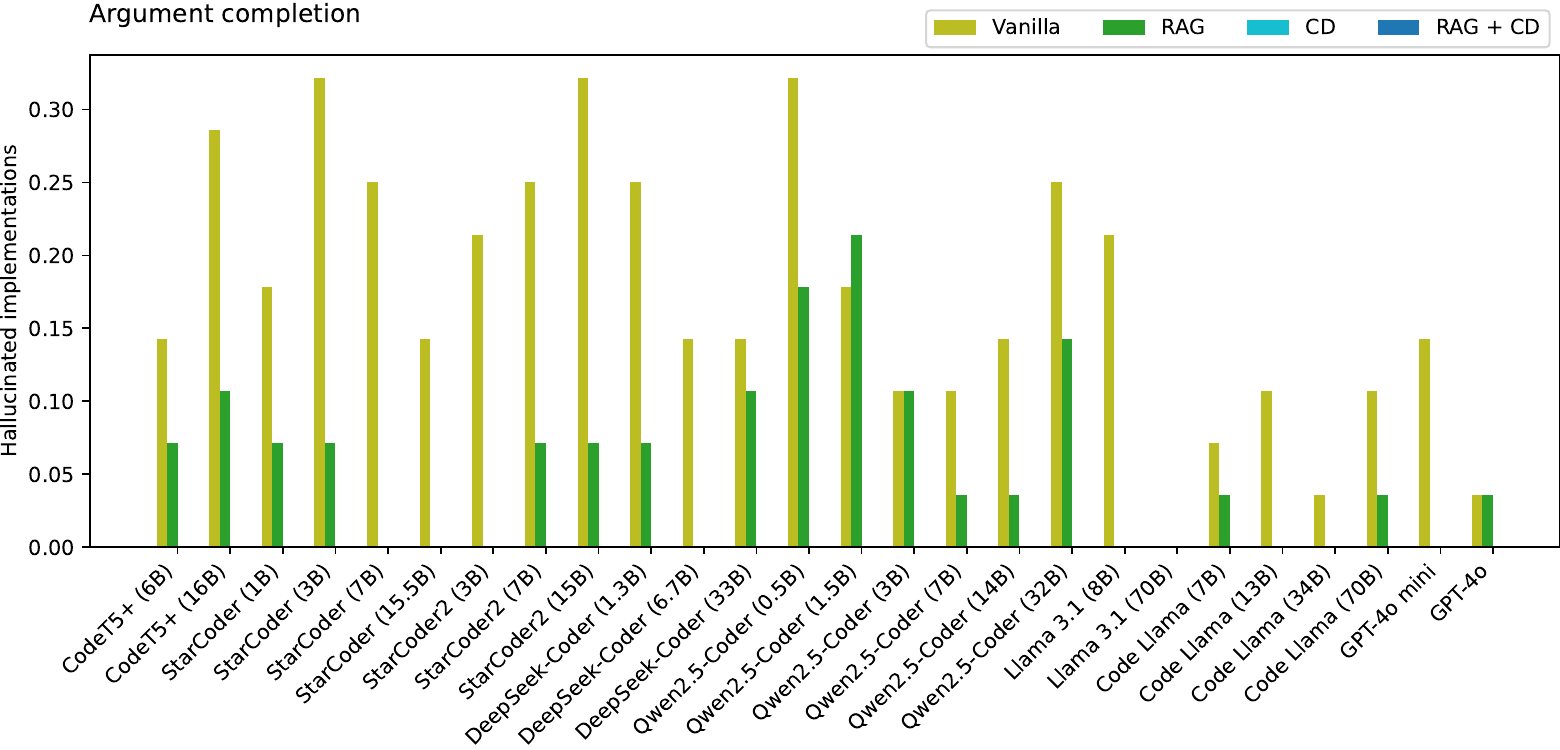}
    \Description{Grouped bar chart visualizing the performance difference between vanilla generation, RAG, constrained decoding, and RAG + constrained decoding.}
    \caption{Comparison of \emph{hallucinated endpoints} between generation settings on the \emph{real-world dataset}}
    \label{fig:res-real-illegal}
\end{figure}

In the following subsections, we elaborate on the experimental results for the four generation settings. The results for the synthetic dataset are presented in Figure~\ref{fig:res-synth-correct}---showing the ratio of correct implementations for full and argument completion across all four generation settings---and Figure~\ref{fig:res-synth-illegal}---showing the ratio of hallucinated endpoints and implementations, respectively, for the two setups. Similarly, Figures~\ref{fig:res-real-correct} and~\ref{fig:res-real-illegal} present the results for the real-world dataset. Comprehensive tables with numeric results for all metrics are provided in Appendix~\ref{app:results}.

\subsection{Vanilla Generation}
\label{subsec:eval-vanilla}

The results for vanilla generation on the synthetic dataset are consistent with our previous findings~\cite{maninger_benchmarking_2025}\footnote{Small quantitative deviations are due to minor adjustments of the prompt and implementation.}. We will focus our analysis here on a few key aspects and refer the reader to our prior work for a broader discussion.

The overall benchmark performance for the full completion setup is low, at an average correctness rate among open-source models of 13\% and Code Llama~(70B) as the best performing model at 30\%. GPT-4o achieves 58\%. All Qwen2.5-Coder and Llama~3.1 models except Qwen2.5-Coder~(32B) do not produce any executable implementation as they are unable to continue the given starter code in a syntactically valid way. The remaining models hallucinate on average 31\% of all endpoints.

On the real-world dataset, we observe a higher baseline performance---47\% correctness on average among open-source models producing executable implementations. We attribute this to the tasks in this dataset being drawn from real-world repositories, and thus having a high likelihood of representing common use cases of wide-spread APIs that are also represented in the models' training corpora. Nonetheless, most Qwen2.5-Coder and Llama~3.1 models that were unable to generate syntactically valid code completions for synthetic tasks fail here as well. The average rate of hallucinated endpoints among the remaining models remains high at 29\%.

On both datasets, models achieve generally higher correctness rates for argument completion than for full completion---26\% and 55\% on average, respectively. On the synthetic dataset, Qwen2.5-Coder~(14B) scores highest at 47\%. Interestingly, GPT-4o-mini performs better than GPT-4o (63\% vs. 59\%). An important caveat for the real-world dataset is that several tasks included do not require parameters in locations other than the path, i.e., there are no further arguments to complete and all the model has to do is recognize this fact and terminate the statement by generating a closing parenthesis.

On average, 39\% and 20\% of implementations contain hallucinations, depending on the dataset. We hypothesize that this increased hallucination rate compared to the full completion setup stems from the fact that for argument completion, the model is forced to generate arguments for a predetermined endpoint that it might not be familiar with. In contrast, for full completion, it is free to select any endpoint it is familiar with---even if it is the wrong choice, it is less likely to be a hallucination.

\begin{tcolorbox}[colback=gray!10, boxrule=0pt, colframe=gray!10, top=2pt, bottom=2pt, left=4pt, right=4pt, breakable, pad at break=2pt]
\textbf{Summary:} Vanilla generation yields low overall correctness, particularly for full completion (13\% average across open-source models on the synthetic dataset), with hallucinated endpoints affecting nearly a 1/3 of all generations. Model size is only a weak predictor of performance, and several recent models fail to produce executable code entirely. Real-world tasks yield higher baseline correctness (47\%), presumably reflecting the prevalence of common API patterns in training data, yet hallucination rates remain substantial. These results confirm that unassisted LLMs are unreliable for web API invocation code generation and motivate the mitigation approaches studied in the following subsections.
\end{tcolorbox}

\subsection{Retrieval-Augmented Generation}
\label{subsec:eval-rag}

In the case of the full completion setup on the synthetic dataset, RAG yields a clear benefit: it more than doubles the average correct implementation rate compared to vanilla generation, raising it to 22\%. The picture changes substantially, however, for the argument completion setup on the synthetic dataset and for both setups on the real-world dataset: in all these cases, RAG's correctness scores fall within 1\% of the vanilla baseline, indicating negligible net impact.

Examining per-model results reveals that the effect of RAG is highly model-dependent rather than uniformly beneficial or harmful. Individual models swing between a +562\% gain in correctness (StarCoder~1B, full completion, synthetic dataset) and complete collapse to 0\% correctness (CodeT5+~6B, full completion, synthetic dataset), underscoring that RAG’s utility is strongly conditioned on the underlying model's capacity to leverage retrieved context.

The ambivalent outcomes stem from several competing effects. On the positive side, RAG helps models select the correct API endpoint---an advantage that, by definition, vanishes in the argument completion setup, where the endpoint is already given.

On the negative side, RAG's potential benefits in selecting correct parameters are overshadowed by a substantial increase in the number of \emph{unnecessary parameters} for most models (see Appendix~F).
These parameters are valid (but optional) for the chosen endpoint according to the API specification yet irrelevant or undesirable for the task at hand. We hypothesize that most LLMs---even those not explicitly instruction-tuned---tend to treat retrieved context as implicit guidance, attempting to incorporate as many listed parameters as possible into the generated API call. This over-inclusion is further amplified when the same parameter is defined  across multiple endpoints and thus appears repeatedly in the retrieved context. A prime example is Asana's optional \code{opt\_pretty} parameter, which is used to format responses in a human-readable way. Intended for debugging, it can be passed to every API endpoint. Accordingly, it appears in every retrieved chunk, and we observe a high frequency of use among models. How susceptible a given model is to this behavior ultimately determines how much RAG hurts its correctness.

For Qwen2.5-Coder and Llama~3.1, RAG has no effect on their tendency to produce only non-executable code. Among the remaining models, however, RAG delivers substantial reductions in hallucinations on average: 82\% fewer on the synthetic dataset and 10\% fewer on the real-world dataset for full completion, and 60\% and 51\% fewer, respectively, for argument completion.

\begin{tcolorbox}[title={\textbf{RQ1:} \textit{To what extent does RAG improve the correctness of LLM-generated web API invocation code?}}, coltitle=black, colbacktitle=gray!10, colback=gray!10, boxrule=0pt, colframe=gray!10, toptitle=2pt, bottom=2pt, left=4pt, right=4pt, breakable, pad at break=2pt]
\textbf{Answer:} RAG effectively reduces hallucinations but does not uniformly improve correctness and can even degrade it. Its benefits are most pronounced for full completion, where it also aids endpoint selection, while its impact on argument completion is negligible. Whether RAG is net-positive depends strongly on the individual model: models prone to incorporating retrieved parameters indiscriminately suffer the greatest correctness losses, making careful model-specific evaluation advisable before deploying RAG in practice.
\end{tcolorbox}

\subsection{Constrained Decoding}
\label{subsec:eval-cd}

On the synthetic dataset, constrained decoding consistently increases correctness for all models and both setups. For full completion---where RAG achieved its highest average gain at +100\%---CD outperforms it with +209\%, leading to an average correctness rate of 28\%. For argument completion, the gain is +143\%, corresponding to an average correctness rate of 57\%. Code Llama~(70B) remains the strongest open-source model at 50\% for full completion, closing the gap to GPT-40 at 58\%. For argument completion, Code Llama~(70B) with CD achieves 72\% correctness, outperforming GPT-4o (59\%) and GPT-4o mini (63\%).

The Qwen2.5-Coder and Llama~3.1 models that were previously unable to generate any executable code in the full completion setup, now constrained into the required syntactic structure, achieve executability rates similar to the other models. However, disregarding models with 0\% executability, the overall rate of non-executable implementations increases slightly compared to the previous generation settings: from 92\% (vanilla) and 86\% (RAG) to 81\% (CD) for full completion and from 98\% (vanilla) and 93\% (RAG) to 85\% (CD) for argument completion. A potential explanation for this is the distortion an LLM's next token probability distribution that CD may cause~\cite{beurer-kellner_guiding_2024}.

On the real-world dataset, we see similar trends, but with a smaller amplitude due to the higher baseline performance. CD increases correctness by +69\% and +18\%, depending on the setup. Unlike on the synthetic dataset, on the real-world dataset, a few models drop in correctness with CD: 6 out of 23 models for full completion and 3 out of 23 for argument completion. The only extreme case is Qwen-2.5-Coder~(0.5B), which drops correctness by -82\% on full completion---in all other cases, correctness drops by less than 30\%. This drop is likely a consequence of the reduced executability rates.

The most striking property of CD---one that holds universally across all models, setups, and datasets---is its guarantee of zero hallucinated endpoints and implementations. This is a direct consequence of how the constraints are constructed (cf. Section~\ref{sec:impl-cd}), and our experiments consistently confirm it.

\begin{tcolorbox}[title={\textbf{RQ2:} \textit{To what extent does CD improve the correctness of LLM-generated web API invocation code?}}, coltitle=black, colbacktitle=gray!10, colback=gray!10, boxrule=0pt, colframe=gray!10, toptitle=2pt, bottom=2pt, left=4pt, right=4pt, breakable, pad at break=2pt]
\textbf{Answer:} CD eliminates hallucinations of URLs, HTTP methods, and arguments by construction, reducing them to zero across all evaluated settings. Beyond this guarantee, CD consistently and substantially boosts the correctness of generated API invocations---in some cases making smaller constrained models competitive with larger unconstrained ones. Given these clear benefits and the negligible risk of correctness degradations, CD can be recommended unconditionally for web API invocation tasks.
\end{tcolorbox}

\subsection{Combining RAG and CD}
\label{subsec:eval-rag+cd}

The combination of RAG and CD retains CD’s zero-hallucination guarantee while additionally leveraging retrieved context to guide generation. This yields strong average correctness gains over the vanilla baselines: +332\% and +111\% on the synthetic dataset, and +74\% and +14\% on the real-world dataset, for full and argument completion, respectively. RAG~+~CD also achieves the highest correctness scores across all setups and datasets: 67\% (DeepSeek-Coder 33B) and 73\% (Qwen2.5-Coder 14B) for full and argument completion on the synthetic dataset, and 71\% (Llama 3.1 70B) and 71\% (Qwen2.5-Coder 32B / Llama 3.1 70B) on the real-world dataset.

Compared to CD alone, RAG~+~CD increases the share of executable implementations by +10\% and +7\% for full and argument completion on the synthetic dataset, with only a +1\% gain on the real-world dataset. Correctness gains over CD, however, are limited to full completion (+43\% synthetic, +11\% real-world); for argument completion, average correctness declines (-14\% synthetic, -3\% real-world), consistent with the parameter overuse effect described in Section~\ref{subsec:eval-rag}.

RAG~+~CD is also less consistent than CD alone regarding the achieved correctness gains. CD improved correctness for all models on the synthetic dataset and reduced it for only 6 (full completion) and 3 (argument completion) models on the real-world dataset. RAG~+~CD, by contrast, reduces correctness for 1 and 2 models on the synthetic dataset, and for 9 and 5 models on the real-world dataset---a wider spread of regressions.

\begin{tcolorbox}[title={\textbf{RQ3:} \textit{To what extent does combining RAG and CD improve the correctness of LLM-generated web API invocation code compared to each technique in isolation?}}, coltitle=black, colbacktitle=gray!10, colback=gray!10, boxrule=0pt, colframe=gray!10, toptitle=2pt, bottom=2pt, left=4pt, right=4pt, breakable, pad at break=2pt]
\textbf{Answer:} RAG~+~CD achieves the highest correctness scores across all setups and datasets, outperforming both techniques in isolation. However, this peak performance comes at the cost of consistency: compared to CD alone, RAG~+~CD produces more correctness regressions across individual models, and its average correctness gain over CD is limited to full completion---argument completion suffers due to parameter overuse. RAG~+~CD is therefore advisable when maximum correctness is the priority and model behavior with retrieved context has been validated in advance. When neither condition holds, CD alone is the safer and more robust choice.

\end{tcolorbox}

\subsection{Efficiency}
\label{subsec:efficiency}

\begin{table}
    \centering
    \caption{Generation speed in \emph{tokens per second} for each API and generation setting}
    \label{tab:efficiency}
    \begin{tabular}{lrrrr}
        \toprule
        API & Vanilla & RAG & CD & RAG + CD \\
        \midrule
        Asana & 29.85 & 29.54 & 7.15 & 6.27 \\
        Google Calendar & 30.94 & 26.54 & 10.36 & 7.25 \\
        Google Sheets & 31.72 & 26.26 & 10.37 & 2.31 \\
        Slack Web & 29.23 & 29.59 & 7.30 & 7.49 \\
        \bottomrule
    \end{tabular}
\end{table}

Depending on the use case, the efficiency of the generation setting may be an important factor as it directly affects inference speed. To estimate the overhead of RAG and CD, we run a series of runtime measurements and calculate the average token per second (TPS) generation speed. The results are shown in Table~\ref{tab:efficiency}.

We selected one representative task for each API of the synthetic dataset and repeated the generation ten times per task on one A100~40GB GPU, using StarCoder2~\cite{lozhkov_starcoder_2024} and the full completion setup. The time to set up and run the retriever for RAG or to create the constraints for CD is excluded from the measurements. Since the variation in runtime between individual runs was low\footnote{The highest recorded standard deviation is 0.6s on a 30s run; most are much lower. The complete measurements are available in our artifact.}, we focus our discussion on the average TPS.

The results show that vanilla generation is largely invariant to the API or task (ca. 30~TPS), which is expected since the model's prompt is almost static except for the API name and the task description, which have a negligible impact on the input length. RAG is overall slightly slower than vanilla due to the increased computational cost of processing the retrieved context. For Asana and Slack, the difference is indistinguishable from measurement noise, for Google Calendar and Sheets, it amounts to a -4~TPS or -5~TPS slowdown, respectively (about -13\% reduction). CD has a clear performance impact compared to vanilla due to constraint checking. For Asana and Slack, it is about -76\%, for Google Calendar and Sheets about -67\%. Naturally, the combination of RAG and CD combines the performance overhead of the two approaches. The strongest reduction can be observed for Google Sheets at roughly -93\%, while Slack shows no measurable reduction.

The impact of RAG on inference speed might be acceptable in most scenarios given the potential improvement of generation results it offers. The only factor affecting RAG's overhead is the truncation threshold of the retrieved context as transformers' inference speed is known to scale with input length (the exact scaling behavior depends on the used attention mechanism and is typically between linear and quadratic). The impact of CD on inference speed, on the other hand, highly depends on the CD implementation. The prototype implementation used for our study (for reasons outlined in Section~\ref{sec:impl-cd}) is clearly not optimized for performance and unsuitable for use cases with real-time requirements. However, existing works show that CD can be implemented in highly efficient ways~\cite{willard_efficient_2023}. If attention masks can be fully or partially precomputed (which is the case for our constraints), CD can run at near-zero overhead~\cite{ugare_syncode_2025, dong_xgrammar_2025}.

\section{Discussion and Limitations}
\label{sec:limitations}

This section discusses practical implications of our results---including implementation challenges of RAG and CD---and limitations that should be considered when interpreting the findings.

\paragraph{Practical Implications}

The conducted experiments demonstrate that retrieval-augmented generation, constrained decoding, or both,
are effective in reducing or eliminating errors and hallucinations of LLMs in writing web API integration code. In direct comparison, constrained decoding offers more consistent gains than RAG and provides hard guarantees on specification-compliance, while also avoiding prompt bloat~\cite{gan_rag-mcp_2025}. Regardless of which approach is chosen, it is advisable to validate that the model in use actually benefits from it.

We were unable to identify meaningful patterns relating model family or size to the degree of benefit from RAG or CD. The correlation between benchmark performance and model size is, for instance, only weak. Manual inspection of generated code suggests that the main reason is the highly idiosyncratic behavior of individual models---even within the same family. Code Llama~(34B) is a prime example: it frequently inserts placeholders such as \code{<body>}, producing syntactically broken code, while its smaller and larger siblings do not exhibit this behavior.

On the implementation side, we experienced different challenges for RAG and CD. 

RAG benefits from a production-grade ecosystem of components, such as vector databases and embedding models, that can be readily built upon. Its main challenge lies in the degrees of freedom at each stage of a RAG pipeline (e.g., preprocessing, chunking, embedding, retrieval, reranking, formatting, etc.). Even when established best practices are followed, many parameters remain to be tuned: chunk size, overlap, number of retrieved chunks, truncation threshold, and others. This creates a risk of overfitting the pipeline to certain models, which is especially problematic in a benchmarking context where all models must be treated equally.

CD’s main challenge lies in constraint design. Care must be taken not to over- or under-constrain a model. In addition, certain rules are difficult or impossible to express with regexes (or grammars), requiring workarounds or approximations. Regarding the completion engine, we implemented a custom one to be able to explore the orchestration of multiple constraint rulesets and to experiment with features of extended regular expressions not supported by off-the-shelf completion engines (cf. Section~\ref{sec:impl-cd}). However, the complexity of the constraint checker is negligible compared to that of the constraint generator.

\paragraph{Limitations}

Our implementations of RAG and CD require the user to provide the name and specification of the API, for which invocations are to be generated. This limitation could be addressed by automatically inferring the API referred to in a task description and retrieving the corresponding specification. However, our work targets use cases in the API integrations industry, where usually the customer knows which APIs they want to be integrated, and the developer only has to decide which endpoints to use in order to connect the APIs in the desired way. In this scenario, having to provide the correct API specification to our system is not a real limitation. Our work is not intended for highly exploratory development scenarios, but future work could fill this gap.

Both RAG and, to a larger extent, CD lead to a small increase in non-executable implementation. Since WAPIIBench's evaluation pipeline is execution-based, non-executable codes cannot be properly evaluated and do not contribute to correctness and hallucination metrics. To provide a complementary view on the evaluation results, we calculate a second variant of each metric after filtering out any non-executable codes---these metrics are presented in the appendix. However, for practical applications like IDEs, sporadic cases of non-executable codes are no big concern, as code suggestions can be generated in the background and discarded or regenerated if they fail a syntax check.

As mentioned earlier, the biggest limitation of our RAG evaluation is the large design space for the retriever implementation. While we tried to make reasonable design decisions and to validate those decisions with preliminary experiments, any change to the RAG pipeline will inevitably influence the evaluation results in a non-uniform way across models. Another minor limitation of our RAG implementation is that during preprocessing of OpenAPI specifications, references in are only resolved up to a certain level to avoid inflating the size of individual chunks. However, this also means that the retriever will be less useful for tasks requiring deeply nested parameters.

The limitations of our CD implementation revolve around what structures the constraints do and do not permit. There is a natural tradeoff between allowing everything that is correct and preventing everything that is incorrect. In consequence, effective constraints will often limit the model's flexibility at the same time. For example, to be able to effectively guide the model through generating a syntactically correct and hallucination-free web API invocation, we constrain the Axios call syntax to the one shown in Listing~\ref{lst:invocation-example}. However, some models may prefer a different syntax (e.g., \code{axios.request(<url>, <method>, <config>)}) and perform better when given the liberty to use it. Another limitation is that our constraints act only locally from the beginning of an API invocation to its end. When (instead of literals) variables are used as parameter values, we have no control over such a variable's value or data type.

The main limitation of the new real-world dataset is its small size due to the high manual effort involved in its creation. However, WAPIIBench's extensibility makes it easy for everyone to contribute tasks, so the dataset may be expanded in the future. The newly implemented support for per-task starter codes is limited to program prefixes, so fill-in-the-middle generation is not possible. Moreover, variable or constant definitions within the starter code and that are used for execution-based evaluation must be primitive data types, objects, or arrays---it is not possible to use imported classes or functions.

Despite these limitations, the consistent empirical trends across 24 models, two datasets, two starter code setups, and four generation settings provide a solid foundation for the practical adoption and further refinement of RAG and CD in web API invocation scenarios.

\section{Related Work}
\label{sec:related-work}

\paragraph{Differentiation of API Invocation Settings}
\label{subsec:rw-invocations}

API invocation has been studied in several settings that differ substantially from the one considered in this work. We distinguish five categories:
\begin{enumerate*}
    \item \emph{general web APIs} (our setting),
    \item \emph{domain-specific web APIs},
    \item \emph{SDK-wrapped web APIs},
    \item \emph{local function APIs}, and
    \item \emph{tool APIs for AI agents}.
\end{enumerate*}

Our work considers general RESTful web APIs described by OpenAPI, where code must explicitly construct complete HTTP requests. Compared to local function calls, this requires identifying operations by HTTP method and endpoint, configuring multiple request components (path, query, headers, and body), and handling externally documented interfaces. We are not aware of other works besides WAPIIBench~\cite{maninger_benchmarking_2025} that systematically analyze the capabilities of LLMs in this specific setting.

Prior work has largely focused on different settings. Domain-specific web APIs target constrained ecosystems such as OData~\cite{su_building_2017, su_natural_2018, hosseini_compositional_2021}, making the task considerably narrower than arbitrary OpenAPI services. SDK-based approaches replace HTTP requests with library-specific abstractions~\cite{shu_dialog2api_2022, jain_mitigating_2025}, shifting the problem toward SDK usage rather than direct API invocation. Research on local function calls~\cite{zan_private-library-oriented_2025, eghbali_-hallucinator_2024, dutta_applying_2024} considers a much simpler interface consisting of a function name and a single argument list. Finally, tool-use benchmarks for LLM agents~\cite{schick_toolformer_2023, song_restgpt_2023, li_api-bank_2023, qin_toolllm_2024, yehudai_survey_2025} typically expose function-like interfaces optimized for agent interaction rather than production-ready HTTP integration code. Consequently, techniques and evaluation results from these settings are not directly transferable to general web API invocation.

\paragraph{Retrieval-Augmented Generation}
\label{subsec:rw-rag}

Retrieval augmented generation~(RAG)~\cite{lewis_retrieval-augmented_2020} is commonly used when a model's memorized knowledge is insufficient to solve a given task. This includes many applications in code generation. To improve correctness in general code generation, either similar code examples are retrieved~\cite{lu_reacc_2022, zhang_repocoder_2023} or excerpts from software documentation~\cite{zhou_docprompting_2023}. Some RAG approaches specifically target local function API invocations, retrieving examples~\cite{chen_towards_2025} or documentation~\cite{patil_gorilla_2024, zan_private-library-oriented_2025}. Work on web APIs is limited to SDK wrappers~\cite{jain_mitigating_2025}. To our knowledge, we are the first to study RAG in the context of general web API invocations as defined at the beginning of this section. Our work falls into the category of retrieval from documentation, concretely OpenAPI specification documents. In this setting, we are facing some special challenges as chunks from OpenAPI specifications must be carefully crafted (i.e., scoping by endpoint, resolution of references, and inlining of certain global definitions) to preserve crucial context. Moreover, unlike other works, we design a custom format for the retriever output to optimize token usage and LLM comprehension.

\paragraph{Constrained Decoding}
\label{subsec:rw-cd}

General-purpose constrained decoding frameworks for code generation~\cite{willard_efficient_2023, beurer-kellner_guiding_2024, ugare_syncode_2025, dong_xgrammar_2025} primarily enforce syntactic validity using regular expressions or context-free grammars. More recent work has extended constrained decoding to semantic correctness, for example by incorporating database schemas~\cite{scholak_picard_2021}, domain-specific semantic rules~\cite{poesia_synchromesh_2022}, or programming language type systems~\cite{mundler_type-constrained_2025}.

Another line of work addresses hallucinated function or method invocations by constraining generation based on information available in the code base, such as language servers or mined program dependencies~\cite{agrawal_monitor-guided_2023, wei_copiloting_2023, chen_towards_2025}. While effective for local APIs, these approaches cannot readily be applied to web APIs, whose interfaces are specified externally and whose invocations are structurally more complex, requiring the generation of complete HTTP requests.

Constrained decoding has also been applied to tool use in AI agents~\cite{zhang_dont_2023, wang_fantastic_2024}, where tools are invoked through comparatively simple JSON or Python-based interfaces. In contrast, our work targets general web APIs described by OpenAPI specifications. To our knowledge, we are the first to explore constrained decoding in this setting, introducing constraints that capture the structure of web API invocations and exploit externally provided API specifications to prevent hallucinations.

\paragraph{Alternative Approaches}
\label{subsec:rw-alternatives}

Besides retrieval-augmented generation and constrained decoding, several other approaches have been proposed to mitigate problems LLMs face when generating API invocations. Most of these works focus on local libraries (primarily in Python), but some also consider web APIs, even though they are not as complex as the ones targeted by our work (cf. discussion at the beginning of this section). These approaches are primarily fine-tuning~\cite{patil_gorilla_2024, zan_private-library-oriented_2025, dutta_applying_2024} and iterative refinement~\cite{zhang_repocoder_2023, eghbali_-hallucinator_2024, ding_toolcoder_2025} to reduce hallucinations and improve correctness. All these approaches come with their own advantages and disadvantages: RAG is well established and relatively easy to implement but risks prompt bloat~\cite{gan_rag-mcp_2025}. Constrained decoding is the only approach to offer guarantees but is relatively inflexible and, depending on the implementation, adds inference-time overhead. Fine-tuning is arguably the most natural way to improve a model but has the highest computational cost and its training data is prone to becoming outdated. Iterative refinement might be more reliable than RAG but can be expensive in terms of token consumption. It is worth exploring all these approaches to assess their suitability for web API invocations. We argue that they should be seen as complementary rather than alternative.

\section{Conclusion and Future Work}
\label{sec:conclusion}

In this paper, we examined  approaches to address the challenges faced by large language models when generating web API invocation code: retrieval-augmented generation~(RAG), constrained decoding~(CD), and a combination thereof. To this end, we implemented a retriever that processes OpenAPI specifications and stores, retrieves, and compactly formats endpoint-level chunks that can be injected into a model's prompt. Furthermore, we developed a custom translation algorithm that derives constraints, represented as regular expressions, from OpenAPI specifications. Both implementations effectively manage the complexity and scale of real-world APIs. 

To evaluate the approaches we use the original synthetic dataset of the WAPIIBench benchmark \cite{maninger_benchmarking_2025} and a new dataset with tasks manually derived from real-world GitHub repositories that we introduce in this paper. Our evaluation results demonstrate that both RAG and CD can reduce hallucinated endpoints and arguments and improve the overall correctness of the generated web API invocations. However, the potential benefit of RAG is much more model- and task-dependent than that of CD. Moreover, only CD can guarantee the absence of hallucinations. The combination of RAG and CD can further boost correctness for some models but degrades correctness for others.

Overall, our work underscores the importance of integrating quality assurance techniques into AI-driven code generation workflows as a practical path toward correctness and reliability.

Looking ahead, we plan to extend RAG and CD to further code generation tasks and to broaden the quality criteria they target beyond correctness and specification-compliance, for instance, to security properties. At the same time, our benchmark lays the foundation for future work to explore other approaches to improving quality of LLM-based code generation. Although our evaluation pipeline is specialized for JavaScript and Axios, the underlying datasets and methodology are language- and library-agnostic and can be transferred to other ecosystems. WAPIIBench is also designed for extensibility: new datasets can be added, and existing ones expanded to cover additional APIs or completion scenarios with minimal effort.

\begin{acks}
    This work was funded by the Deutsche Forschungsgemeinschaft (DFG, German Research Foundation) under Germany's Excellence Strategy (EXC-3057/1 \textit{Reasonable Artificial Intelligence}, Project No. 533677015), the Hessian Ministry of Higher Education, Research, Science and the Arts within the cluster project \textit{The Third Wave of Artificial Intelligence} (3AI), by the National Research Center for Applied Cybersecurity ATHENE within the project \textit{Foundational Models for Secure Software Development}, and by the LOEWE initiative (Hesse, Germany) [LOEWE/4a//519/05/00.002(0013)/95].
\end{acks}

\bibliographystyle{ACM-Reference-Format}
\bibliography{bibliography}

\clearpage

\appendix

\section{Data Availability}
\label{app:artifact}

WAPIIBench is available on GitHub\footnote{\url{https://github.com/stg-tud/WAPIIBench}}. In addition, we provide all model-generated codes and evaluation results in our artifact\footnote{\url{https://doi.org/10.5281/zenodo.13758414}}.

\section{Constrained Decoding Details}
\label{app:cd-details}

We implemented our constrained decoding engine as an extension of the Hugging~Face Transformers library (cf. Appendix~\ref{app:tech}) by providing a \code{LogitsProcessor} subclass that manages the whole constrained decoding process. The implementation is model-agnostic and compatible with all encoder-decoder and decoder-only architectures available on Hugging~Face.

Our constraints ensure for each argument that its name is defined in the specification for the location it appears in, and that the associated expression is either a valid variable name or a value that matches the specified schema and data type. Complex data structures such as arrays and objects are constrained deeply, i.e., constraints are constructed for their members as well, recursively. Moreover, we ensure that an argument is not passed more than once in the same location and that required arguments are indeed passed. While the general call syntax is predetermined in favor of the most common way to use Axios (even though alternative syntaxes exist), whitespace is handled flexibly by our constraints to permit different code formattings.

When using greedy decoding, where always the most likely next token is generated, we optimize the search for allowed next tokens by checking candidate tokens in order of their probability and stop once the first matching token is found. Further performance optimizations are out of scope for this work, as related work has already proposed effective measures to improve it (cf. Section~\ref{subsec:background-cd}), which could be incorporated into our implementation.

Our approach requires the user to select the API he or she wants to use and provide its specification. For an automated API selection, more research on processing the task description and classifying the implied API would be necessary. The evaluation in this work was limited to one API at a time. Enforcing constraints for multiple different APIs at once requires only implementing OR-semantics in addition to the current AND-semantics for generation rules.

The generation rules we implemented are activated when \code{axios.} is generated and deactivated when the end of a statement (e.g., by \code{);} or \code{).then(}) is generated. This makes it feasible to generate whole programs while only constraining Axios invocations. However, both the start and the stop conditions are heuristics and may fail, e.g., if a matching string is found in a comment or nested expression. Improving the detection of start and stop conditions is only possible to a certain extent due to the fundamental limitations of regular expressions (e.g., it is impossible to decide if parentheses are balanced). When generating whole programs, we also cannot constrain the data types and values of variables that are declared before and used within the Axios invocation, as it cannot be foreseen which variable will be used for what (and if it is relevant to the API invocation in the first place). More research on this problem is needed for constrained decoding approaches in general.

Our constraints are designed to match a code snippet if it contains an Axios invocation that respects JavaScript syntax and API usage constraints from an OpenAPI specification. However, due to some simplifications we made to keep the complexity of the generated regexes in bounds, the opposite is not true, i.e., there are syntactically invalid Axios invocations that are matched by our constraints. One example are single, double, and backtick quotation marks, for which we do not always ensure that the opening and the closing one match. However, we did not encounter any cases where a model made such a mistake. Hence, we did not consider it worthwhile to implement such constraints, which would add even more complexity to the regexes. Another OpenAPI-related example is the \code{schema} property of any parameter, which provides additional constraints beyond its data type, e.g., that a string value should be an email address or that an integer value should be in the range from 0 to 100. Although it would be feasible to respect such schema constraints (at least approximately), we did not implement them so far, since by constraining the data type, we already cover the majority of parameter-value-related mistakes. Finally, OpenAPI specifications sometimes contain usage constraints that are only explained in a parameter's \code{description} property, which contains a free form textual description of an operation. For instance, the \code{maxResults} parameter in the Google Calendar API contains a passage ``The page size can never be larger than 250 entries'', but this information is not reflected in the parameter's \code{schema} (even though \code{schema}s support specifying maximum and minimum values for a parameter). More research is needed to enable extracting such constraints from \code{description}s and using them for constrained decoding and evaluation purposes.

For API parameters that are nested objects or arrays, some simplifications were necessary. Since OpenAPI permits recursive schemas, we prevent an infinite regress while generating constraints from self-recursive schemas by limiting the recursion depth to three. This depth is sufficient for all samples in our dataset but could be adapted if necessary. For arrays, we had to set the maximum number of elements to five, due to an incompatibility with the way we prevent duplicate arguments. This is, however, no fundamental limitation. Again, the chosen maximum is sufficient for our dataset, and we argue that arrays of more than five elements should anyway not be defined inline in most cases.

\section{Technologies}
\label{app:tech}

We implemented WAPIIBench using the following technologies:
\begin{itemize}[noitemsep]
	\item \href{https://www.openapis.org/}{OpenAPI}
	\item \href{https://github.com/manchenkoff/openapi3-parser}{OpenAPI 3 parser}
	\item \href{https://axios-http.com/}{Axios}
	\item \href{https://github.com/ctimmerm/axios-mock-adapter}{axios-mock-adapter}
	\item \href{https://huggingface.co/docs/transformers/index}{Hugging Face Transformers}
    \item \href{https://reference.langchain.com/}{LangChain}
    \item \href{https://docs.trychroma.com/docs/overview/introduction}{Chroma}
	\item \href{https://github.com/mrabarnett/mrab-regex}{regex}
\end{itemize}

\section{Models}
\label{app:models}

These are the exact names of the models we evaluated:
\begin{itemize}[noitemsep]
    \item \href{https://huggingface.co/bigcode/starcoderbase-1b}{\code{bigcode/starcoderbase-1b}}
	\item \href{https://huggingface.co/bigcode/starcoderbase-3b}{\code{bigcode/starcoderbase-3b}}
    \item \href{https://huggingface.co/bigcode/starcoderbase-7b}{\code{bigcode/starcoderbase-7b}}
    \item \href{https://huggingface.co/bigcode/starcoderbase}{\code{bigcode/starcoderbase}}
	\item \href{https://huggingface.co/bigcode/starcoder2-3b}{\code{bigcode/starcoder2-3b}}
    \item \href{https://huggingface.co/bigcode/starcoder2-7b}{\code{bigcode/starcoder2-7b}}
    \item \href{https://huggingface.co/bigcode/starcoder2-15b}{\code{bigcode/starcoder2-15b}}
	\item \href{https://huggingface.co/deepseek-ai/deepseek-coder-1.3b-base}{\code{deepseek-ai/deepseek-coder-1.3b-base}}
    \item \href{https://huggingface.co/deepseek-ai/deepseek-coder-6.7b-base}{\code{deepseek-ai/deepseek-coder-6.7b-base}}
	\item \href{https://huggingface.co/deepseek-ai/deepseek-coder-33b-base}{\code{deepseek-ai/deepseek-coder-33b-base}}
	\item \href{https://huggingface.co/codellama/CodeLlama-7b-hf}{\code{meta-llama/CodeLlama-7b-hf}}
	\item \href{https://huggingface.co/codellama/CodeLlama-13b-hf}{\code{meta-llama/CodeLlama-13b-hf}}
    \item \href{https://huggingface.co/codellama/CodeLlama-34b-hf}{\code{meta-llama/CodeLlama-34b-hf}}
	\item \href{https://huggingface.co/codellama/CodeLlama-70b-hf}{\code{meta-llama/CodeLlama-70b-hf}}
	\item \href{https://huggingface.co/meta-llama/Llama-3.1-8B}{\code{meta-llama/Llama-3.1-8B}}
	\item \href{https://huggingface.co/meta-llama/Llama-3.1-70B}{\code{meta-llama/Llama-3.1-70B}}
	\item \href{https://platform.openai.com/docs/models/gpt-4o}{\code{openai/gpt-4o}}
	\item \href{https://platform.openai.com/docs/models/gpt-4o-mini}{\code{openai/gpt-4o-mini}}
	\item \href{https://huggingface.co/Qwen/Qwen2.5-Coder-0.5B}{\code{Qwen/Qwen2.5-Coder-0.5B}}
	\item \href{https://huggingface.co/Qwen/Qwen2.5-Coder-1.5B}{\code{Qwen/Qwen2.5-Coder-1.5B}}
	\item \href{https://huggingface.co/Qwen/Qwen2.5-Coder-3B}{\code{Qwen/Qwen2.5-Coder-3B}}
	\item \href{https://huggingface.co/Qwen/Qwen2.5-Coder-7B}{\code{Qwen/Qwen2.5-Coder-7B}}
	\item \href{https://huggingface.co/Qwen/Qwen2.5-Coder-14B}{\code{Qwen/Qwen2.5-Coder-14B}}
	\item \href{https://huggingface.co/Qwen/Qwen2.5-Coder-32B}{\code{Qwen/Qwen2.5-Coder-32B}}
    \item \href{https://huggingface.co/Salesforce/codet5p-6b}{\code{Salesforce/codet5p-6b}}
	\item \href{https://huggingface.co/Salesforce/codet5p-16b}{\code{Salesforce/codet5p-16b}}
\end{itemize}

We use only base (i.e., non-instruction-tuned) models. While instruction-tuned models tend to perform better on coding tasks\footnote{Cf., e.g., the EvalPlus leaderboard (\url{https://evalplus.github.io/leaderboard.html})}, we considered them to be inappropriate for our setting, which is based on \emph{code completion}. Performing code completion on instruction-tuned models leads to rather unnatural results, as the models often do not directly return the completion and instead start with some response (``Here is your completed code ...'') followed by a code snipped that may contain
\begin{enumerate*}
    \item only the completion,
    \item the starter code followed by the completion, or
    \item an arbitrarily modified version of the starter code and corresponding completion.    
\end{enumerate*}
These factors make the model output very hard to parse reliably. Therefore, we decided to focus our evaluation on base models.

The following hyperparameters were used:
\begin{itemize}[noitemsep]
    \item floating point precision = 16 bit
    \item \# beams = 1
    \item temperature = 0.0
\end{itemize}

\section{Prompts}
\label{app:prompts}

\begin{lstlisting}[caption={Prompt for generating API invocations used for the synthetic dataset. Words in curly braces are placeholders.}, label=lst:prompt-synth, frame=none, float]
{spec}
You are an AI programming assistant that helps users write API requests. You are given a comment that describes what the user wants to achieve and are supposed to implement it using the Axios library in JavaScript. For this, write a single call to Axios (using the syntax `axios.<method>(url[, config])`) that does exactly what was described in the comment.

* Make sure to include all parameters in `config` that are required to solve the given task, but do not include any unnecessary parameters.
* Insert all values directly into the place where they belong rather than using intermediate variables.
* If the API requires authentication, prefer OAuth2 over other schemes and use `<token>` as a placeholder for the authorization token.
* If a request body requires a media type other than `text/json`, explicitly set the `Content-Type` header to the respective type, and Axios will automatically serialize the request body accordingly.
* Always use HTTPS.

Your next task is about the {api} API. Complete the following code snippet:

```javascript
// {task}
const axios = require('axios');

axios.{method}('{url}',
\end{lstlisting}

\begin{lstlisting}[caption={Prompt for generating API invocations used for the real-world dataset. Words in curly braces are placeholders.}, label=lst:prompt-real, frame=none, float]
{spec}
You are an AI programming assistant that helps users write API requests. You are given a comment that describes what the user wants to achieve and are supposed to implement it using the Axios library in JavaScript. For this, write a single call to Axios (using the syntax `axios.<method>(url[, config])`) that does exactly what was described in the comment.

* Make sure to include all parameters in `config` that are required to solve the given task, but do not include any unnecessary parameters.
* If a request body requires a media type other than `text/json`, explicitly set the `Content-Type` header to the respective type, and Axios will automatically serialize the request body accordingly.
* Always use HTTPS.

Your next task is about the {api} API. Complete the following code snippet:

```javascript
{starter_code}
\end{lstlisting}

Listing~\ref{lst:prompt-synth} shows the prompt used in WAPIIBench's evaluation pipeline when generating API invocation code based on tasks from the synthetic dataset. Listing~\ref{lst:prompt-real} shows the prompt adapted for the new real-world dataset.

\section{Extended Results}
\label{app:results}

\begin{table}
    \centering
    \caption{Complete evaluation metrics for correctness and specification-compliance of API invocations}
    \label{tab:metrics-full}
    \begin{tabularx}{\linewidth}{p{0.30\linewidth}X}
        \toprule
        Metric & Description \\
        \midrule
        Executable implementations & Generated code is complete and contains no syntax or runtime error \\
        Correct implementations & Generated executable code matches the ground-truth configuration \\
        Hallucinated implementations & Generated code contains at least one violation of the API specification \\
        Correct endpoints & Generated URL--method combination matches the ground-truth URL \\
        Hallucinated endpoints & Generated URL--method combination is not defined in the API specification \\
        Correct URLs & Generated URL matches the ground-truth URL \\
        Hallucinated URLs & Generated URL is not defined in the API specification \\
        Correct methods & Generated HTTP method matches the ground-truth HTTP method \\
        Hallucinated methods & Generated HTTP method is not defined for the generated URL in the API specification \\
        Correct argument names & Generated arguments are correct \\
        Correct argument values & Generated argument values are correct \\
        Missing arguments & Expected arguments are not generated \\
        Unexpected arguments & Generated arguments are not expected \\
        Unnecessary arguments & Redundant arguments are generated for an API endpoint \\
        Hallucinated arguments & Generated arguments are not permitted for the generated API endpoint \\
        Mean argument precision & Probability that the generated arguments are correct \\
        Mean argument recall & Probability that the correct arguments are generated \\
        Mean argument Jaccard index & Overlap between generated and correct arguments \\
        Mean argument value conditional accuracy & Probability that an argument value is correct if the argument name is correct \\
        Total errors & Any type of error prevented execution \\
        Incomplete implementations & Generated code did not contain a complete API invocation \\
        Runtime errors & Generated code produced an error when trying to execute it \\
        Timeouts & Constrained decoding was canceled because finding allowed next tokens took too long \\
        Unsatisfiable constraints & Constrained decoding was canceled because no allowed next token existed \\
        \bottomrule
    \end{tabularx}
\end{table}

The full set of metrics calculated by WAPIIBench is explained in Table~\ref{tab:metrics-full}. Tables~\ref{tab:res-full-synth-vanilla-inv} to~\ref{tab:res-full-real-rag+cd-end} provide a comprehensive summary of these metrics for all evaluated models (cf. Appendix~\ref{app:models}) and both datasets. The best values are in bold. Tables~\ref{tab:res-gain-correct-rag} to~\ref{tab:res-gain-correct-rag+cd} compare the correctness rates of all models between vanilla generation and the other generation settings (RAG, CD, and RAG~+~CD). Likewise, Tables~\ref{tab:res-gain-illegal-rag} to~\ref{tab:res-gain-illegal-rag+cd} compare the hallucination rates.

Note that some metrics can only be calculated either for \emph{full completion} or for \emph{argument completion} and are therefore not shown in all tables. Values marked with \emph{(t)} are ratios relative to the \emph{total samples} (395 for the synthetic dataset, 28 for the real-world dataset), while values marked with \emph{(e)} are relative to the subset of \emph{executable implementations}\footnote{To be exact, the calculation for (t) and (e) values is identical except for a filtering step to remove non-executable samples before calculating the latter.}, which varies from experiment to experiment. The complete raw data these results are based on can be found in our artifact (cf. Appendix~\ref{app:artifact}). Small discrepancies to previously published results~\cite{maninger_benchmarking_2025} are due to minor prompt adjustments and improvements to the evaluation pipeline.

\begin{landscape}
\begin{table}[p]
\centering
\caption{Complete evaluation results for \emph{full completion} with \emph{vanilla} generation on the \emph{synthetic dataset}}
\label{tab:res-full-synth-vanilla-inv}
\makebox[1\textwidth][c]{
\resizebox{\tablescale\textwidth}{!}{
\begin{tabular}{lrrrrrrrrrrrrrrrrrrrrrrrrrr}
\makebox[20pt][l]{\rotatebox{45}{}} & \makebox[20pt][l]{\rotatebox{45}{CodeT5+ (6B)}} & \makebox[20pt][l]{\rotatebox{45}{CodeT5+ (16B)}} & \makebox[20pt][l]{\rotatebox{45}{StarCoder (1B)}} & \makebox[20pt][l]{\rotatebox{45}{StarCoder (3B)}} & \makebox[20pt][l]{\rotatebox{45}{StarCoder (7B)}} & \makebox[20pt][l]{\rotatebox{45}{StarCoder (15.5B)}} & \makebox[20pt][l]{\rotatebox{45}{StarCoder2 (3B)}} & \makebox[20pt][l]{\rotatebox{45}{StarCoder2 (7B)}} & \makebox[20pt][l]{\rotatebox{45}{StarCoder2 (15B)}} & \makebox[20pt][l]{\rotatebox{45}{DeepSeek-Coder (1.3B)}} & \makebox[20pt][l]{\rotatebox{45}{DeepSeek-Coder (6.7B)}} & \makebox[20pt][l]{\rotatebox{45}{DeepSeek-Coder (33B)}} & \makebox[20pt][l]{\rotatebox{45}{Qwen2.5-Coder (0.5B)}} & \makebox[20pt][l]{\rotatebox{45}{Qwen2.5-Coder (1.5B)}} & \makebox[20pt][l]{\rotatebox{45}{Qwen2.5-Coder (3B)}} & \makebox[20pt][l]{\rotatebox{45}{Qwen2.5-Coder (7B)}} & \makebox[20pt][l]{\rotatebox{45}{Qwen2.5-Coder (14B)}} & \makebox[20pt][l]{\rotatebox{45}{Qwen2.5-Coder (32B)}} & \makebox[20pt][l]{\rotatebox{45}{Llama 3.1 (8B)}} & \makebox[20pt][l]{\rotatebox{45}{Llama 3.1 (70B)}} & \makebox[20pt][l]{\rotatebox{45}{Code Llama (7B)}} & \makebox[20pt][l]{\rotatebox{45}{Code Llama (13B)}} & \makebox[20pt][l]{\rotatebox{45}{Code Llama (34B)}} & \makebox[20pt][l]{\rotatebox{45}{Code Llama (70B)}} & \makebox[20pt][l]{\rotatebox{45}{GPT-4o mini}} & \makebox[20pt][l]{\rotatebox{45}{GPT-4o}} \\
\midrule
Executable implementations (t) & \phantom{00}0.96 & \phantom{00}0.96 & \phantom{00}0.96 & \phantom{00}0.96 & \phantom{00}0.98 & \phantom{00}0.99 & \phantom{00}0.97 & \phantom{00}0.99 & \phantom{00}0.98 & \phantom{00}0.97 & \phantom{00}0.64 & \phantom{00}0.97 & \phantom{00}0.00 & \phantom{00}0.00 & \phantom{00}0.00 & \phantom{00}0.00 & \phantom{00}0.00 & \phantom{00}0.43 & \phantom{00}0.00 & \phantom{00}0.00 & \phantom{00}0.99 & \phantom{00}0.98 & \phantom{00}0.96 & \textbf{1.00} & \textbf{1.00} & \textbf{1.00} \\
\bottomrule
\end{tabular}

}}
\end{table}
\end{landscape}

\begin{landscape}
\begin{table}[p]
\centering
\caption{Complete evaluation results for \emph{argument completion} with \emph{vanilla} generation on the \emph{synthetic dataset}}
\label{tab:res-full-synth-vanilla-end}
\makebox[1\textwidth][c]{
\resizebox{\tablescale\textwidth}{!}{
\begin{tabular}{lrrrrrrrrrrrrrrrrrrrrrrrrrr}
\makebox[20pt][l]{\rotatebox{45}{}} & \makebox[20pt][l]{\rotatebox{45}{CodeT5+ (6B)}} & \makebox[20pt][l]{\rotatebox{45}{CodeT5+ (16B)}} & \makebox[20pt][l]{\rotatebox{45}{StarCoder (1B)}} & \makebox[20pt][l]{\rotatebox{45}{StarCoder (3B)}} & \makebox[20pt][l]{\rotatebox{45}{StarCoder (7B)}} & \makebox[20pt][l]{\rotatebox{45}{StarCoder (15.5B)}} & \makebox[20pt][l]{\rotatebox{45}{StarCoder2 (3B)}} & \makebox[20pt][l]{\rotatebox{45}{StarCoder2 (7B)}} & \makebox[20pt][l]{\rotatebox{45}{StarCoder2 (15B)}} & \makebox[20pt][l]{\rotatebox{45}{DeepSeek-Coder (1.3B)}} & \makebox[20pt][l]{\rotatebox{45}{DeepSeek-Coder (6.7B)}} & \makebox[20pt][l]{\rotatebox{45}{DeepSeek-Coder (33B)}} & \makebox[20pt][l]{\rotatebox{45}{Qwen2.5-Coder (0.5B)}} & \makebox[20pt][l]{\rotatebox{45}{Qwen2.5-Coder (1.5B)}} & \makebox[20pt][l]{\rotatebox{45}{Qwen2.5-Coder (3B)}} & \makebox[20pt][l]{\rotatebox{45}{Qwen2.5-Coder (7B)}} & \makebox[20pt][l]{\rotatebox{45}{Qwen2.5-Coder (14B)}} & \makebox[20pt][l]{\rotatebox{45}{Qwen2.5-Coder (32B)}} & \makebox[20pt][l]{\rotatebox{45}{Llama 3.1 (8B)}} & \makebox[20pt][l]{\rotatebox{45}{Llama 3.1 (70B)}} & \makebox[20pt][l]{\rotatebox{45}{Code Llama (7B)}} & \makebox[20pt][l]{\rotatebox{45}{Code Llama (13B)}} & \makebox[20pt][l]{\rotatebox{45}{Code Llama (34B)}} & \makebox[20pt][l]{\rotatebox{45}{Code Llama (70B)}} & \makebox[20pt][l]{\rotatebox{45}{GPT-4o mini}} & \makebox[20pt][l]{\rotatebox{45}{GPT-4o}} \\
\midrule
Executable implementations (t) & \phantom{00}0.98 & \phantom{00}0.98 & \phantom{00}0.97 & \phantom{00}0.97 & \phantom{00}0.98 & \phantom{00}0.99 & \phantom{00}0.99 & \phantom{00}0.99 & \phantom{00}0.99 & \phantom{00}0.98 & \phantom{00}0.99 & \phantom{00}0.99 & \phantom{00}0.90 & \phantom{00}0.96 & \phantom{00}0.98 & \phantom{00}0.99 & \phantom{00}0.98 & \phantom{00}0.99 & \phantom{00}0.98 & \phantom{00}0.99 & \textbf{1.00} & \phantom{00}0.99 & \phantom{00}0.99 & \textbf{1.00} & \textbf{1.00} & \textbf{1.00} \\
\bottomrule
\end{tabular}

}}
\end{table}
\end{landscape}

\begin{landscape}
\begin{table}[p]
\centering
\caption{Complete evaluation results for \emph{full completion} with \emph{retrieval-augmented generation} on the \emph{synthetic dataset}}
\label{tab:res-full-synth-rag-inv}
\makebox[1\textwidth][c]{
\resizebox{\tablescale\textwidth}{!}{
\begin{tabular}{lrrrrrrrrrrrrrrrrrrrrrrrrrr}
\makebox[20pt][l]{\rotatebox{45}{}} & \makebox[20pt][l]{\rotatebox{45}{CodeT5+ (6B)}} & \makebox[20pt][l]{\rotatebox{45}{CodeT5+ (16B)}} & \makebox[20pt][l]{\rotatebox{45}{StarCoder (1B)}} & \makebox[20pt][l]{\rotatebox{45}{StarCoder (3B)}} & \makebox[20pt][l]{\rotatebox{45}{StarCoder (7B)}} & \makebox[20pt][l]{\rotatebox{45}{StarCoder (15.5B)}} & \makebox[20pt][l]{\rotatebox{45}{StarCoder2 (3B)}} & \makebox[20pt][l]{\rotatebox{45}{StarCoder2 (7B)}} & \makebox[20pt][l]{\rotatebox{45}{StarCoder2 (15B)}} & \makebox[20pt][l]{\rotatebox{45}{DeepSeek-Coder (1.3B)}} & \makebox[20pt][l]{\rotatebox{45}{DeepSeek-Coder (6.7B)}} & \makebox[20pt][l]{\rotatebox{45}{DeepSeek-Coder (33B)}} & \makebox[20pt][l]{\rotatebox{45}{Qwen2.5-Coder (0.5B)}} & \makebox[20pt][l]{\rotatebox{45}{Qwen2.5-Coder (1.5B)}} & \makebox[20pt][l]{\rotatebox{45}{Qwen2.5-Coder (3B)}} & \makebox[20pt][l]{\rotatebox{45}{Qwen2.5-Coder (7B)}} & \makebox[20pt][l]{\rotatebox{45}{Qwen2.5-Coder (14B)}} & \makebox[20pt][l]{\rotatebox{45}{Qwen2.5-Coder (32B)}} & \makebox[20pt][l]{\rotatebox{45}{Llama 3.1 (8B)}} & \makebox[20pt][l]{\rotatebox{45}{Llama 3.1 (70B)}} & \makebox[20pt][l]{\rotatebox{45}{Code Llama (7B)}} & \makebox[20pt][l]{\rotatebox{45}{Code Llama (13B)}} & \makebox[20pt][l]{\rotatebox{45}{Code Llama (34B)}} & \makebox[20pt][l]{\rotatebox{45}{Code Llama (70B)}} & \makebox[20pt][l]{\rotatebox{45}{GPT-4o mini}} & \makebox[20pt][l]{\rotatebox{45}{GPT-4o}} \\
\midrule
Executable implementations (t) & \phantom{00}0.18 & \phantom{00}0.43 & \phantom{00}0.95 & \phantom{00}0.97 & \phantom{00}0.99 & \textbf{1.00} & \phantom{00}0.99 & \textbf{1.00} & \phantom{00}0.98 & \phantom{00}0.95 & \phantom{00}0.90 & \phantom{00}0.99 & \phantom{00}0.00 & \phantom{00}0.00 & \phantom{00}0.00 & \phantom{00}0.00 & \phantom{00}0.00 & \phantom{00}0.28 & \phantom{00}0.00 & \phantom{00}0.00 & \textbf{1.00} & \phantom{00}0.98 & \phantom{00}0.97 & \textbf{1.00} & \phantom{00}0.99 & \phantom{00}0.99 \\
\bottomrule
\end{tabular}

}}
\end{table}
\end{landscape}

\begin{landscape}
\begin{table}[p]
\centering
\caption{Complete evaluation results for \emph{argument completion} with \emph{retrieval-augmented generation} on the \emph{synthetic dataset}}
\label{tab:res-full-synth-rag-end}
\makebox[1\textwidth][c]{
\resizebox{\tablescale\textwidth}{!}{
\begin{tabular}{lrrrrrrrrrrrrrrrrrrrrrrrrrr}
\makebox[20pt][l]{\rotatebox{45}{}} & \makebox[20pt][l]{\rotatebox{45}{CodeT5+ (6B)}} & \makebox[20pt][l]{\rotatebox{45}{CodeT5+ (16B)}} & \makebox[20pt][l]{\rotatebox{45}{StarCoder (1B)}} & \makebox[20pt][l]{\rotatebox{45}{StarCoder (3B)}} & \makebox[20pt][l]{\rotatebox{45}{StarCoder (7B)}} & \makebox[20pt][l]{\rotatebox{45}{StarCoder (15.5B)}} & \makebox[20pt][l]{\rotatebox{45}{StarCoder2 (3B)}} & \makebox[20pt][l]{\rotatebox{45}{StarCoder2 (7B)}} & \makebox[20pt][l]{\rotatebox{45}{StarCoder2 (15B)}} & \makebox[20pt][l]{\rotatebox{45}{DeepSeek-Coder (1.3B)}} & \makebox[20pt][l]{\rotatebox{45}{DeepSeek-Coder (6.7B)}} & \makebox[20pt][l]{\rotatebox{45}{DeepSeek-Coder (33B)}} & \makebox[20pt][l]{\rotatebox{45}{Qwen2.5-Coder (0.5B)}} & \makebox[20pt][l]{\rotatebox{45}{Qwen2.5-Coder (1.5B)}} & \makebox[20pt][l]{\rotatebox{45}{Qwen2.5-Coder (3B)}} & \makebox[20pt][l]{\rotatebox{45}{Qwen2.5-Coder (7B)}} & \makebox[20pt][l]{\rotatebox{45}{Qwen2.5-Coder (14B)}} & \makebox[20pt][l]{\rotatebox{45}{Qwen2.5-Coder (32B)}} & \makebox[20pt][l]{\rotatebox{45}{Llama 3.1 (8B)}} & \makebox[20pt][l]{\rotatebox{45}{Llama 3.1 (70B)}} & \makebox[20pt][l]{\rotatebox{45}{Code Llama (7B)}} & \makebox[20pt][l]{\rotatebox{45}{Code Llama (13B)}} & \makebox[20pt][l]{\rotatebox{45}{Code Llama (34B)}} & \makebox[20pt][l]{\rotatebox{45}{Code Llama (70B)}} & \makebox[20pt][l]{\rotatebox{45}{GPT-4o mini}} & \makebox[20pt][l]{\rotatebox{45}{GPT-4o}} \\
\midrule
Executable implementations (t) & \phantom{00}0.20 & \phantom{00}0.40 & \phantom{00}0.95 & \phantom{00}0.97 & \phantom{00}0.99 & \textbf{1.00} & \phantom{00}0.99 & \textbf{1.00} & \phantom{00}0.99 & \phantom{00}0.98 & \phantom{00}0.98 & \phantom{00}0.99 & \phantom{00}0.94 & \phantom{00}0.99 & \phantom{00}0.99 & \textbf{1.00} & \textbf{1.00} & \phantom{00}0.97 & \phantom{00}0.99 & \textbf{1.00} & \phantom{00}0.99 & \phantom{00}0.99 & \phantom{00}0.99 & \textbf{1.00} & \phantom{00}0.99 & \textbf{1.00} \\
\bottomrule
\end{tabular}

}}
\end{table}
\end{landscape}

\begin{landscape}
\begin{table}[p]
\centering
\caption{Complete evaluation results for \emph{full completion} with \emph{constrained decoding} on the \emph{synthetic dataset}}
\label{tab:res-full-synth-cd-inv}
\makebox[1\textwidth][c]{
\resizebox{\tablescale\textwidth}{!}{
\begin{tabular}{lrrrrrrrrrrrrrrrrrrrrrrr}
\makebox[20pt][l]{\rotatebox{45}{}} & \makebox[20pt][l]{\rotatebox{45}{CodeT5+ (6B)}} & \makebox[20pt][l]{\rotatebox{45}{CodeT5+ (16B)}} & \makebox[20pt][l]{\rotatebox{45}{StarCoder (1B)}} & \makebox[20pt][l]{\rotatebox{45}{StarCoder (3B)}} & \makebox[20pt][l]{\rotatebox{45}{StarCoder (7B)}} & \makebox[20pt][l]{\rotatebox{45}{StarCoder (15.5B)}} & \makebox[20pt][l]{\rotatebox{45}{StarCoder2 (3B)}} & \makebox[20pt][l]{\rotatebox{45}{StarCoder2 (7B)}} & \makebox[20pt][l]{\rotatebox{45}{StarCoder2 (15B)}} & \makebox[20pt][l]{\rotatebox{45}{DeepSeek-Coder (1.3B)}} & \makebox[20pt][l]{\rotatebox{45}{DeepSeek-Coder (6.7B)}} & \makebox[20pt][l]{\rotatebox{45}{DeepSeek-Coder (33B)}} & \makebox[20pt][l]{\rotatebox{45}{Qwen2.5-Coder (0.5B)}} & \makebox[20pt][l]{\rotatebox{45}{Qwen2.5-Coder (1.5B)}} & \makebox[20pt][l]{\rotatebox{45}{Qwen2.5-Coder (3B)}} & \makebox[20pt][l]{\rotatebox{45}{Qwen2.5-Coder (7B)}} & \makebox[20pt][l]{\rotatebox{45}{Qwen2.5-Coder (14B)}} & \makebox[20pt][l]{\rotatebox{45}{Qwen2.5-Coder (32B)}} & \makebox[20pt][l]{\rotatebox{45}{Llama 3.1 (8B)}} & \makebox[20pt][l]{\rotatebox{45}{Llama 3.1 (70B)}} & \makebox[20pt][l]{\rotatebox{45}{Code Llama (7B)}} & \makebox[20pt][l]{\rotatebox{45}{Code Llama (13B)}} & \makebox[20pt][l]{\rotatebox{45}{Code Llama (70B)}} \\
\midrule
Executable implementations (t) & \phantom{00}0.58 & \phantom{00}0.75 & \textbf{\phantom{00}0.93} & \phantom{00}0.85 & \phantom{00}0.87 & \phantom{00}0.86 & \phantom{00}0.81 & \phantom{00}0.81 & \phantom{00}0.85 & \phantom{00}0.84 & \phantom{00}0.79 & \phantom{00}0.79 & \phantom{00}0.69 & \phantom{00}0.76 & \phantom{00}0.74 & \phantom{00}0.81 & \phantom{00}0.80 & \phantom{00}0.86 & \phantom{00}0.80 & \phantom{00}0.80 & \phantom{00}0.73 & \phantom{00}0.85 & \phantom{00}0.81 \\
\bottomrule
\end{tabular}

}}
\end{table}
\end{landscape}

\begin{landscape}
\begin{table}[p]
\centering
\caption{Complete evaluation results for \emph{argument completion} with \emph{constrained decoding} on the \emph{synthetic dataset}}
\label{tab:res-full-synth-cd-end}
\makebox[1\textwidth][c]{
\resizebox{\tablescale\textwidth}{!}{
\begin{tabular}{lrrrrrrrrrrrrrrrrrrrrrrr}
\makebox[20pt][l]{\rotatebox{45}{}} & \makebox[20pt][l]{\rotatebox{45}{CodeT5+ (6B)}} & \makebox[20pt][l]{\rotatebox{45}{CodeT5+ (16B)}} & \makebox[20pt][l]{\rotatebox{45}{StarCoder (1B)}} & \makebox[20pt][l]{\rotatebox{45}{StarCoder (3B)}} & \makebox[20pt][l]{\rotatebox{45}{StarCoder (7B)}} & \makebox[20pt][l]{\rotatebox{45}{StarCoder (15.5B)}} & \makebox[20pt][l]{\rotatebox{45}{StarCoder2 (3B)}} & \makebox[20pt][l]{\rotatebox{45}{StarCoder2 (7B)}} & \makebox[20pt][l]{\rotatebox{45}{StarCoder2 (15B)}} & \makebox[20pt][l]{\rotatebox{45}{DeepSeek-Coder (1.3B)}} & \makebox[20pt][l]{\rotatebox{45}{DeepSeek-Coder (6.7B)}} & \makebox[20pt][l]{\rotatebox{45}{DeepSeek-Coder (33B)}} & \makebox[20pt][l]{\rotatebox{45}{Qwen2.5-Coder (0.5B)}} & \makebox[20pt][l]{\rotatebox{45}{Qwen2.5-Coder (1.5B)}} & \makebox[20pt][l]{\rotatebox{45}{Qwen2.5-Coder (3B)}} & \makebox[20pt][l]{\rotatebox{45}{Qwen2.5-Coder (7B)}} & \makebox[20pt][l]{\rotatebox{45}{Qwen2.5-Coder (14B)}} & \makebox[20pt][l]{\rotatebox{45}{Qwen2.5-Coder (32B)}} & \makebox[20pt][l]{\rotatebox{45}{Llama 3.1 (8B)}} & \makebox[20pt][l]{\rotatebox{45}{Llama 3.1 (70B)}} & \makebox[20pt][l]{\rotatebox{45}{Code Llama (7B)}} & \makebox[20pt][l]{\rotatebox{45}{Code Llama (13B)}} & \makebox[20pt][l]{\rotatebox{45}{Code Llama (70B)}} \\
\midrule
Executable implementations (t) & \phantom{00}0.55 & \phantom{00}0.74 & \phantom{00}0.89 & \phantom{00}0.84 & \textbf{\phantom{00}0.91} & \phantom{00}0.87 & \phantom{00}0.89 & \phantom{00}0.89 & \phantom{00}0.86 & \phantom{00}0.90 & \phantom{00}0.81 & \phantom{00}0.81 & \phantom{00}0.88 & \phantom{00}0.89 & \phantom{00}0.85 & \phantom{00}0.87 & \phantom{00}0.90 & \phantom{00}0.89 & \phantom{00}0.84 & \phantom{00}0.85 & \phantom{00}0.78 & \phantom{00}0.88 & \phantom{00}0.83 \\
\bottomrule
\end{tabular}

}}
\end{table}
\end{landscape}

\begin{landscape}
\begin{table}[p]
\centering
\caption{Complete evaluation results for \emph{full completion} with \emph{retrieval-augmented generation} plus \emph{constrained decoding} on the \emph{synthetic dataset}}
\label{tab:res-full-synth-rag+cd-inv}
\makebox[1\textwidth][c]{
\resizebox{\tablescale\textwidth}{!}{

}}
\end{table}
\end{landscape}

\begin{landscape}
\begin{table}[p]
\centering
\caption{Complete evaluation results for \emph{argument completion} with \emph{retrieval-augmented generation} plus \emph{constrained decoding} on the \emph{synthetic dataset}}
\label{tab:res-full-synth-rag+cd-end}
\makebox[1\textwidth][c]{
\resizebox{\tablescale\textwidth}{!}{

}}
\end{table}
\end{landscape}

\begin{landscape}
\begin{table}[p]
\centering
\caption{Complete evaluation results for \emph{full completion} with \emph{vanilla} generation on the \emph{real-world dataset}}
\label{tab:res-full-real-vanilla-inv}
\makebox[1\textwidth][c]{
\resizebox{\tablescale\textwidth}{!}{
\begin{tabular}{lrrrrrrrrrrrrrrrrrrrrrrrrrr}
\makebox[20pt][l]{\rotatebox{45}{}} & \makebox[20pt][l]{\rotatebox{45}{CodeT5+ (6B)}} & \makebox[20pt][l]{\rotatebox{45}{CodeT5+ (16B)}} & \makebox[20pt][l]{\rotatebox{45}{StarCoder (1B)}} & \makebox[20pt][l]{\rotatebox{45}{StarCoder (3B)}} & \makebox[20pt][l]{\rotatebox{45}{StarCoder (7B)}} & \makebox[20pt][l]{\rotatebox{45}{StarCoder (15.5B)}} & \makebox[20pt][l]{\rotatebox{45}{StarCoder2 (3B)}} & \makebox[20pt][l]{\rotatebox{45}{StarCoder2 (7B)}} & \makebox[20pt][l]{\rotatebox{45}{StarCoder2 (15B)}} & \makebox[20pt][l]{\rotatebox{45}{DeepSeek-Coder (1.3B)}} & \makebox[20pt][l]{\rotatebox{45}{DeepSeek-Coder (6.7B)}} & \makebox[20pt][l]{\rotatebox{45}{DeepSeek-Coder (33B)}} & \makebox[20pt][l]{\rotatebox{45}{Qwen2.5-Coder (0.5B)}} & \makebox[20pt][l]{\rotatebox{45}{Qwen2.5-Coder (1.5B)}} & \makebox[20pt][l]{\rotatebox{45}{Qwen2.5-Coder (3B)}} & \makebox[20pt][l]{\rotatebox{45}{Qwen2.5-Coder (7B)}} & \makebox[20pt][l]{\rotatebox{45}{Qwen2.5-Coder (14B)}} & \makebox[20pt][l]{\rotatebox{45}{Qwen2.5-Coder (32B)}} & \makebox[20pt][l]{\rotatebox{45}{Llama 3.1 (8B)}} & \makebox[20pt][l]{\rotatebox{45}{Llama 3.1 (70B)}} & \makebox[20pt][l]{\rotatebox{45}{Code Llama (7B)}} & \makebox[20pt][l]{\rotatebox{45}{Code Llama (13B)}} & \makebox[20pt][l]{\rotatebox{45}{Code Llama (34B)}} & \makebox[20pt][l]{\rotatebox{45}{Code Llama (70B)}} & \makebox[20pt][l]{\rotatebox{45}{GPT-4o mini}} & \makebox[20pt][l]{\rotatebox{45}{GPT-4o}} \\
\midrule
Executable implementations (t) & \phantom{00}0.89 & \phantom{00}0.96 & \textbf{1.00} & \textbf{1.00} & \textbf{1.00} & \textbf{1.00} & \textbf{1.00} & \textbf{1.00} & \textbf{1.00} & \textbf{1.00} & \textbf{1.00} & \textbf{1.00} & \phantom{00}0.61 & \phantom{00}0.00 & \phantom{00}0.00 & \phantom{00}0.14 & \phantom{00}0.00 & \phantom{00}0.25 & \phantom{00}0.14 & \phantom{00}0.00 & \textbf{1.00} & \textbf{1.00} & \textbf{1.00} & \textbf{1.00} & \textbf{1.00} & \phantom{00}0.79 \\
\bottomrule
\end{tabular}

}}
\end{table}
\end{landscape}

\begin{landscape}
\begin{table}[p]
\centering
\caption{Complete evaluation results for \emph{argument completion} with \emph{vanilla} generation on the \emph{real-world dataset}}
\label{tab:res-full-real-vanilla-end}
\makebox[1\textwidth][c]{
\resizebox{\tablescale\textwidth}{!}{
\begin{tabular}{lrrrrrrrrrrrrrrrrrrrrrrrrrr}
\makebox[20pt][l]{\rotatebox{45}{}} & \makebox[20pt][l]{\rotatebox{45}{CodeT5+ (6B)}} & \makebox[20pt][l]{\rotatebox{45}{CodeT5+ (16B)}} & \makebox[20pt][l]{\rotatebox{45}{StarCoder (1B)}} & \makebox[20pt][l]{\rotatebox{45}{StarCoder (3B)}} & \makebox[20pt][l]{\rotatebox{45}{StarCoder (7B)}} & \makebox[20pt][l]{\rotatebox{45}{StarCoder (15.5B)}} & \makebox[20pt][l]{\rotatebox{45}{StarCoder2 (3B)}} & \makebox[20pt][l]{\rotatebox{45}{StarCoder2 (7B)}} & \makebox[20pt][l]{\rotatebox{45}{StarCoder2 (15B)}} & \makebox[20pt][l]{\rotatebox{45}{DeepSeek-Coder (1.3B)}} & \makebox[20pt][l]{\rotatebox{45}{DeepSeek-Coder (6.7B)}} & \makebox[20pt][l]{\rotatebox{45}{DeepSeek-Coder (33B)}} & \makebox[20pt][l]{\rotatebox{45}{Qwen2.5-Coder (0.5B)}} & \makebox[20pt][l]{\rotatebox{45}{Qwen2.5-Coder (1.5B)}} & \makebox[20pt][l]{\rotatebox{45}{Qwen2.5-Coder (3B)}} & \makebox[20pt][l]{\rotatebox{45}{Qwen2.5-Coder (7B)}} & \makebox[20pt][l]{\rotatebox{45}{Qwen2.5-Coder (14B)}} & \makebox[20pt][l]{\rotatebox{45}{Qwen2.5-Coder (32B)}} & \makebox[20pt][l]{\rotatebox{45}{Llama 3.1 (8B)}} & \makebox[20pt][l]{\rotatebox{45}{Llama 3.1 (70B)}} & \makebox[20pt][l]{\rotatebox{45}{Code Llama (7B)}} & \makebox[20pt][l]{\rotatebox{45}{Code Llama (13B)}} & \makebox[20pt][l]{\rotatebox{45}{Code Llama (34B)}} & \makebox[20pt][l]{\rotatebox{45}{Code Llama (70B)}} & \makebox[20pt][l]{\rotatebox{45}{GPT-4o mini}} & \makebox[20pt][l]{\rotatebox{45}{GPT-4o}} \\
\bottomrule
\end{tabular}

}}
\end{table}
\end{landscape}

\begin{landscape}
\begin{table}[p]
\centering
\caption{Complete evaluation results for \emph{full completion} with \emph{retrieval-augmented generation} on the \emph{real-world dataset}}
\label{tab:res-full-real-rag-inv}
\makebox[1\textwidth][c]{
\resizebox{\tablescale\textwidth}{!}{
\begin{tabular}{lrrrrrrrrrrrrrrrrrrrrrrrrrr}
\makebox[20pt][l]{\rotatebox{45}{}} & \makebox[20pt][l]{\rotatebox{45}{CodeT5+ (6B)}} & \makebox[20pt][l]{\rotatebox{45}{CodeT5+ (16B)}} & \makebox[20pt][l]{\rotatebox{45}{StarCoder (1B)}} & \makebox[20pt][l]{\rotatebox{45}{StarCoder (3B)}} & \makebox[20pt][l]{\rotatebox{45}{StarCoder (7B)}} & \makebox[20pt][l]{\rotatebox{45}{StarCoder (15.5B)}} & \makebox[20pt][l]{\rotatebox{45}{StarCoder2 (3B)}} & \makebox[20pt][l]{\rotatebox{45}{StarCoder2 (7B)}} & \makebox[20pt][l]{\rotatebox{45}{StarCoder2 (15B)}} & \makebox[20pt][l]{\rotatebox{45}{DeepSeek-Coder (1.3B)}} & \makebox[20pt][l]{\rotatebox{45}{DeepSeek-Coder (6.7B)}} & \makebox[20pt][l]{\rotatebox{45}{DeepSeek-Coder (33B)}} & \makebox[20pt][l]{\rotatebox{45}{Qwen2.5-Coder (0.5B)}} & \makebox[20pt][l]{\rotatebox{45}{Qwen2.5-Coder (1.5B)}} & \makebox[20pt][l]{\rotatebox{45}{Qwen2.5-Coder (3B)}} & \makebox[20pt][l]{\rotatebox{45}{Qwen2.5-Coder (7B)}} & \makebox[20pt][l]{\rotatebox{45}{Qwen2.5-Coder (14B)}} & \makebox[20pt][l]{\rotatebox{45}{Qwen2.5-Coder (32B)}} & \makebox[20pt][l]{\rotatebox{45}{Llama 3.1 (8B)}} & \makebox[20pt][l]{\rotatebox{45}{Llama 3.1 (70B)}} & \makebox[20pt][l]{\rotatebox{45}{Code Llama (7B)}} & \makebox[20pt][l]{\rotatebox{45}{Code Llama (13B)}} & \makebox[20pt][l]{\rotatebox{45}{Code Llama (34B)}} & \makebox[20pt][l]{\rotatebox{45}{Code Llama (70B)}} & \makebox[20pt][l]{\rotatebox{45}{GPT-4o mini}} & \makebox[20pt][l]{\rotatebox{45}{GPT-4o}} \\
\midrule
Executable implementations (t) & \phantom{00}0.61 & \phantom{00}0.75 & \textbf{1.00} & \phantom{00}0.96 & \textbf{1.00} & \phantom{00}0.93 & \phantom{00}0.96 & \textbf{1.00} & \textbf{1.00} & \textbf{1.00} & \phantom{00}0.96 & \textbf{1.00} & \phantom{00}0.50 & \phantom{00}0.00 & \phantom{00}0.00 & \phantom{00}0.07 & \phantom{00}0.04 & \phantom{00}0.21 & \phantom{00}0.21 & \phantom{00}0.00 & \phantom{00}0.89 & \textbf{1.00} & \textbf{1.00} & \textbf{1.00} & \textbf{1.00} & \phantom{00}0.89 \\
\bottomrule
\end{tabular}

}}
\end{table}
\end{landscape}

\begin{landscape}
\begin{table}[p]
\centering
\caption{Complete evaluation results for \emph{argument completion} with \emph{retrieval-augmented generation} on the \emph{real-world dataset}}
\label{tab:res-full-real-rag-end}
\makebox[1\textwidth][c]{
\resizebox{\tablescale\textwidth}{!}{
\begin{tabular}{lrrrrrrrrrrrrrrrrrrrrrrrrrr}
\makebox[20pt][l]{\rotatebox{45}{}} & \makebox[20pt][l]{\rotatebox{45}{CodeT5+ (6B)}} & \makebox[20pt][l]{\rotatebox{45}{CodeT5+ (16B)}} & \makebox[20pt][l]{\rotatebox{45}{StarCoder (1B)}} & \makebox[20pt][l]{\rotatebox{45}{StarCoder (3B)}} & \makebox[20pt][l]{\rotatebox{45}{StarCoder (7B)}} & \makebox[20pt][l]{\rotatebox{45}{StarCoder (15.5B)}} & \makebox[20pt][l]{\rotatebox{45}{StarCoder2 (3B)}} & \makebox[20pt][l]{\rotatebox{45}{StarCoder2 (7B)}} & \makebox[20pt][l]{\rotatebox{45}{StarCoder2 (15B)}} & \makebox[20pt][l]{\rotatebox{45}{DeepSeek-Coder (1.3B)}} & \makebox[20pt][l]{\rotatebox{45}{DeepSeek-Coder (6.7B)}} & \makebox[20pt][l]{\rotatebox{45}{DeepSeek-Coder (33B)}} & \makebox[20pt][l]{\rotatebox{45}{Qwen2.5-Coder (0.5B)}} & \makebox[20pt][l]{\rotatebox{45}{Qwen2.5-Coder (1.5B)}} & \makebox[20pt][l]{\rotatebox{45}{Qwen2.5-Coder (3B)}} & \makebox[20pt][l]{\rotatebox{45}{Qwen2.5-Coder (7B)}} & \makebox[20pt][l]{\rotatebox{45}{Qwen2.5-Coder (14B)}} & \makebox[20pt][l]{\rotatebox{45}{Qwen2.5-Coder (32B)}} & \makebox[20pt][l]{\rotatebox{45}{Llama 3.1 (8B)}} & \makebox[20pt][l]{\rotatebox{45}{Llama 3.1 (70B)}} & \makebox[20pt][l]{\rotatebox{45}{Code Llama (7B)}} & \makebox[20pt][l]{\rotatebox{45}{Code Llama (13B)}} & \makebox[20pt][l]{\rotatebox{45}{Code Llama (34B)}} & \makebox[20pt][l]{\rotatebox{45}{Code Llama (70B)}} & \makebox[20pt][l]{\rotatebox{45}{GPT-4o mini}} & \makebox[20pt][l]{\rotatebox{45}{GPT-4o}} \\
\midrule
Executable implementations (t) & \phantom{00}0.75 & \phantom{00}0.79 & \phantom{00}0.89 & \phantom{00}0.89 & \phantom{00}0.96 & \phantom{00}0.93 & \phantom{00}0.96 & \phantom{00}0.86 & \phantom{00}0.89 & \phantom{00}0.96 & \phantom{00}0.89 & \textbf{1.00} & \phantom{00}0.82 & \phantom{00}0.93 & \phantom{00}0.86 & \textbf{1.00} & \phantom{00}0.89 & \phantom{00}0.89 & \phantom{00}0.68 & \textbf{1.00} & \textbf{1.00} & \phantom{00}0.71 & \phantom{00}0.96 & \phantom{00}0.86 & \phantom{00}0.96 & \phantom{00}0.89 \\
\bottomrule
\end{tabular}

}}
\end{table}
\end{landscape}

\begin{landscape}
\begin{table}[p]
\centering
\caption{Complete evaluation results for \emph{full completion} with \emph{constrained decoding} on the \emph{real-world dataset}}
\label{tab:res-full-real-cd-inv}
\makebox[1\textwidth][c]{
\resizebox{\tablescale\textwidth}{!}{
\begin{tabular}{lrrrrrrrrrrrrrrrrrrrrrrr}
\makebox[20pt][l]{\rotatebox{45}{}} & \makebox[20pt][l]{\rotatebox{45}{CodeT5+ (6B)}} & \makebox[20pt][l]{\rotatebox{45}{CodeT5+ (16B)}} & \makebox[20pt][l]{\rotatebox{45}{StarCoder (1B)}} & \makebox[20pt][l]{\rotatebox{45}{StarCoder (3B)}} & \makebox[20pt][l]{\rotatebox{45}{StarCoder (7B)}} & \makebox[20pt][l]{\rotatebox{45}{StarCoder (15.5B)}} & \makebox[20pt][l]{\rotatebox{45}{StarCoder2 (3B)}} & \makebox[20pt][l]{\rotatebox{45}{StarCoder2 (7B)}} & \makebox[20pt][l]{\rotatebox{45}{StarCoder2 (15B)}} & \makebox[20pt][l]{\rotatebox{45}{DeepSeek-Coder (1.3B)}} & \makebox[20pt][l]{\rotatebox{45}{DeepSeek-Coder (6.7B)}} & \makebox[20pt][l]{\rotatebox{45}{DeepSeek-Coder (33B)}} & \makebox[20pt][l]{\rotatebox{45}{Qwen2.5-Coder (0.5B)}} & \makebox[20pt][l]{\rotatebox{45}{Qwen2.5-Coder (1.5B)}} & \makebox[20pt][l]{\rotatebox{45}{Qwen2.5-Coder (3B)}} & \makebox[20pt][l]{\rotatebox{45}{Qwen2.5-Coder (7B)}} & \makebox[20pt][l]{\rotatebox{45}{Qwen2.5-Coder (14B)}} & \makebox[20pt][l]{\rotatebox{45}{Qwen2.5-Coder (32B)}} & \makebox[20pt][l]{\rotatebox{45}{Llama 3.1 (8B)}} & \makebox[20pt][l]{\rotatebox{45}{Llama 3.1 (70B)}} & \makebox[20pt][l]{\rotatebox{45}{Code Llama (7B)}} & \makebox[20pt][l]{\rotatebox{45}{Code Llama (13B)}} & \makebox[20pt][l]{\rotatebox{45}{Code Llama (70B)}} \\
\midrule
Executable implementations (t) & \phantom{00}0.89 & \phantom{00}0.86 & \phantom{00}0.89 & \textbf{\phantom{00}0.96} & \textbf{\phantom{00}0.96} & \phantom{00}0.89 & \phantom{00}0.93 & \phantom{00}0.93 & \phantom{00}0.89 & \textbf{\phantom{00}0.96} & \phantom{00}0.89 & \phantom{00}0.93 & \phantom{00}0.68 & \phantom{00}0.68 & \phantom{00}0.75 & \phantom{00}0.93 & \phantom{00}0.86 & \phantom{00}0.86 & \phantom{00}0.89 & \phantom{00}0.89 & \phantom{00}0.93 & \phantom{00}0.89 & \phantom{00}0.82 \\
\bottomrule
\end{tabular}

}}
\end{table}
\end{landscape}

\begin{landscape}
\begin{table}[p]
\centering
\caption{Complete evaluation results for \emph{argument completion} with \emph{constrained decoding} on the \emph{real-world dataset}}
\label{tab:res-full-real-cd-end}
\makebox[1\textwidth][c]{
\resizebox{\tablescale\textwidth}{!}{
\begin{tabular}{lrrrrrrrrrrrrrrrrrrrrrrr}
\makebox[20pt][l]{\rotatebox{45}{}} & \makebox[20pt][l]{\rotatebox{45}{CodeT5+ (6B)}} & \makebox[20pt][l]{\rotatebox{45}{CodeT5+ (16B)}} & \makebox[20pt][l]{\rotatebox{45}{StarCoder (1B)}} & \makebox[20pt][l]{\rotatebox{45}{StarCoder (3B)}} & \makebox[20pt][l]{\rotatebox{45}{StarCoder (7B)}} & \makebox[20pt][l]{\rotatebox{45}{StarCoder (15.5B)}} & \makebox[20pt][l]{\rotatebox{45}{StarCoder2 (3B)}} & \makebox[20pt][l]{\rotatebox{45}{StarCoder2 (7B)}} & \makebox[20pt][l]{\rotatebox{45}{StarCoder2 (15B)}} & \makebox[20pt][l]{\rotatebox{45}{DeepSeek-Coder (1.3B)}} & \makebox[20pt][l]{\rotatebox{45}{DeepSeek-Coder (6.7B)}} & \makebox[20pt][l]{\rotatebox{45}{DeepSeek-Coder (33B)}} & \makebox[20pt][l]{\rotatebox{45}{Qwen2.5-Coder (0.5B)}} & \makebox[20pt][l]{\rotatebox{45}{Qwen2.5-Coder (1.5B)}} & \makebox[20pt][l]{\rotatebox{45}{Qwen2.5-Coder (3B)}} & \makebox[20pt][l]{\rotatebox{45}{Qwen2.5-Coder (7B)}} & \makebox[20pt][l]{\rotatebox{45}{Qwen2.5-Coder (14B)}} & \makebox[20pt][l]{\rotatebox{45}{Qwen2.5-Coder (32B)}} & \makebox[20pt][l]{\rotatebox{45}{Llama 3.1 (8B)}} & \makebox[20pt][l]{\rotatebox{45}{Llama 3.1 (70B)}} & \makebox[20pt][l]{\rotatebox{45}{Code Llama (7B)}} & \makebox[20pt][l]{\rotatebox{45}{Code Llama (13B)}} & \makebox[20pt][l]{\rotatebox{45}{Code Llama (70B)}} \\
\midrule
Executable implementations (t) & \phantom{00}0.89 & \phantom{00}0.86 & \phantom{00}0.89 & \phantom{00}0.96 & \phantom{00}0.93 & \phantom{00}0.93 & \textbf{1.00} & \phantom{00}0.96 & \phantom{00}0.89 & \phantom{00}0.96 & \phantom{00}0.93 & \phantom{00}0.93 & \phantom{00}0.79 & \phantom{00}0.96 & \phantom{00}0.89 & \phantom{00}0.96 & \phantom{00}0.96 & \phantom{00}0.96 & \phantom{00}0.93 & \phantom{00}0.93 & \phantom{00}0.86 & \phantom{00}0.82 & \phantom{00}0.79 \\
\bottomrule
\end{tabular}

}}
\end{table}
\end{landscape}

\begin{landscape}
\begin{table}[p]
\centering
\caption{Complete evaluation results for \emph{full completion} with \emph{retrieval-augmented generation} plus \emph{constrained decoding} on the \emph{real-world dataset}}
\label{tab:res-full-real-rag+cd-inv}
\makebox[1\textwidth][c]{
\resizebox{\tablescale\textwidth}{!}{

}}
\end{table}
\end{landscape}

\begin{landscape}
\begin{table}[p]
\centering
\caption{Complete evaluation results for \emph{argument completion} with \emph{retrieval-augmented generation} plus \emph{constrained decoding} on the \emph{real-world dataset}}
\label{tab:res-full-real-rag+cd-end}
\makebox[1\textwidth][c]{
\resizebox{\tablescale\textwidth}{!}{

}}
\end{table}
\end{landscape}

\begin{table}[p]
\centering
\caption{Relative change of \emph{correct implementations} from vanilla generation to \emph{retrieval-augmented generation}~(RAG) for both setups and both datasets. Models with a 0\% baseline score are excluded from the calculation of the average gain.}
\label{tab:res-gain-correct-rag}
\footnotesize
\begin{subtable}[b]{0.49\textwidth}
\centering
\caption{Full completion, synthetic dataset}
\label{tab:res-gain-correct-rag-inv-synth}
\begin{tabular}{llll}
\toprule
 & Vanilla & RAG & Gain \\
\midrule
CodeT5+ (6B) & 0.04 & 0.00 & -100\% \\
CodeT5+ (16B) & 0.07 & 0.08 & +19\% \\
StarCoder (1B) & 0.03 & 0.22 & +562\% \\
StarCoder (3B) & 0.05 & 0.28 & +474\% \\
StarCoder (7B) & 0.12 & 0.34 & +179\% \\
StarCoder (15.5B) & 0.13 & 0.36 & +166\% \\
StarCoder2 (3B) & 0.13 & 0.33 & +150\% \\
StarCoder2 (7B) & 0.10 & 0.14 & +32\% \\
StarCoder2 (15B) & 0.25 & 0.33 & +33\% \\
DeepSeek-Coder (1.3B) & 0.07 & 0.06 & -21\% \\
DeepSeek-Coder (6.7B) & 0.14 & 0.20 & +39\% \\
DeepSeek-Coder (33B) & 0.17 & 0.24 & +39\% \\
Qwen2.5-Coder (0.5B) & 0.00 & 0.00 & N/A \\
Qwen2.5-Coder (1.5B) & 0.00 & 0.00 & N/A \\
Qwen2.5-Coder (3B) & 0.00 & 0.00 & N/A \\
Qwen2.5-Coder (7B) & 0.00 & 0.00 & N/A \\
Qwen2.5-Coder (14B) & 0.00 & 0.00 & N/A \\
Qwen2.5-Coder (32B) & 0.17 & 0.13 & -20\% \\
Llama 3.1 (8B) & 0.00 & 0.00 & N/A \\
Llama 3.1 (70B) & 0.00 & 0.00 & N/A \\
Code Llama (7B) & 0.03 & 0.07 & +155\% \\
Code Llama (13B) & 0.18 & 0.04 & -80\% \\
Code Llama (34B) & 0.17 & 0.23 & +39\% \\
Code Llama (70B) & 0.30 & 0.42 & +39\% \\
\midrule
Average & 0.09 & 0.14 & +100\% \\
Minimum & 0.00 & 0.00 & -100\% \\
Maximum & 0.30 & 0.42 & +562\% \\
\bottomrule
\end{tabular}

\end{subtable}
\hfill
\begin{subtable}[b]{0.49\textwidth}
\centering
\caption{Argument completion, synthetic dataset}
\label{tab:res-gain-correct-rag-end-synth}
\begin{tabular}{llll}
\toprule
 & Vanilla & RAG & Gain \\
\midrule
CodeT5+ (6B) & 0.24 & 0.05 & -78\% \\
CodeT5+ (16B) & 0.21 & 0.09 & -59\% \\
StarCoder (1B) & 0.22 & 0.26 & +15\% \\
StarCoder (3B) & 0.23 & 0.30 & +32\% \\
StarCoder (7B) & 0.29 & 0.38 & +28\% \\
StarCoder (15.5B) & 0.28 & 0.39 & +37\% \\
StarCoder2 (3B) & 0.22 & 0.35 & +56\% \\
StarCoder2 (7B) & 0.19 & 0.15 & -23\% \\
StarCoder2 (15B) & 0.24 & 0.29 & +21\% \\
DeepSeek-Coder (1.3B) & 0.19 & 0.09 & -55\% \\
DeepSeek-Coder (6.7B) & 0.29 & 0.23 & -19\% \\
DeepSeek-Coder (33B) & 0.26 & 0.28 & +6\% \\
Qwen2.5-Coder (0.5B) & 0.17 & 0.19 & +10\% \\
Qwen2.5-Coder (1.5B) & 0.24 & 0.07 & -70\% \\
Qwen2.5-Coder (3B) & 0.27 & 0.37 & +38\% \\
Qwen2.5-Coder (7B) & 0.32 & 0.44 & +38\% \\
Qwen2.5-Coder (14B) & 0.47 & 0.46 & -2\% \\
Qwen2.5-Coder (32B) & 0.43 & 0.45 & +6\% \\
Llama 3.1 (8B) & 0.10 & 0.09 & -13\% \\
Llama 3.1 (70B) & 0.30 & 0.29 & -6\% \\
Code Llama (7B) & 0.05 & 0.08 & +63\% \\
Code Llama (13B) & 0.26 & 0.05 & -83\% \\
Code Llama (34B) & 0.40 & 0.29 & -28\% \\
Code Llama (70B) & 0.42 & 0.46 & +9\% \\
\midrule
Average & 0.26 & 0.25 & -3\% \\
Minimum & 0.05 & 0.05 & -83\% \\
Maximum & 0.47 & 0.46 & +63\% \\
\bottomrule
\end{tabular}

\end{subtable}
\newline
\vspace*{1em}
\newline
\begin{subtable}[b]{0.49\textwidth}
\centering
\caption{Full completion, real-world dataset}
\label{tab:res-gain-correct-rag-inv-real}
\begin{tabular}{llll}
\toprule
 & Vanilla & RAG & Gain \\
\midrule
CodeT5+ (6B) & 0.39 & 0.25 & -36\% \\
CodeT5+ (16B) & 0.54 & 0.39 & -27\% \\
StarCoder (1B) & 0.46 & 0.46 & +0\% \\
StarCoder (3B) & 0.57 & 0.54 & -6\% \\
StarCoder (7B) & 0.61 & 0.64 & +6\% \\
StarCoder (15.5B) & 0.50 & 0.54 & +7\% \\
StarCoder2 (3B) & 0.64 & 0.61 & -6\% \\
StarCoder2 (7B) & 0.46 & 0.46 & +0\% \\
StarCoder2 (15B) & 0.54 & 0.61 & +13\% \\
DeepSeek-Coder (1.3B) & 0.57 & 0.68 & +19\% \\
DeepSeek-Coder (6.7B) & 0.54 & 0.46 & -13\% \\
DeepSeek-Coder (33B) & 0.54 & 0.68 & +27\% \\
Qwen2.5-Coder (0.5B) & 0.39 & 0.29 & -27\% \\
Qwen2.5-Coder (1.5B) & 0.00 & 0.00 & N/A \\
Qwen2.5-Coder (3B) & 0.00 & 0.00 & N/A \\
Qwen2.5-Coder (7B) & 0.07 & 0.04 & -50\% \\
Qwen2.5-Coder (14B) & 0.00 & 0.04 & N/A \\
Qwen2.5-Coder (32B) & 0.18 & 0.21 & +20\% \\
Llama 3.1 (8B) & 0.11 & 0.07 & -33\% \\
Llama 3.1 (70B) & 0.00 & 0.00 & N/A \\
Code Llama (7B) & 0.61 & 0.57 & -6\% \\
Code Llama (13B) & 0.68 & 0.61 & -11\% \\
Code Llama (34B) & 0.39 & 0.43 & +9\% \\
Code Llama (70B) & 0.71 & 0.71 & +0\% \\
\midrule
Average & 0.40 & 0.39 & -6\% \\
Minimum & 0.00 & 0.00 & -50\% \\
Maximum & 0.71 & 0.71 & +27\% \\
\bottomrule
\end{tabular}

\end{subtable}
\hfill
\begin{subtable}[b]{0.49\textwidth}
\centering
\caption{Argument completion, real-world dataset}
\label{tab:res-gain-correct-rag-end-real}
\begin{tabular}{llll}
\toprule
 & Vanilla & RAG & Gain \\
\midrule
CodeT5+ (6B) & 0.54 & 0.46 & -13\% \\
CodeT5+ (16B) & 0.57 & 0.54 & -6\% \\
StarCoder (1B) & 0.68 & 0.57 & -16\% \\
StarCoder (3B) & 0.50 & 0.46 & -7\% \\
StarCoder (7B) & 0.57 & 0.57 & +0\% \\
StarCoder (15.5B) & 0.57 & 0.64 & +13\% \\
StarCoder2 (3B) & 0.50 & 0.61 & +21\% \\
StarCoder2 (7B) & 0.39 & 0.46 & +18\% \\
StarCoder2 (15B) & 0.50 & 0.54 & +7\% \\
DeepSeek-Coder (1.3B) & 0.50 & 0.61 & +21\% \\
DeepSeek-Coder (6.7B) & 0.71 & 0.75 & +5\% \\
DeepSeek-Coder (33B) & 0.64 & 0.68 & +6\% \\
Qwen2.5-Coder (0.5B) & 0.43 & 0.46 & +8\% \\
Qwen2.5-Coder (1.5B) & 0.61 & 0.54 & -12\% \\
Qwen2.5-Coder (3B) & 0.61 & 0.50 & -18\% \\
Qwen2.5-Coder (7B) & 0.61 & 0.57 & -6\% \\
Qwen2.5-Coder (14B) & 0.68 & 0.46 & -32\% \\
Qwen2.5-Coder (32B) & 0.46 & 0.54 & +15\% \\
Llama 3.1 (8B) & 0.39 & 0.46 & +18\% \\
Llama 3.1 (70B) & 0.00 & 0.64 & N/A \\
Code Llama (7B) & 0.57 & 0.68 & +19\% \\
Code Llama (13B) & 0.68 & 0.57 & -16\% \\
Code Llama (34B) & 0.57 & 0.57 & +0\% \\
Code Llama (70B) & 0.43 & 0.46 & +8\% \\
\midrule
Average & 0.53 & 0.56 & +2\% \\
Minimum & 0.00 & 0.46 & -32\% \\
Maximum & 0.71 & 0.75 & +21\% \\
\bottomrule
\end{tabular}

\end{subtable}
\end{table}

\begin{table}[p]
\centering
\caption{Relative change of \emph{correct implementations} from vanilla generation to \emph{constrained decoding}~(CD) for both setups and both datasets. Models with a 0\% baseline score are excluded from the calculation of the average gain.}
\label{tab:res-gain-correct-cd}
\footnotesize
\begin{subtable}[b]{0.49\textwidth}
\centering
\caption{Full completion, synthetic dataset}
\label{tab:res-gain-correct-cd-inv-synth}
\begin{tabular}{llll}
\toprule
 & Vanilla & CD & Gain \\
\midrule
CodeT5+ (6B) & 0.04 & 0.17 & +294\% \\
CodeT5+ (16B) & 0.07 & 0.18 & +159\% \\
StarCoder (1B) & 0.03 & 0.19 & +477\% \\
StarCoder (3B) & 0.05 & 0.32 & +558\% \\
StarCoder (7B) & 0.12 & 0.37 & +206\% \\
StarCoder (15.5B) & 0.13 & 0.30 & +121\% \\
StarCoder2 (3B) & 0.13 & 0.35 & +163\% \\
StarCoder2 (7B) & 0.10 & 0.36 & +246\% \\
StarCoder2 (15B) & 0.25 & 0.46 & +85\% \\
DeepSeek-Coder (1.3B) & 0.07 & 0.21 & +190\% \\
DeepSeek-Coder (6.7B) & 0.14 & 0.42 & +189\% \\
DeepSeek-Coder (33B) & 0.17 & 0.46 & +162\% \\
Qwen2.5-Coder (0.5B) & 0.00 & 0.07 & N/A \\
Qwen2.5-Coder (1.5B) & 0.00 & 0.10 & N/A \\
Qwen2.5-Coder (3B) & 0.00 & 0.17 & N/A \\
Qwen2.5-Coder (7B) & 0.00 & 0.34 & N/A \\
Qwen2.5-Coder (14B) & 0.00 & 0.19 & N/A \\
Qwen2.5-Coder (32B) & 0.17 & 0.42 & +152\% \\
Llama 3.1 (8B) & 0.00 & 0.23 & N/A \\
Llama 3.1 (70B) & 0.00 & 0.32 & N/A \\
Code Llama (7B) & 0.03 & 0.09 & +217\% \\
Code Llama (13B) & 0.18 & 0.28 & +59\% \\
Code Llama (70B) & 0.30 & 0.50 & +65\% \\
\midrule
Average & 0.09 & 0.28 & +209\% \\
Minimum & 0.00 & 0.07 & +59\% \\
Maximum & 0.30 & 0.50 & +558\% \\
\bottomrule
\end{tabular}

\end{subtable}
\hfill
\begin{subtable}[b]{0.49\textwidth}
\centering
\caption{Argument completion, synthetic dataset}
\label{tab:res-gain-correct-cd-end-synth}
\begin{tabular}{llll}
\toprule
 & Vanilla & CD & Gain \\
\midrule
CodeT5+ (6B) & 0.24 & 0.36 & +53\% \\
CodeT5+ (16B) & 0.21 & 0.52 & +149\% \\
StarCoder (1B) & 0.22 & 0.56 & +150\% \\
StarCoder (3B) & 0.23 & 0.57 & +152\% \\
StarCoder (7B) & 0.29 & 0.66 & +124\% \\
StarCoder (15.5B) & 0.28 & 0.46 & +63\% \\
StarCoder2 (3B) & 0.22 & 0.63 & +183\% \\
StarCoder2 (7B) & 0.19 & 0.67 & +242\% \\
StarCoder2 (15B) & 0.24 & 0.69 & +185\% \\
DeepSeek-Coder (1.3B) & 0.19 & 0.42 & +116\% \\
DeepSeek-Coder (6.7B) & 0.29 & 0.59 & +108\% \\
DeepSeek-Coder (33B) & 0.26 & 0.66 & +153\% \\
Qwen2.5-Coder (0.5B) & 0.17 & 0.39 & +130\% \\
Qwen2.5-Coder (1.5B) & 0.24 & 0.65 & +171\% \\
Qwen2.5-Coder (3B) & 0.27 & 0.62 & +134\% \\
Qwen2.5-Coder (7B) & 0.32 & 0.63 & +98\% \\
Qwen2.5-Coder (14B) & 0.47 & 0.67 & +43\% \\
Qwen2.5-Coder (32B) & 0.43 & 0.66 & +55\% \\
Llama 3.1 (8B) & 0.10 & 0.60 & +508\% \\
Llama 3.1 (70B) & 0.30 & 0.65 & +115\% \\
Code Llama (7B) & 0.05 & 0.14 & +200\% \\
Code Llama (13B) & 0.26 & 0.49 & +86\% \\
Code Llama (70B) & 0.42 & 0.72 & +70\% \\
\midrule
Average & 0.26 & 0.57 & +143\% \\
Minimum & 0.05 & 0.14 & +43\% \\
Maximum & 0.47 & 0.72 & +508\% \\
\bottomrule
\end{tabular}

\end{subtable}
\newline
\vspace*{1em}
\newline
\begin{subtable}[b]{0.49\textwidth}
\centering
\caption{Full completion, real-world dataset}
\label{tab:res-gain-correct-cd-inv-real}
\begin{tabular}{llll}
\toprule
 & Vanilla & CD & Gain \\
\midrule
CodeT5+ (6B) & 0.39 & 0.29 & -27\% \\
CodeT5+ (16B) & 0.54 & 0.43 & -20\% \\
StarCoder (1B) & 0.46 & 0.43 & -8\% \\
StarCoder (3B) & 0.57 & 0.57 & +0\% \\
StarCoder (7B) & 0.61 & 0.64 & +6\% \\
StarCoder (15.5B) & 0.50 & 0.54 & +7\% \\
StarCoder2 (3B) & 0.64 & 0.64 & +0\% \\
StarCoder2 (7B) & 0.46 & 0.50 & +8\% \\
StarCoder2 (15B) & 0.54 & 0.57 & +7\% \\
DeepSeek-Coder (1.3B) & 0.57 & 0.57 & +0\% \\
DeepSeek-Coder (6.7B) & 0.54 & 0.61 & +13\% \\
DeepSeek-Coder (33B) & 0.54 & 0.68 & +27\% \\
Qwen2.5-Coder (0.5B) & 0.39 & 0.07 & -82\% \\
Qwen2.5-Coder (1.5B) & 0.00 & 0.04 & N/A \\
Qwen2.5-Coder (3B) & 0.00 & 0.25 & N/A \\
Qwen2.5-Coder (7B) & 0.07 & 0.64 & +800\% \\
Qwen2.5-Coder (14B) & 0.00 & 0.57 & N/A \\
Qwen2.5-Coder (32B) & 0.18 & 0.54 & +200\% \\
Llama 3.1 (8B) & 0.11 & 0.54 & +400\% \\
Llama 3.1 (70B) & 0.00 & 0.61 & N/A \\
Code Llama (7B) & 0.61 & 0.61 & +0\% \\
Code Llama (13B) & 0.68 & 0.61 & -11\% \\
Code Llama (70B) & 0.71 & 0.61 & -15\% \\
\midrule
Average & 0.40 & 0.50 & +69\% \\
Minimum & 0.00 & 0.04 & -82\% \\
Maximum & 0.71 & 0.68 & +800\% \\
\bottomrule
\end{tabular}

\end{subtable}
\hfill
\begin{subtable}[b]{0.49\textwidth}
\centering
\caption{Argument completion, real-world dataset}
\label{tab:res-gain-correct-cd-end-real}
\begin{tabular}{llll}
\toprule
 & Vanilla & CD & Gain \\
\midrule
CodeT5+ (6B) & 0.54 & 0.57 & +7\% \\
CodeT5+ (16B) & 0.57 & 0.57 & +0\% \\
StarCoder (1B) & 0.68 & 0.61 & -11\% \\
StarCoder (3B) & 0.50 & 0.68 & +36\% \\
StarCoder (7B) & 0.57 & 0.64 & +13\% \\
StarCoder (15.5B) & 0.57 & 0.64 & +13\% \\
StarCoder2 (3B) & 0.50 & 0.68 & +36\% \\
StarCoder2 (7B) & 0.39 & 0.64 & +64\% \\
StarCoder2 (15B) & 0.50 & 0.64 & +29\% \\
DeepSeek-Coder (1.3B) & 0.50 & 0.61 & +21\% \\
DeepSeek-Coder (6.7B) & 0.71 & 0.68 & -5\% \\
DeepSeek-Coder (33B) & 0.64 & 0.68 & +6\% \\
Qwen2.5-Coder (0.5B) & 0.43 & 0.57 & +33\% \\
Qwen2.5-Coder (1.5B) & 0.61 & 0.64 & +6\% \\
Qwen2.5-Coder (3B) & 0.61 & 0.61 & +0\% \\
Qwen2.5-Coder (7B) & 0.61 & 0.64 & +6\% \\
Qwen2.5-Coder (14B) & 0.68 & 0.68 & +0\% \\
Qwen2.5-Coder (32B) & 0.46 & 0.68 & +46\% \\
Llama 3.1 (8B) & 0.39 & 0.64 & +64\% \\
Llama 3.1 (70B) & 0.00 & 0.64 & N/A \\
Code Llama (7B) & 0.57 & 0.61 & +6\% \\
Code Llama (13B) & 0.68 & 0.61 & -11\% \\
Code Llama (70B) & 0.43 & 0.61 & +42\% \\
\midrule
Average & 0.53 & 0.63 & +18\% \\
Minimum & 0.00 & 0.57 & -11\% \\
Maximum & 0.71 & 0.68 & +64\% \\
\bottomrule
\end{tabular}

\end{subtable}
\end{table}

\begin{table}[p]
\centering
\caption{Relative change of \emph{correct implementations} from vanilla generation to \emph{retrieval-augmented generation}~(RAG) combined with \emph{constrained decoding}~(CD) for both setups and both datasets. Models with a 0\% baseline score are excluded from the calculation of the average gain.}
\label{tab:res-gain-correct-rag+cd}
\footnotesize
\begin{subtable}[b]{0.49\textwidth}
\centering
\caption{Full completion, synthetic dataset}
\label{tab:res-gain-correct-rag+cd-inv-synth}
\begin{tabular}{llll}
\toprule
 & Vanilla & RAG + CD & Gain \\
\midrule
CodeT5+ (6B) & 0.04 & 0.18 & +324\% \\
CodeT5+ (16B) & 0.07 & 0.30 & +333\% \\
StarCoder (1B) & 0.03 & 0.46 & +1308\% \\
StarCoder (3B) & 0.05 & 0.47 & +884\% \\
StarCoder (7B) & 0.12 & 0.51 & +321\% \\
StarCoder (15.5B) & 0.13 & 0.43 & +223\% \\
StarCoder2 (3B) & 0.13 & 0.54 & +308\% \\
StarCoder2 (7B) & 0.10 & 0.26 & +151\% \\
StarCoder2 (15B) & 0.25 & 0.63 & +152\% \\
DeepSeek-Coder (1.3B) & 0.07 & 0.08 & +10\% \\
DeepSeek-Coder (6.7B) & 0.14 & 0.36 & +151\% \\
DeepSeek-Coder (33B) & 0.17 & 0.67 & +281\% \\
Qwen2.5-Coder (0.5B) & 0.00 & 0.17 & N/A \\
Qwen2.5-Coder (1.5B) & 0.00 & 0.11 & N/A \\
Qwen2.5-Coder (3B) & 0.00 & 0.38 & N/A \\
Qwen2.5-Coder (7B) & 0.00 & 0.46 & N/A \\
Qwen2.5-Coder (14B) & 0.00 & 0.34 & N/A \\
Qwen2.5-Coder (32B) & 0.17 & 0.42 & +148\% \\
Llama 3.1 (8B) & 0.00 & 0.42 & N/A \\
Llama 3.1 (70B) & 0.00 & 0.47 & N/A \\
Code Llama (7B) & 0.03 & 0.22 & +671\% \\
Code Llama (13B) & 0.18 & 0.07 & -61\% \\
Code Llama (70B) & 0.30 & 0.63 & +108\% \\
\midrule
Average & 0.09 & 0.37 & +332\% \\
Minimum & 0.00 & 0.07 & -61\% \\
Maximum & 0.30 & 0.67 & +1308\% \\
\bottomrule
\end{tabular}

\end{subtable}
\hfill
\begin{subtable}[b]{0.49\textwidth}
\centering
\caption{Argument completion, synthetic dataset}
\label{tab:res-gain-correct-rag+cd-end-synth}
\begin{tabular}{llll}
\toprule
 & Vanilla & RAG + CD & Gain \\
\midrule
CodeT5+ (6B) & 0.24 & 0.25 & +8\% \\
CodeT5+ (16B) & 0.21 & 0.36 & +74\% \\
StarCoder (1B) & 0.22 & 0.55 & +148\% \\
StarCoder (3B) & 0.23 & 0.51 & +126\% \\
StarCoder (7B) & 0.29 & 0.58 & +97\% \\
StarCoder (15.5B) & 0.28 & 0.48 & +68\% \\
StarCoder2 (3B) & 0.22 & 0.60 & +169\% \\
StarCoder2 (7B) & 0.19 & 0.28 & +43\% \\
StarCoder2 (15B) & 0.24 & 0.71 & +195\% \\
DeepSeek-Coder (1.3B) & 0.19 & 0.11 & -44\% \\
DeepSeek-Coder (6.7B) & 0.29 & 0.37 & +31\% \\
DeepSeek-Coder (33B) & 0.26 & 0.72 & +177\% \\
Qwen2.5-Coder (0.5B) & 0.17 & 0.40 & +137\% \\
Qwen2.5-Coder (1.5B) & 0.24 & 0.29 & +23\% \\
Qwen2.5-Coder (3B) & 0.27 & 0.61 & +128\% \\
Qwen2.5-Coder (7B) & 0.32 & 0.66 & +107\% \\
Qwen2.5-Coder (14B) & 0.47 & 0.73 & +57\% \\
Qwen2.5-Coder (32B) & 0.43 & 0.65 & +52\% \\
Llama 3.1 (8B) & 0.10 & 0.58 & +487\% \\
Llama 3.1 (70B) & 0.30 & 0.57 & +88\% \\
Code Llama (7B) & 0.05 & 0.23 & +384\% \\
Code Llama (13B) & 0.26 & 0.08 & -71\% \\
Code Llama (70B) & 0.42 & 0.70 & +66\% \\
\midrule
Average & 0.26 & 0.48 & +111\% \\
Minimum & 0.05 & 0.08 & -71\% \\
Maximum & 0.47 & 0.73 & +487\% \\
\bottomrule
\end{tabular}

\end{subtable}
\newline
\vspace*{1em}
\newline
\begin{subtable}[b]{0.49\textwidth}
\centering
\caption{Full completion, real-world dataset}
\label{tab:res-gain-correct-rag+cd-inv-real}
\begin{tabular}{llll}
\toprule
 & Vanilla & RAG + CD & Gain \\
\midrule
CodeT5+ (6B) & 0.39 & 0.36 & -9\% \\
CodeT5+ (16B) & 0.54 & 0.39 & -27\% \\
StarCoder (1B) & 0.46 & 0.43 & -8\% \\
StarCoder (3B) & 0.57 & 0.46 & -19\% \\
StarCoder (7B) & 0.61 & 0.61 & +0\% \\
StarCoder (15.5B) & 0.50 & 0.50 & +0\% \\
StarCoder2 (3B) & 0.64 & 0.50 & -22\% \\
StarCoder2 (7B) & 0.46 & 0.46 & +0\% \\
StarCoder2 (15B) & 0.54 & 0.54 & +0\% \\
DeepSeek-Coder (1.3B) & 0.57 & 0.57 & +0\% \\
DeepSeek-Coder (6.7B) & 0.54 & 0.64 & +20\% \\
DeepSeek-Coder (33B) & 0.54 & 0.68 & +27\% \\
Qwen2.5-Coder (0.5B) & 0.39 & 0.21 & -45\% \\
Qwen2.5-Coder (1.5B) & 0.00 & 0.04 & N/A \\
Qwen2.5-Coder (3B) & 0.00 & 0.43 & N/A \\
Qwen2.5-Coder (7B) & 0.07 & 0.68 & +850\% \\
Qwen2.5-Coder (14B) & 0.00 & 0.50 & N/A \\
Qwen2.5-Coder (32B) & 0.18 & 0.61 & +240\% \\
Llama 3.1 (8B) & 0.11 & 0.57 & +433\% \\
Llama 3.1 (70B) & 0.00 & 0.71 & N/A \\
Code Llama (7B) & 0.61 & 0.57 & -6\% \\
Code Llama (13B) & 0.68 & 0.57 & -16\% \\
Code Llama (70B) & 0.71 & 0.64 & -10\% \\
\midrule
Average & 0.40 & 0.51 & +74\% \\
Minimum & 0.00 & 0.04 & -45\% \\
Maximum & 0.71 & 0.71 & +850\% \\
\bottomrule
\end{tabular}

\end{subtable}
\hfill
\begin{subtable}[b]{0.49\textwidth}
\centering
\caption{Argument completion, real-world dataset}
\label{tab:res-gain-correct-rag+cd-end-real}
\begin{tabular}{llll}
\toprule
 & Vanilla & RAG + CD & Gain \\
\midrule
CodeT5+ (6B) & 0.54 & 0.50 & -7\% \\
CodeT5+ (16B) & 0.57 & 0.57 & +0\% \\
StarCoder (1B) & 0.68 & 0.61 & -11\% \\
StarCoder (3B) & 0.50 & 0.61 & +21\% \\
StarCoder (7B) & 0.57 & 0.61 & +6\% \\
StarCoder (15.5B) & 0.57 & 0.61 & +6\% \\
StarCoder2 (3B) & 0.50 & 0.61 & +21\% \\
StarCoder2 (7B) & 0.39 & 0.61 & +55\% \\
StarCoder2 (15B) & 0.50 & 0.61 & +21\% \\
DeepSeek-Coder (1.3B) & 0.50 & 0.57 & +14\% \\
DeepSeek-Coder (6.7B) & 0.71 & 0.64 & -10\% \\
DeepSeek-Coder (33B) & 0.64 & 0.64 & +0\% \\
Qwen2.5-Coder (0.5B) & 0.43 & 0.61 & +42\% \\
Qwen2.5-Coder (1.5B) & 0.61 & 0.61 & +0\% \\
Qwen2.5-Coder (3B) & 0.61 & 0.61 & +0\% \\
Qwen2.5-Coder (7B) & 0.61 & 0.64 & +6\% \\
Qwen2.5-Coder (14B) & 0.68 & 0.64 & -5\% \\
Qwen2.5-Coder (32B) & 0.46 & 0.71 & +54\% \\
Llama 3.1 (8B) & 0.39 & 0.64 & +64\% \\
Llama 3.1 (70B) & 0.00 & 0.71 & N/A \\
Code Llama (7B) & 0.57 & 0.57 & +0\% \\
Code Llama (13B) & 0.68 & 0.64 & -5\% \\
Code Llama (70B) & 0.43 & 0.61 & +42\% \\
\midrule
Average & 0.53 & 0.62 & +14\% \\
Minimum & 0.00 & 0.50 & -11\% \\
Maximum & 0.71 & 0.71 & +64\% \\
\bottomrule
\end{tabular}

\end{subtable}
\end{table}

\begin{table}[p]
\centering
\caption{Relative change of \emph{hallucinated endpoints} (full completion) and \emph{hallucinated implementations} (argument completion) from vanilla generation to \emph{retrieval-augmented generation}~(RAG) for both setups and both datasets. Models with a 0\% baseline score are excluded from the calculation of the average gain.}
\label{tab:res-gain-illegal-rag}
\footnotesize
\begin{subtable}[b]{0.49\textwidth}
\centering
\caption{Full completion, synthetic dataset}
\label{tab:res-gain-illegal-rag-inv-synth}
\begin{tabular}{llll}
\toprule
 & Vanilla & RAG & Gain \\
\midrule
CodeT5+ (6B) & 0.70 & 0.17 & -76\% \\
CodeT5+ (16B) & 0.46 & 0.05 & -90\% \\
StarCoder (1B) & 0.30 & 0.10 & -68\% \\
StarCoder (3B) & 0.33 & 0.07 & -80\% \\
StarCoder (7B) & 0.36 & 0.08 & -78\% \\
StarCoder (15.5B) & 0.23 & 0.05 & -80\% \\
StarCoder2 (3B) & 0.28 & 0.06 & -78\% \\
StarCoder2 (7B) & 0.33 & 0.06 & -83\% \\
StarCoder2 (15B) & 0.26 & 0.05 & -81\% \\
DeepSeek-Coder (1.3B) & 0.38 & 0.04 & -89\% \\
DeepSeek-Coder (6.7B) & 0.15 & 0.01 & -92\% \\
DeepSeek-Coder (33B) & 0.20 & 0.03 & -86\% \\
Qwen2.5-Coder (0.5B) & 0.00 & 0.00 & N/A \\
Qwen2.5-Coder (1.5B) & 0.00 & 0.00 & N/A \\
Qwen2.5-Coder (3B) & 0.00 & 0.00 & N/A \\
Qwen2.5-Coder (7B) & 0.00 & 0.00 & N/A \\
Qwen2.5-Coder (14B) & 0.00 & 0.00 & N/A \\
Qwen2.5-Coder (32B) & 0.10 & 0.01 & -90\% \\
Llama 3.1 (8B) & 0.00 & 0.00 & N/A \\
Llama 3.1 (70B) & 0.00 & 0.00 & N/A \\
Code Llama (7B) & 0.38 & 0.04 & -89\% \\
Code Llama (13B) & 0.30 & 0.05 & -85\% \\
Code Llama (34B) & 0.27 & 0.08 & -70\% \\
Code Llama (70B) & 0.27 & 0.04 & -85\% \\
\midrule
Average & 0.22 & 0.04 & -82\% \\
Minimum & 0.00 & 0.00 & -92\% \\
Maximum & 0.70 & 0.17 & -68\% \\
\bottomrule
\end{tabular}

\end{subtable}
\hfill
\begin{subtable}[b]{0.49\textwidth}
\centering
\caption{Argument completion, synthetic dataset}
\label{tab:res-gain-illegal-rag-end-synth}
\begin{tabular}{llll}
\toprule
 & Vanilla & RAG & Gain \\
\midrule
CodeT5+ (6B) & 0.57 & 0.03 & -95\% \\
CodeT5+ (16B) & 0.54 & 0.06 & -90\% \\
StarCoder (1B) & 0.48 & 0.25 & -47\% \\
StarCoder (3B) & 0.49 & 0.21 & -56\% \\
StarCoder (7B) & 0.35 & 0.16 & -55\% \\
StarCoder (15.5B) & 0.28 & 0.21 & -26\% \\
StarCoder2 (3B) & 0.45 & 0.20 & -56\% \\
StarCoder2 (7B) & 0.48 & 0.14 & -70\% \\
StarCoder2 (15B) & 0.35 & 0.11 & -68\% \\
DeepSeek-Coder (1.3B) & 0.50 & 0.19 & -61\% \\
DeepSeek-Coder (6.7B) & 0.30 & 0.16 & -47\% \\
DeepSeek-Coder (33B) & 0.40 & 0.07 & -84\% \\
Qwen2.5-Coder (0.5B) & 0.54 & 0.38 & -30\% \\
Qwen2.5-Coder (1.5B) & 0.52 & 0.19 & -62\% \\
Qwen2.5-Coder (3B) & 0.44 & 0.27 & -40\% \\
Qwen2.5-Coder (7B) & 0.41 & 0.12 & -70\% \\
Qwen2.5-Coder (14B) & 0.26 & 0.04 & -84\% \\
Qwen2.5-Coder (32B) & 0.28 & 0.13 & -54\% \\
Llama 3.1 (8B) & 0.34 & 0.06 & -83\% \\
Llama 3.1 (70B) & 0.26 & 0.11 & -60\% \\
Code Llama (7B) & 0.19 & 0.10 & -46\% \\
Code Llama (13B) & 0.31 & 0.07 & -76\% \\
Code Llama (34B) & 0.26 & 0.21 & -19\% \\
Code Llama (70B) & 0.24 & 0.07 & -69\% \\
\midrule
Average & 0.39 & 0.15 & -60\% \\
Minimum & 0.19 & 0.03 & -95\% \\
Maximum & 0.57 & 0.38 & -19\% \\
\bottomrule
\end{tabular}

\end{subtable}
\newline
\vspace*{1em}
\newline
\begin{subtable}[b]{0.49\textwidth}
\centering
\caption{Full completion, real-world dataset}
\label{tab:res-gain-illegal-rag-inv-real}
\begin{tabular}{llll}
\toprule
 & Vanilla & RAG & Gain \\
\midrule
CodeT5+ (6B) & 0.36 & 0.25 & -30\% \\
CodeT5+ (16B) & 0.43 & 0.18 & -58\% \\
StarCoder (1B) & 0.43 & 0.32 & -25\% \\
StarCoder (3B) & 0.25 & 0.25 & +0\% \\
StarCoder (7B) & 0.29 & 0.29 & +0\% \\
StarCoder (15.5B) & 0.29 & 0.18 & -37\% \\
StarCoder2 (3B) & 0.29 & 0.18 & -37\% \\
StarCoder2 (7B) & 0.36 & 0.32 & -10\% \\
StarCoder2 (15B) & 0.32 & 0.29 & -11\% \\
DeepSeek-Coder (1.3B) & 0.32 & 0.29 & -11\% \\
DeepSeek-Coder (6.7B) & 0.25 & 0.36 & +43\% \\
DeepSeek-Coder (33B) & 0.36 & 0.18 & -50\% \\
Qwen2.5-Coder (0.5B) & 0.11 & 0.18 & +67\% \\
Qwen2.5-Coder (1.5B) & 0.00 & 0.00 & N/A \\
Qwen2.5-Coder (3B) & 0.00 & 0.00 & N/A \\
Qwen2.5-Coder (7B) & 0.04 & 0.00 & -100\% \\
Qwen2.5-Coder (14B) & 0.00 & 0.00 & N/A \\
Qwen2.5-Coder (32B) & 0.00 & 0.00 & N/A \\
Llama 3.1 (8B) & 0.00 & 0.07 & N/A \\
Llama 3.1 (70B) & 0.00 & 0.00 & N/A \\
Code Llama (7B) & 0.18 & 0.14 & -20\% \\
Code Llama (13B) & 0.25 & 0.29 & +14\% \\
Code Llama (34B) & 0.36 & 0.36 & +0\% \\
Code Llama (70B) & 0.18 & 0.18 & +0\% \\
\midrule
Average & 0.21 & 0.18 & -15\% \\
Minimum & 0.00 & 0.00 & -100\% \\
Maximum & 0.43 & 0.36 & +67\% \\
\bottomrule
\end{tabular}

\end{subtable}
\hfill
\begin{subtable}[b]{0.49\textwidth}
\centering
\caption{Argument completion, real-world dataset}
\label{tab:res-gain-illegal-rag-end-real}
\begin{tabular}{llll}
\toprule
 & Vanilla & RAG & Gain \\
\midrule
CodeT5+ (6B) & 0.14 & 0.07 & -50\% \\
CodeT5+ (16B) & 0.29 & 0.11 & -62\% \\
StarCoder (1B) & 0.18 & 0.07 & -60\% \\
StarCoder (3B) & 0.32 & 0.07 & -78\% \\
StarCoder (7B) & 0.25 & 0.00 & -100\% \\
StarCoder (15.5B) & 0.14 & 0.00 & -100\% \\
StarCoder2 (3B) & 0.21 & 0.00 & -100\% \\
StarCoder2 (7B) & 0.25 & 0.07 & -71\% \\
StarCoder2 (15B) & 0.32 & 0.07 & -78\% \\
DeepSeek-Coder (1.3B) & 0.25 & 0.07 & -71\% \\
DeepSeek-Coder (6.7B) & 0.14 & 0.00 & -100\% \\
DeepSeek-Coder (33B) & 0.14 & 0.11 & -25\% \\
Qwen2.5-Coder (0.5B) & 0.32 & 0.18 & -44\% \\
Qwen2.5-Coder (1.5B) & 0.18 & 0.21 & +20\% \\
Qwen2.5-Coder (3B) & 0.11 & 0.11 & +0\% \\
Qwen2.5-Coder (7B) & 0.11 & 0.04 & -67\% \\
Qwen2.5-Coder (14B) & 0.14 & 0.04 & -75\% \\
Qwen2.5-Coder (32B) & 0.25 & 0.14 & -43\% \\
Llama 3.1 (8B) & 0.21 & 0.00 & -100\% \\
Llama 3.1 (70B) & 0.00 & 0.00 & N/A \\
Code Llama (7B) & 0.07 & 0.04 & -50\% \\
Code Llama (13B) & 0.11 & 0.00 & -100\% \\
Code Llama (34B) & 0.04 & 0.00 & -100\% \\
Code Llama (70B) & 0.11 & 0.04 & -67\% \\
\midrule
Average & 0.18 & 0.06 & -66\% \\
Minimum & 0.00 & 0.00 & -100\% \\
Maximum & 0.32 & 0.21 & +20\% \\
\bottomrule
\end{tabular}

\end{subtable}
\end{table}

\begin{table}[p]
\centering
\caption{Relative change of \emph{hallucinated endpoints} (full completion) and \emph{hallucinated implementations} (argument completion) from vanilla generation to \emph{constrained decoding}~(CD) for both setups and both datasets. Models with a 0\% baseline score are excluded from the calculation of the average gain.}
\label{tab:res-gain-illegal-cd}
\footnotesize
\begin{subtable}[b]{0.49\textwidth}
\centering
\caption{Full completion, synthetic dataset}
\label{tab:res-gain-illegal-cd-inv-synth}
\begin{tabular}{llll}
\toprule
 & Vanilla & CD & Gain \\
\midrule
CodeT5+ (6B) & 0.70 & 0.00 & -100\% \\
CodeT5+ (16B) & 0.46 & 0.00 & -100\% \\
StarCoder (1B) & 0.30 & 0.00 & -100\% \\
StarCoder (3B) & 0.33 & 0.00 & -100\% \\
StarCoder (7B) & 0.36 & 0.00 & -100\% \\
StarCoder (15.5B) & 0.23 & 0.00 & -100\% \\
StarCoder2 (3B) & 0.28 & 0.00 & -100\% \\
StarCoder2 (7B) & 0.33 & 0.00 & -100\% \\
StarCoder2 (15B) & 0.26 & 0.00 & -100\% \\
DeepSeek-Coder (1.3B) & 0.38 & 0.00 & -100\% \\
DeepSeek-Coder (6.7B) & 0.15 & 0.00 & -100\% \\
DeepSeek-Coder (33B) & 0.20 & 0.00 & -100\% \\
Qwen2.5-Coder (0.5B) & 0.00 & 0.00 & N/A \\
Qwen2.5-Coder (1.5B) & 0.00 & 0.00 & N/A \\
Qwen2.5-Coder (3B) & 0.00 & 0.00 & N/A \\
Qwen2.5-Coder (7B) & 0.00 & 0.00 & N/A \\
Qwen2.5-Coder (14B) & 0.00 & 0.00 & N/A \\
Qwen2.5-Coder (32B) & 0.10 & 0.00 & -100\% \\
Llama 3.1 (8B) & 0.00 & 0.00 & N/A \\
Llama 3.1 (70B) & 0.00 & 0.00 & N/A \\
Code Llama (7B) & 0.38 & 0.00 & -100\% \\
Code Llama (13B) & 0.30 & 0.00 & -100\% \\
Code Llama (70B) & 0.27 & 0.00 & -100\% \\
\midrule
Average & 0.22 & 0.00 & -100\% \\
Minimum & 0.00 & 0.00 & -100\% \\
Maximum & 0.70 & 0.00 & -100\% \\
\bottomrule
\end{tabular}

\end{subtable}
\hfill
\begin{subtable}[b]{0.49\textwidth}
\centering
\caption{Argument completion, synthetic dataset}
\label{tab:res-gain-illegal-cd-end-synth}
\begin{tabular}{llll}
\toprule
 & Vanilla & CD & Gain \\
\midrule
CodeT5+ (6B) & 0.57 & 0.00 & -100\% \\
CodeT5+ (16B) & 0.54 & 0.00 & -100\% \\
StarCoder (1B) & 0.48 & 0.00 & -100\% \\
StarCoder (3B) & 0.49 & 0.00 & -100\% \\
StarCoder (7B) & 0.35 & 0.00 & -100\% \\
StarCoder (15.5B) & 0.28 & 0.00 & -100\% \\
StarCoder2 (3B) & 0.45 & 0.00 & -100\% \\
StarCoder2 (7B) & 0.48 & 0.00 & -100\% \\
StarCoder2 (15B) & 0.35 & 0.00 & -100\% \\
DeepSeek-Coder (1.3B) & 0.50 & 0.00 & -100\% \\
DeepSeek-Coder (6.7B) & 0.30 & 0.00 & -100\% \\
DeepSeek-Coder (33B) & 0.40 & 0.00 & -100\% \\
Qwen2.5-Coder (0.5B) & 0.54 & 0.00 & -100\% \\
Qwen2.5-Coder (1.5B) & 0.52 & 0.00 & -100\% \\
Qwen2.5-Coder (3B) & 0.44 & 0.00 & -100\% \\
Qwen2.5-Coder (7B) & 0.41 & 0.00 & -100\% \\
Qwen2.5-Coder (14B) & 0.26 & 0.00 & -100\% \\
Qwen2.5-Coder (32B) & 0.28 & 0.00 & -100\% \\
Llama 3.1 (8B) & 0.34 & 0.00 & -100\% \\
Llama 3.1 (70B) & 0.26 & 0.00 & -100\% \\
Code Llama (7B) & 0.19 & 0.00 & -100\% \\
Code Llama (13B) & 0.31 & 0.00 & -100\% \\
Code Llama (70B) & 0.24 & 0.00 & -100\% \\
\midrule
Average & 0.39 & 0.00 & -100\% \\
Minimum & 0.19 & 0.00 & -100\% \\
Maximum & 0.57 & 0.00 & -100\% \\
\bottomrule
\end{tabular}

\end{subtable}
\newline
\vspace*{1em}
\newline
\begin{subtable}[b]{0.49\textwidth}
\centering
\caption{Full completion, real-world dataset}
\label{tab:res-gain-illegal-cd-inv-real}
\begin{tabular}{llll}
\toprule
 & Vanilla & CD & Gain \\
\midrule
CodeT5+ (6B) & 0.36 & 0.00 & -100\% \\
CodeT5+ (16B) & 0.43 & 0.00 & -100\% \\
StarCoder (1B) & 0.43 & 0.00 & -100\% \\
StarCoder (3B) & 0.25 & 0.00 & -100\% \\
StarCoder (7B) & 0.29 & 0.00 & -100\% \\
StarCoder (15.5B) & 0.29 & 0.00 & -100\% \\
StarCoder2 (3B) & 0.29 & 0.00 & -100\% \\
StarCoder2 (7B) & 0.36 & 0.00 & -100\% \\
StarCoder2 (15B) & 0.32 & 0.00 & -100\% \\
DeepSeek-Coder (1.3B) & 0.32 & 0.00 & -100\% \\
DeepSeek-Coder (6.7B) & 0.25 & 0.00 & -100\% \\
DeepSeek-Coder (33B) & 0.36 & 0.00 & -100\% \\
Qwen2.5-Coder (0.5B) & 0.11 & 0.00 & -100\% \\
Qwen2.5-Coder (1.5B) & 0.00 & 0.00 & N/A \\
Qwen2.5-Coder (3B) & 0.00 & 0.00 & N/A \\
Qwen2.5-Coder (7B) & 0.04 & 0.00 & -100\% \\
Qwen2.5-Coder (14B) & 0.00 & 0.00 & N/A \\
Qwen2.5-Coder (32B) & 0.00 & 0.00 & N/A \\
Llama 3.1 (8B) & 0.00 & 0.00 & N/A \\
Llama 3.1 (70B) & 0.00 & 0.00 & N/A \\
Code Llama (7B) & 0.18 & 0.00 & -100\% \\
Code Llama (13B) & 0.25 & 0.00 & -100\% \\
Code Llama (70B) & 0.18 & 0.00 & -100\% \\
\midrule
Average & 0.20 & 0.00 & -100\% \\
Minimum & 0.00 & 0.00 & -100\% \\
Maximum & 0.43 & 0.00 & -100\% \\
\bottomrule
\end{tabular}

\end{subtable}
\hfill
\begin{subtable}[b]{0.49\textwidth}
\centering
\caption{Argument completion, real-world dataset}
\label{tab:res-gain-illegal-cd-end-real}
\begin{tabular}{llll}
\toprule
 & Vanilla & CD & Gain \\
\midrule
CodeT5+ (6B) & 0.14 & 0.00 & -100\% \\
CodeT5+ (16B) & 0.29 & 0.00 & -100\% \\
StarCoder (1B) & 0.18 & 0.00 & -100\% \\
StarCoder (3B) & 0.32 & 0.00 & -100\% \\
StarCoder (7B) & 0.25 & 0.00 & -100\% \\
StarCoder (15.5B) & 0.14 & 0.00 & -100\% \\
StarCoder2 (3B) & 0.21 & 0.00 & -100\% \\
StarCoder2 (7B) & 0.25 & 0.00 & -100\% \\
StarCoder2 (15B) & 0.32 & 0.00 & -100\% \\
DeepSeek-Coder (1.3B) & 0.25 & 0.00 & -100\% \\
DeepSeek-Coder (6.7B) & 0.14 & 0.00 & -100\% \\
DeepSeek-Coder (33B) & 0.14 & 0.00 & -100\% \\
Qwen2.5-Coder (0.5B) & 0.32 & 0.00 & -100\% \\
Qwen2.5-Coder (1.5B) & 0.18 & 0.00 & -100\% \\
Qwen2.5-Coder (3B) & 0.11 & 0.00 & -100\% \\
Qwen2.5-Coder (7B) & 0.11 & 0.00 & -100\% \\
Qwen2.5-Coder (14B) & 0.14 & 0.00 & -100\% \\
Qwen2.5-Coder (32B) & 0.25 & 0.00 & -100\% \\
Llama 3.1 (8B) & 0.21 & 0.00 & -100\% \\
Llama 3.1 (70B) & 0.00 & 0.00 & N/A \\
Code Llama (7B) & 0.07 & 0.00 & -100\% \\
Code Llama (13B) & 0.11 & 0.00 & -100\% \\
Code Llama (70B) & 0.11 & 0.00 & -100\% \\
\midrule
Average & 0.18 & 0.00 & -100\% \\
Minimum & 0.00 & 0.00 & -100\% \\
Maximum & 0.32 & 0.00 & -100\% \\
\bottomrule
\end{tabular}

\end{subtable}
\end{table}

\begin{table}[p]
\centering
\caption{Relative change of \emph{hallucinated endpoints} (full completion) and \emph{hallucinated implementations} (argument completion) from vanilla generation to \emph{retrieval-augmented generation}~(RAG) combined with \emph{constrained decoding}~(CD) for both setups and both datasets. Models with a 0\% baseline score are excluded from the calculation of the average gain.}
\label{tab:res-gain-illegal-rag+cd}
\footnotesize
\begin{subtable}[b]{0.49\textwidth}
\centering
\caption{Full completion, synthetic dataset}
\label{tab:res-gain-illegal-rag+cd-inv-synth}
\begin{tabular}{llll}
\toprule
 & Vanilla & RAG + CD & Gain \\
\midrule
CodeT5+ (6B) & 0.70 & 0.00 & -100\% \\
CodeT5+ (16B) & 0.46 & 0.00 & -100\% \\
StarCoder (1B) & 0.30 & 0.00 & -100\% \\
StarCoder (3B) & 0.33 & 0.00 & -100\% \\
StarCoder (7B) & 0.36 & 0.00 & -100\% \\
StarCoder (15.5B) & 0.23 & 0.00 & -100\% \\
StarCoder2 (3B) & 0.28 & 0.00 & -100\% \\
StarCoder2 (7B) & 0.33 & 0.00 & -100\% \\
StarCoder2 (15B) & 0.26 & 0.00 & -100\% \\
DeepSeek-Coder (1.3B) & 0.38 & 0.00 & -100\% \\
DeepSeek-Coder (6.7B) & 0.15 & 0.00 & -100\% \\
DeepSeek-Coder (33B) & 0.20 & 0.00 & -100\% \\
Qwen2.5-Coder (0.5B) & 0.00 & 0.00 & N/A \\
Qwen2.5-Coder (1.5B) & 0.00 & 0.00 & N/A \\
Qwen2.5-Coder (3B) & 0.00 & 0.00 & N/A \\
Qwen2.5-Coder (7B) & 0.00 & 0.00 & N/A \\
Qwen2.5-Coder (14B) & 0.00 & 0.00 & N/A \\
Qwen2.5-Coder (32B) & 0.10 & 0.00 & -100\% \\
Llama 3.1 (8B) & 0.00 & 0.00 & N/A \\
Llama 3.1 (70B) & 0.00 & 0.00 & N/A \\
Code Llama (7B) & 0.38 & 0.00 & -100\% \\
Code Llama (13B) & 0.30 & 0.00 & -100\% \\
Code Llama (70B) & 0.27 & 0.00 & -100\% \\
\midrule
Average & 0.22 & 0.00 & -100\% \\
Minimum & 0.00 & 0.00 & -100\% \\
Maximum & 0.70 & 0.00 & -100\% \\
\bottomrule
\end{tabular}

\end{subtable}
\hfill
\begin{subtable}[b]{0.49\textwidth}
\centering
\caption{Argument completion, synthetic dataset}
\label{tab:res-gain-illegal-rag+cd-end-synth}
\begin{tabular}{llll}
\toprule
 & Vanilla & RAG + CD & Gain \\
\midrule
CodeT5+ (6B) & 0.57 & 0.00 & -100\% \\
CodeT5+ (16B) & 0.54 & 0.00 & -100\% \\
StarCoder (1B) & 0.48 & 0.00 & -100\% \\
StarCoder (3B) & 0.49 & 0.00 & -100\% \\
StarCoder (7B) & 0.35 & 0.00 & -100\% \\
StarCoder (15.5B) & 0.28 & 0.00 & -100\% \\
StarCoder2 (3B) & 0.45 & 0.00 & -100\% \\
StarCoder2 (7B) & 0.48 & 0.00 & -100\% \\
StarCoder2 (15B) & 0.35 & 0.00 & -100\% \\
DeepSeek-Coder (1.3B) & 0.50 & 0.00 & -100\% \\
DeepSeek-Coder (6.7B) & 0.30 & 0.00 & -100\% \\
DeepSeek-Coder (33B) & 0.40 & 0.00 & -100\% \\
Qwen2.5-Coder (0.5B) & 0.54 & 0.00 & -100\% \\
Qwen2.5-Coder (1.5B) & 0.52 & 0.00 & -100\% \\
Qwen2.5-Coder (3B) & 0.44 & 0.00 & -100\% \\
Qwen2.5-Coder (7B) & 0.41 & 0.00 & -100\% \\
Qwen2.5-Coder (14B) & 0.26 & 0.00 & -100\% \\
Qwen2.5-Coder (32B) & 0.28 & 0.00 & -100\% \\
Llama 3.1 (8B) & 0.34 & 0.00 & -100\% \\
Llama 3.1 (70B) & 0.26 & 0.00 & -100\% \\
Code Llama (7B) & 0.19 & 0.00 & -100\% \\
Code Llama (13B) & 0.31 & 0.00 & -100\% \\
Code Llama (70B) & 0.24 & 0.00 & -100\% \\
\midrule
Average & 0.39 & 0.00 & -100\% \\
Minimum & 0.19 & 0.00 & -100\% \\
Maximum & 0.57 & 0.00 & -100\% \\
\bottomrule
\end{tabular}

\end{subtable}
\newline
\vspace*{1em}
\newline
\begin{subtable}[b]{0.49\textwidth}
\centering
\caption{Full completion, real-world dataset}
\label{tab:res-gain-illegal-rag+cd-inv-real}
\begin{tabular}{llll}
\toprule
 & Vanilla & RAG + CD & Gain \\
\midrule
CodeT5+ (6B) & 0.36 & 0.00 & -100\% \\
CodeT5+ (16B) & 0.43 & 0.00 & -100\% \\
StarCoder (1B) & 0.43 & 0.00 & -100\% \\
StarCoder (3B) & 0.25 & 0.00 & -100\% \\
StarCoder (7B) & 0.29 & 0.00 & -100\% \\
StarCoder (15.5B) & 0.29 & 0.00 & -100\% \\
StarCoder2 (3B) & 0.29 & 0.00 & -100\% \\
StarCoder2 (7B) & 0.36 & 0.00 & -100\% \\
StarCoder2 (15B) & 0.32 & 0.00 & -100\% \\
DeepSeek-Coder (1.3B) & 0.32 & 0.00 & -100\% \\
DeepSeek-Coder (6.7B) & 0.25 & 0.00 & -100\% \\
DeepSeek-Coder (33B) & 0.36 & 0.00 & -100\% \\
Qwen2.5-Coder (0.5B) & 0.11 & 0.00 & -100\% \\
Qwen2.5-Coder (1.5B) & 0.00 & 0.00 & N/A \\
Qwen2.5-Coder (3B) & 0.00 & 0.00 & N/A \\
Qwen2.5-Coder (7B) & 0.04 & 0.00 & -100\% \\
Qwen2.5-Coder (14B) & 0.00 & 0.00 & N/A \\
Qwen2.5-Coder (32B) & 0.00 & 0.00 & N/A \\
Llama 3.1 (8B) & 0.00 & 0.00 & N/A \\
Llama 3.1 (70B) & 0.00 & 0.00 & N/A \\
Code Llama (7B) & 0.18 & 0.00 & -100\% \\
Code Llama (13B) & 0.25 & 0.00 & -100\% \\
Code Llama (70B) & 0.18 & 0.00 & -100\% \\
\midrule
Average & 0.20 & 0.00 & -100\% \\
Minimum & 0.00 & 0.00 & -100\% \\
Maximum & 0.43 & 0.00 & -100\% \\
\bottomrule
\end{tabular}

\end{subtable}
\hfill
\begin{subtable}[b]{0.49\textwidth}
\centering
\caption{Argument completion, real-world dataset}
\label{tab:res-gain-illegal-rag+cd-end-real}
\begin{tabular}{llll}
\toprule
 & Vanilla & RAG + CD & Gain \\
\midrule
CodeT5+ (6B) & 0.14 & 0.00 & -100\% \\
CodeT5+ (16B) & 0.29 & 0.00 & -100\% \\
StarCoder (1B) & 0.18 & 0.00 & -100\% \\
StarCoder (3B) & 0.32 & 0.00 & -100\% \\
StarCoder (7B) & 0.25 & 0.00 & -100\% \\
StarCoder (15.5B) & 0.14 & 0.00 & -100\% \\
StarCoder2 (3B) & 0.21 & 0.00 & -100\% \\
StarCoder2 (7B) & 0.25 & 0.00 & -100\% \\
StarCoder2 (15B) & 0.32 & 0.00 & -100\% \\
DeepSeek-Coder (1.3B) & 0.25 & 0.00 & -100\% \\
DeepSeek-Coder (6.7B) & 0.14 & 0.00 & -100\% \\
DeepSeek-Coder (33B) & 0.14 & 0.00 & -100\% \\
Qwen2.5-Coder (0.5B) & 0.32 & 0.00 & -100\% \\
Qwen2.5-Coder (1.5B) & 0.18 & 0.00 & -100\% \\
Qwen2.5-Coder (3B) & 0.11 & 0.00 & -100\% \\
Qwen2.5-Coder (7B) & 0.11 & 0.00 & -100\% \\
Qwen2.5-Coder (14B) & 0.14 & 0.00 & -100\% \\
Qwen2.5-Coder (32B) & 0.25 & 0.00 & -100\% \\
Llama 3.1 (8B) & 0.21 & 0.00 & -100\% \\
Llama 3.1 (70B) & 0.00 & 0.00 & N/A \\
Code Llama (7B) & 0.07 & 0.00 & -100\% \\
Code Llama (13B) & 0.11 & 0.00 & -100\% \\
Code Llama (70B) & 0.11 & 0.00 & -100\% \\
\midrule
Average & 0.18 & 0.00 & -100\% \\
Minimum & 0.00 & 0.00 & -100\% \\
Maximum & 0.32 & 0.00 & -100\% \\
\bottomrule
\end{tabular}

\end{subtable}
\end{table}

\end{document}